\documentclass[aps,prx,twocolumn,superscriptaddress,showpacs,floatfix]{revtex4-2}
\usepackage{comment}
\usepackage{physics}
\usepackage{amsmath,amsthm,amssymb,amsfonts, color, comment, graphicx}
\usepackage{xcolor}
\usepackage{float}
\usepackage{graphicx}
\usepackage{dcolumn}
\usepackage{bm}
\UseRawInputEncoding 
\usepackage{dsfont}
\usepackage{hyperref}

\newcommand{\bea}{\begin{eqnarray}}
\newcommand{\eea}{\end{eqnarray}}

\newcommand{\bb}{\mathbf{b}}
\newcommand{\br}{\mathbf{r}}

\newcommand{\bGa}{\boldsymbol{\Gamma}}

\def\bcao{BaCo$_2$(AsO$_4$)$_2$}
\def\ncpo{Na$_2$BaCo(PO$_4$)$_2$}
\def\ybmg{YbMgGaO$_4$}
\def\kcpo{K$_2$Co(SeO$_3$)$_2$}
\def\ncto{Na$_2$Co$_2$TeO$_6$}
\def\bcpo{BaCo$_2$(PO$_4$)$_2$}
\def\ncso{Na$_3$Co$_2$SbO$_6$}
\def\ybzn{YbZn$_2$GaO$_5$}

\newcommand{\be}{\begin{equation}}
\newcommand{\ee}{\end{equation}}
\newcommand{\bk}{{{\bf{k}}}}

\newcommand{\bK}{{{\bf{K}}}}

\newcommand{\bQ}{{{\bf{Q}}}}
\newcommand{\bM}{{{\bf{M}}}}

\newcommand{\bd}{{{\boldsymbol{\delta}}}}

\newcommand{\brf}{{{\bar{f}}}}
\newcommand{\brw}{{{\bar{w}}}}

\newcommand{\beal}{\begin{align}}
\newcommand{\eeal}{\end{align}}

\newcommand{\upa}{\uparrow}
\newcommand{\dna}{\downarrow}

\def\l{\ell}

\newcommand{\btjstrw}{\mathrel{{\rotatebox[origin=c]{90}
{$\bowtie$}}\kern-0.18em\raisebox{-.95ex}{$\bullet$}
\kern-0.5em\raisebox{.97ex}{$\bullet$}
\kern-1.12em\raisebox{.97ex}{$\bullet$}
\kern-0.52em\raisebox{-.95ex}{$\bullet$}}}

\newcommand{\btjnbrR}{{\mathrel{\rotatebox[origin=c]{90}
{$\bowtie$}}\kern-0.22em\raisebox{.9ex}{$\bullet$}
\kern-1.em\raisebox{-.8ex}{$\bullet$}}}
\newcommand{\btjnbrL}{{\mathrel{\rotatebox[origin=c]{90}
{$\bowtie$}}\kern-0.22em\raisebox{-.8ex}{$\bullet$}
\kern-1.em\raisebox{+.9ex}{$\bullet$}}}

\def\a{\alpha}
\def\b{\beta}
\def\c{\chi}
\def\d{\delta}
\def\e{\epsilon}

\def\g{\gamma}

\def\l{\lambda}

\def\s{\sigma}
\def\t{\tau}

\def\w{\omega}

\def\D{\Delta}

\def\W{\Omega}

\def\md{{\mathcal{D}}}

\def\mg{{\mathcal{G}}}
\def\mh{{\mathcal{H}}}

\def\mj{{\mathcal{J}}}

\def\mo{{\mathcal{O}}}

\def\ms{{\mathcal{S}}}

\def\mz{{\mathcal{Z}}}

\def\ua{\uparrow}
\def\da{\downarrow}

\usepackage{graphicx}
                                                         
\usepackage[caption=false]{subfig}

\newcolumntype{P}[1]{>{\centering\arraybackslash}p{#1}}

\begin{document}
\title{Spin dynamics of an easy-plane Dirac spin liquid in a frustrated XY model: Application to honeycomb cobaltates}

\author{Anjishnu Bose}
\email{anjishnu.bose@mail.utoronto.ca}
\affiliation{Department of Physics, University of Toronto, 60 St. George Street, Toronto, ON, M5S 1A7 Canada}

\author{Arun Paramekanti}
\email{arun.paramekanti@utoronto.ca}
\affiliation{Department of Physics, University of Toronto, 60 St. George Street, Toronto, ON, M5S 1A7 Canada}

\date{\today}
\begin{abstract}
Recent work has shown that the honeycomb lattice spin-$1/2$ $J_1$-$J_3$ XY model, with  nearest-neighbor ferromagnetic 
exchange $J_1$ and frustration induced by third-neighbor antiferromagnetic exchange $J_3$, may be relevant to a wide
range of cobaltate materials. We explore a variational Monte Carlo study of Gutzwiller projected wavefunctions for 
this model and show that an easy-plane Dirac spin liquid (DSL) is a viable `parent' state for the
competing magnetic orders observed in these materials, including ferromagnetic, zig-zag, spiral, and double zig-zag 
orders at intermediate frustration, and show that such broken symmetry states can be easily polarized by a weak in-plane 
magnetic field consistent with experiments. We formulate a modified parton theory 
for such frustrated spin models, and explore the potential instabilities of the DSL due to residual parton interactions 
within a random phase approximation (RPA), both at zero magnetic field and in a nonzero in-plane field.
The broken symmetry states which emerge in the vicinity of this Dirac spin liquid include 
ferromagnetic, zig-zag, and incommensurate spiral orders, with a phase diagram which is consistent with VMC and density matrix 
renormalization group studies. We calculate the dynamical spin response of the easy-plane DSL, including RPA corrections, near the boundary 
of the ordered states, and present results for THz spectroscopy and inelastic neutron scattering, at zero field as well as 
in an in-plane magnetic field, and discuss experimental implications.
\end{abstract}
\maketitle

\section{Introduction}

The exploration of frustrated quantum magnets has led to a wide variety of materials which exhibit magnetic ordering at a temperature $T_N \! \ll \! \theta_{CW}$, where the Curie-Weiss temperature $\theta_{CW}$ is the scale at which local spin correlations are established \cite{QSL_Moessner2019, QSLreview_Broholm2020, QSLreview_McQueen2021, QSLreview_Savary_2017}. However, understanding the spin dynamics of weakly ordered quantum spin-$1/2$ systems in the ordered phase $T \!<\! T_N$ as well as in the disordered regime $T_N \!<\! T \! < \! \theta_{CW}$ remains a challenge.

The conventional approach to study spin dynamics in the ordered state is to use linear spin-wave theory (LSWT) 
to capture sharp magnon modes, and incorporate magnon interactions to go beyond LSWT \cite{magnon1992, magnon2024, Wang2024, Chubukov_largeS1994, Chubokov_largeSHubbard, Chubokov_largeS1992, Chernyshev2017}. While such an approach may be formally justified within a $1/S$ expansion, it is not controlled for spin-$1/2$ systems and thus not necessarily always applicable to quantum magnets with weak magnetic order. At temperatures $T \!>\! T_N$, a semiclassical approach using Landau-Lifshitz equations \cite{Lakshmanan_2011}
averaged over a thermal ensemble of initial conditions generated by classical Monte Carlo simulations has been used to model neutron scattering data \cite{YBK_magnon2023, BCAO_Broholm2023, YBK_spinice2022}. One might reasonably question the validity of this approach to describe spin-$1/2$ systems. Moreover, it is unclear where to draw the boundary, in temperature, between quantum fluctuation versus thermal fluctuation dominated regimes. 
Finally, numerical many-body methods such as density-matrix renormalization group (DMRG) \cite{dmrg_schwol, dmrg_Catarina_2023}, variational Monte Carlo (VMC) \cite{becca_sorella_2017, song2024neuralquantumstatesvariational}, or pseudo-fermion functional renormalization group (pfFRG) \cite{pffrg_Bfield, pffrg_rev, pffrg_pyrocholore} techniques provide useful insights but are computationally expensive to study dynamics.
It is thus useful to ask if there is an alternative route to exploring such systems by starting from a
quantum spin liquid viewpoint even if the ground state is eventually not a quantum spin liquid. Philosophically, such an approach is akin to using a Fermi liquid ground state to explore weak superconducting instabilities in metals \cite{Coleman_2015, BCS1957, Anderson1966}, 
although of course, Fermi liquid theory is on much firmer footing than the theory of gapless spin liquids \cite{DSL_Wen2004, ASL_Hermele2005, DiracMonopoles_PRX2020, Song2019}. Here we ask if there is a possible underlying quantum spin liquid as we extrapolate $T \!\to\! 0$ from temperatures $T \!>\! T_N$, 
which, when perturbed by residual interactions, yields the weakly ordered ground state. We do not aim for a controlled calculation at this stage, noting it as a useful direction for future work.

We explore an approach where we take the spin Hamiltonian $H$ and split it into two parts $H = (1-\alpha) H + \alpha H$, with the understanding that the first term will be treated `classically' within Weiss mean-field like approach, while the second term will be treated `quantum mechanically' using a spin liquid parton theory - for concreteness we consider the partons to be Abrikosov fermions \cite{QSLgauge_XGWenBook}. In a path integral approach, this amounts to Hubbard-Stratonovitch decomposition of the spin exchange Hamiltonian in two channels. In the limit $\alpha=0$, this mean field decomposition recovers the classical spin model phase diagram of $H$ while in the limit $\alpha=1$ the mean field solution is a spin liquid ground state with no magnetic order. The phenomenological parameter $\alpha$ thus controls the strength of quantum fluctuations, akin to $1/S$ in LSWT, and tuning
$\alpha$ interpolates between classically ordered states and a quantum spin liquid.
Going beyond mean field theory for the first term amounts to incorporating order parameter fluctuations, e.g. within a random phase approximation (RPA), while going beyond mean field theory in the second term leads to a 
theory of matter coupled to compact gauge fields \cite{QSLgauge_Lee2014, QSLgauge_XGWenBook, QSLreview_Broholm2020, QSLreview_McQueen2021, QSLreview_Savary_2017}. For a fixed $0 < \alpha < 1$, the full theory including both contributions describes 
partons coupled to a fluctuating order parameter and gauge fields. 

An exact calculation of the partition function in the path integral framework should yield the final answer for any observable which cannot depend on the arbitrary parameter $\alpha$. However, an approximate calculation can result in the free energy depending on $\alpha$, suggesting an optimal decomposition parameter $\alpha^*$ at which this free energy is minimized. In this approach, one can naturally tune between pure classical order ($\alpha=0$) to a purely quantum liquid ($\alpha=1$), with an intermediate optimal \(\a^*\) incorporating both aspects. Here, we will simply fix an optimal $\alpha^*$ to recover salient features of the ground state phase diagram of the frustrated spin model. We then use this to compute the dynamical spin response of the system,
showing that it captures many of the qualitative features such as sharp peaks coexisting with
broad continuum scattering observed in Terahertz (THz) spectroscopy and inelastic neutron scattering.
A similar renormalization of quantum versus classical effects of spin exchange has been considered in old work studying spin dynamics in the doped cuprate superconductors \cite{Lee1997, Lee2001, Lee1999_boson}. In the absence of gauge fluctuations, our model is similar to spin-fermion models which have been previously explored for magnetic quantum critical points in metals
\cite{Chubokov_spinFermion2000, Chubokov_spinFermion2003, Starykh_2020}.

In this paper, we apply this technique to a class of honeycomb cobaltate materials which include {\bcao} \cite{BCAO_regnault1977,BCAO_Regnault2018,BCAO_Zhong2019,Armitage2022,BCAO_Broholm2023,BCAO_thermalcond_Li2022}, 
{\bcpo} \cite{BCPO_Nair2018}, {\ncto} \cite{ncto_simonet2016,ncto_ncso_stock2020,Kappaxy_NCoTeO_Park2022,LANL_ncto2022,kappa_NCTO_Sun_PRB_2023}, 
and {\ncso} \cite{ncso_mcguire2019,ncto_ncso_stock2020}. The \(d^7\) materials were initially proposed to host the elusive Kitaev spin liquid \cite{CoKitaev_Liu2018, CoKitaev_Liu2020}. However, a series of \textit{ab initio} studies \cite{Abinitio_Das2021, Abinitio_HSKim2022, Abinitio_Streltsov2022} and analysis of neutron scattering spectra using semiclassical spin dynamics 
on \bcao \cite{BCAO_Broholm2023} concluded that a frustrated easy-plane \(J_1\)-\(J_3\) XXZ model was a better starting point, with ferromagnetic \(J_1\) and antiferromagnetic \(J_3\). The quantum spin-\(1/2\) XXZ model has been studied using DMRG, VMC, and pfFRG, and the resulting phase diagrams from all methods agree in the small and large \(J_3/J_1\) limit, where they find ferromagnet and zig-zag phases respectively. 
However, the intermediate regime is still unclear and shows signs of weak orderings like Ising Neel \cite{Chernyshev2023, spinwaveXXZ_chernyshev_PRB2022}, incommensurate \cite{j1j3_watanabe2022frustrated, j1j2j3_Fouet2001}, double zig-zag \cite{Asim2024}, but also displays gapless spin-liquid like characteristics \cite{bose_dirac}. On the experimental side, incommensurate ordering has been reported through magnetic Bragg peaks \cite{BCAO_Broholm2023}. However a spin-liquid like continuum has also been reported in the THz response \cite{Armitage2022}, as well as a residual metallic-like thermal conductivity 
\(\kappa \propto T\) \cite{kappa_NCTO_Sun_PRB_2023, Kappaxy_NCoTeO_Park2022, BCAO_thermalcond_Li2022}, indicative of gapless excitations.
Our main result in this paper is the success of such a modified RPA enhanced mean-field theory approach in capturing the spin dynamics of this frustrated honeycomb spin model in a weakly ordered phase. We first build on previous VMC analysis to show the existence of a regime in the phase diagram where the ground state is weakly ordered, but with numerous other competing states in proximity, one of them being the DSL. Studying the RPA instabilities of this DSL reproduces a qualitatively similar phase diagram. Furthermore, we show the RPA corrected dynamics of such a DSL also replicate many salient features observed in THz spectroscopy \cite{Armitage2022} and neutron scattering experiments \cite{BCAO_Broholm2023} on \bcao, while also providing insights into possible nuclear magnetic resonance (NMR) experiments. Lastly, we repeat the analysis in the presence of an applied in-plane Zeeman field, and find good agreement of the saturation field strength, as well as the field THz and neutron scattering measurements done in a magnetic field. We expect that in symmetry broken
phases with magnetic order which gaps out the partons of the DSL (e.g., XY ferromagnetic order), the monopoles of the gauge field will proliferate and
lead to confinement \cite{polyakov}, while in the spin liquid or symmetry broken phases with gapless partons, the impact of gauge fluctuations and 
monopoles has to be considered on a case by case basis which we do not do in this paper.

\section{Methodology} \label{methodology}

Consider a general lattice model of spin-\(1/2\) 
\begin{equation}
    \label{spin ham main}
    \mh = \sum_{\bd}J_{ij}^{ab}(\bd)\sum_{\br}  S_i^a(\br)S_j^b(\br+\bd)\,,
\end{equation}
where \(i, j\) represent sublattices, \(\br\) marks the unit cell \textit{s.t.} \((\br, i)\) and \((\br+\bd, j)\) are neighboring sites, while \(a, b\) indicate spin-directions. Now, the spin-operator can be decomposed in terms of spin-\(1/2\) fermionic spinons \(S_i^a(\br, \t) = \frac{1}{2}f^{\dagger}_{i, \a}(\br, \t)\cdot\s^{a}_{\a\b}\cdot f_{i, \b}(\br, \t)\), under the constraint that \(\sum_{\a}f_{i, \a}^{\dagger}(\br, \t)f_{i, \a}(\br, \t)=1\) which can be imposed using a Lagrange multiplier \(a_0(\br, \t)\). Hence, The total parton action looks like (replacing \(\mo^{\dagger}\) with \(\bar{\mo}\) in the path-integral)
\begin{equation}
    \label{parton action main}
    \ms[\brf, f, a_0]\!=\!\int\!\!d\t \left\{\sum_{\br}\brf_{i, \a}\left[\partial_{\t}-a_0\right]f_{i, \a} + \mh\right\}\,,
\end{equation}
and the partition function being \(\mz = \int \md[\brf, f\,,a_0] \:e^{-\ms[\brf, f, a_0]}\). Such a parton action involves a Hamiltonian with a four-fermion interaction, which can be decoupled using Hubbard-Stratonovich (HS) fields.
However, this decoupling can be done in multiple channels : on-site magnetic Weiss fields \(h_i^a(\br)\), or parton hopping fields living on bonds \(w_{ij}^{\a\b}(\br+\bd/2)\). The magnetic channel Hubbard Stratonovich transformation on a single bond \((i, j)\) looks like
\begin{equation}
    \label{magnetic HS bond}
    \exp\left(-\mh_{ij}\right) = \int dh \exp\left(h_i^a (J^{-1})_{ij}^{ab} h_j^b - 2S_i^a h_i^a\right)\,,
\end{equation}
while in the hopping channel it looks like
\begin{equation}
    \label{hopping HS bond}
    \begin{split}
     \exp\left(-\mh_{ij}\right)& = \int d\bar{w} dw\exp\bigg(\frac{1}{4} J_{ij}^{ab} \s^a_{\a\b}\s^{b}_{\g\d}\bigg[\\
     &\brf_{i, \a}f_{j, \d}\bar{w}_{ij}^{\b\g} + \brf_{j, \g}f_{i, \b}w_{ij}^{\a\d} - w_{ij}^{\a\d}\bar{w}_{ij}^{\b\g} \bigg]\bigg)\,.   
    \end{split}
\end{equation}
When \(h_i^a(\br, \t)\) condenses breaking spin-rotation symmetry, the spins gain an expectation value in the ground state with the same pattern as the Weiss fields \(\sum_{\bd}J_{ij}^{ab}(\bd)\expval{S_j^b(\br)} = \expval{h_i^a(\br)}\). Meanwhile, if \(w_{ij}^{\a\b}(\br, \t)\) condense, we get the fermions hopping on the lattice in some frozen background gauge flux (the same gauge redundancy which arises when writing the spins using fermions) determined by the phases of \(\expval{w_{ij}^{\a\b}}\) \cite{QSLgauge_XGWenBook, QSLreview_Savary_2017}. These ground state expectation values of the condensates can equivalently be calculated in self-consistent mean-field theory as well.\par 

Now, in general, one can perform this decomposition on each interacting bond with some relative ratios as \(\mh_{ij} \xrightarrow{MFT} (1-\a_{ij})\mh^{w}_{ij} + \a_{ij}\mh^{h}_{ij}\) where \(\a_{ij}\in [0, 1]\) captures increasing quantum corrections (\(\a_{ij}>0\)) starting from purely classical physics (\(\a_{ij}=0\)). The full path integral is over the fermionic spinons \(f_{i, \a}\), the Lagrange multiplier \(a_0\) (forming the temporal part of the gauge field), the hopping fields \(w_{ij}^{\a\b}\) (which can be split into an absolute value and a phase, with the phase providing the spatial component of the gauge field), and the Weiss fields \(h_i^a\). As noted earlier, if calculated \emph{exactly}, the final partition function should be independent of the set of all  \(\a_{ij}\) parameters denoted by \(\{\a\}\). However, if we simplify by assuming that the \(w_{ij}^{\a\b}\) fields are condensed around their mean-field values (thereby suppressing gauge fluctuations altogether), then the resulting effective action in general will depend on \(\{\a\}\), \(\mz_{\mathrm{eff}}[\{\a\}] = e^{-\ms_{\mathrm{eff}}[\{\a\}]}\) such that
\begin{equation}
    \label{effective action main}
    \begin{split}
       e^{-\ms_{\mathrm{eff}}[\{\a\}]} & = e^{-\ms_0[\brw, w ; \{\a\}]}\!\int\! \md[\brf, f, h]e^{-\ms[\brf, f, h ; \{\a\}]},
    \end{split}
\end{equation}
where \(\ms_0[\brw, w ; \{\a\}]\) is just the mean-field free energy of the spin liquid. The action being integrated over \(\ms[\brf, f, h ; \{\a\}]\), consists of three parts : (i) free fermions hopping at half-filling on a lattice with the hoppings determined by the mean-field values of \(\expval{w_{ij}^{\a\b}}\) for a given spin-interaction \(J_{ij}^{ab}(\bd)\), and chosen value of \(\a_{ij}(\bd)\), denoted by \(\ms_0[\brf, f ; \{\a\}]\), (ii) quadratic action of the Weiss fields \(h_i^a\) denoted by \(\ms_0[h; \{\a\} ]\), and (iii) the spin-spin interaction between fermions and the Weiss field denoted by \(\ms_{\mathrm{int}}[\brf, f, h ; \{\a\}]\) (details given in Appendix.\ref{appendix:path integral}). One can now integrate out the fermions as they only appear to quadratic order in all the terms, and get an effective action of the Weiss fields, \(\ms_{\mathrm{eff}}[h ; \{\a\}]\). To quadratic order (equivalent to RPA), this looks like (switching to momentum and imaginary frequency space)
\begin{eqnarray}
    \label{effective action h main}
      \ms_{\mathrm{eff}}[h ; \{\a\}] &=& \sum_{i\W}\sum_{\bQ}h_i^a(-\bQ, -i\W) h_j^b(\bQ, i\W) \nonumber\\
      &\times& \big[\mj^{-1}(\bQ ; \{\a\})-\c_0(\bQ, i\W ; \{\a\})\big]_{ij}^{ab},  
\end{eqnarray}
where \(\mj_{ij}^{ab}(\bQ ; \{\a\})\equiv (1-\a_{ij}(\bd))J_{ij}^{ab}(\bd)\) is the re-scaled spin-interaction
written in momentum space, and \(\c_0\) is the bare magnetic susceptibility of the fermions (which depends on \(\{\a\}\) through the hoppings as well). One can then integrate out the Weiss fields and get the free energy
\begin{equation}
    \label{total free energy}
    \begin{split}
      {\cal F}_{\mathrm{eff}}[\{\a\}] &= \ms_0[\brw, w ; \{\a\}]+\\
      &\sum_{i\W,\bQ}\mathrm{Tr}\ln\left(\mj^{-1}(\bQ ; \{\a\})-\c_0(\bQ, i\W ; \{\a\})\right)  
    \end{split}
\end{equation}
The total free energy can now in principle be minimized with respect to \(\{\a\}\) to provide optimal values \(\{\a^{*}\}\). However, one can also treat them as tuning parameters which may be constrained such that the RPA phase diagram reproduces a true many-body numerical calculation of the ground state using any other method (like DMRG or VMC). As we will see in the later sections, the dynamic response functions also obtained in this RPA process exhibits continuum features expected from $f$-spinon excitation of the spin-liquid. Moreover, it also demonstrates well defined features expected from magnon-like excitations 
of the classical order, which emerge as two-particle bound states of the spinons under RPA, tracked by the $h$ Weiss fields.\par

\section{Model}

In this, and the following sections, we use the methodology introduced in Sec.\ref{methodology} and apply it to a specific spin model. Namely, we look at a 
\(J_1\)-\(J_3\) XXZ spin model on the honeycomb lattice which is relevant to a class of material known as the honeycomb cobaltates \cite{BCAO_Regnault2018,BCAO_Zhong2019,Armitage2022,BCAO_Broholm2023,BCAO_thermalcond_Li2022, BCPO_Nair2018, ncto_simonet2016,ncto_ncso_stock2020,Kappaxy_NCoTeO_Park2022,LANL_ncto2022,kappa_NCTO_Sun_PRB_2023}. The model consists of a ferromagnetic easy-plane interaction between nearest-neighbour spins, along with a frustrating antiferromagnetic easy-plane interaction between third-nearest neighbouring spins. 
The Hamiltonian on a bond has the form
\begin{equation}
    \label{easy plane Hamiltonian}
    \mh_{ij} = \left(S_i^x S_j^x + S_i^y S_j^y + \l_{ij} S_i^z S_j^z\right)\,,
\end{equation}
where \(\l_{ij}<<1\) represents the strong in-plane anisotropy. The full Hamiltonian looks like
\begin{equation}
    \label{easy plane Hamiltonian full}
    \mh = -J_1\sum_{\expval{i, j}_1}\mh_{ij} + J_3\sum_{\expval{i,j}_3}\mh_{ij}\,,\:\:\:J_1, J_3>0\,.
\end{equation}
where \(\expval{i, j}_1\) and \(\expval{i, j}_3\) is a shorthand for first nearest and third nearest neighbours on the honeycomb lattice. In this paper, our results will focus on the \(XY\) limit where \(\l=0\).

\subsection{Spinon mean field ground state}
We first analyze the spin Hamiltonian at mean-field level keeping only the hopping channel (\(\a_{ij}=1\)) and find a possible 
spin-liquid ground state. We find that the ground state has opposite hoppings for spin-\(\ua\) and spin-\(\da\) fermions, and has Dirac nodes at the \(\bK\) and \(\bK'\) points in the Brillouin zone \cite{bose_dirac}. The hoppings in this Dirac spin liquid (DSL) can be determined self-consistently in terms of the exchange interactions and are found to be \(t_1 \approx 0.13 J_1\) for nearest neighbour, and \(t_3/t_1\approx0.1\) for third nearest-neighbour. Refer to Appendix.\ref{appendix:MFT} for more details.\par
\begin{figure}[!ht]
 \centering
 \includegraphics[width=0.48\textwidth]{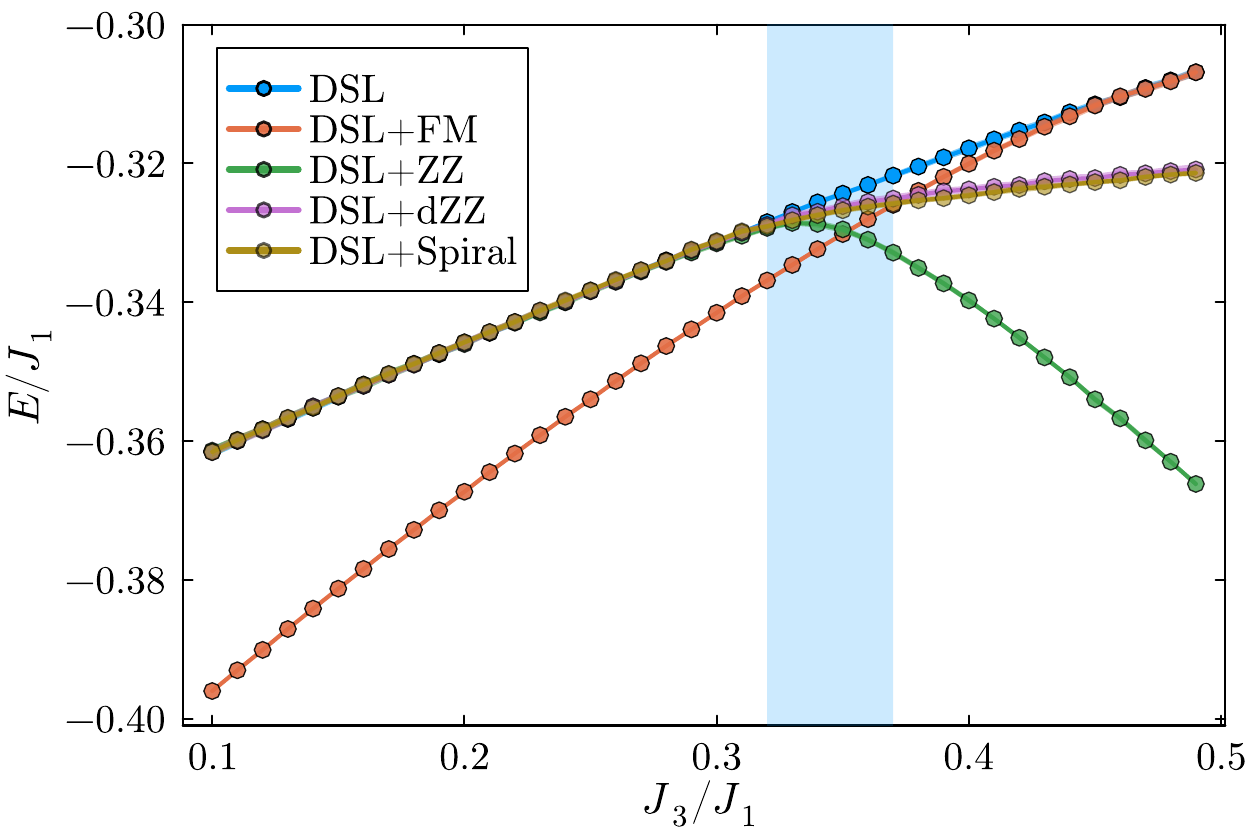}
 \caption{Optimized Gutzwiller projected wavefunction energies for different ansatzes on a finite size system computed using VMC with stochastic optimization of variational parameters including Jastrow factors. The easy-plane Dirac spin liquid (DSL) ansatz has hoppings $\pm t_1$ and $\pm t_3$ (with $\pm$ for spin-$\upa$ and spin-$\dna$ respectively), and Jastrow factors. The states marked ``DSL+order'' incorporate additional Weiss fields as variational parameters which fixed the specific type of broken spin symmetry (FM: ferromagnetic, ZZ: zig-zag, dZZ: double zig-zag, Spiral: coplanar spiral state). Shaded region is the intermediate window \(0.32 \lesssim J_3/J_1 \lesssim 0.37\) where various weakly ordered broken symmetry states are energetically competitive 
 (to within \(1\)-\(2\%\)), and where large scale DMRG finds evidence for a weakly ordered phase. 
 See Appendix \ref{appendix:VMC} for details including schematic patterns of broken symmetry states. DMRG numerics appear to yield different results in
 this intermediate regime depending on small changes to the Hamiltonian \cite{Chernyshev2023,Asim2024}, consistent with many competing orders in this
 window of $J_3/J_1$.}
\label{fig:VMC energies}
\end{figure}

\subsection{Gutzwiller projected wavefunction}

Going beyond mean field theory, we also studied the phase diagram of the $J_1$-$J_3$ XY model using Gutzwiller projected parton wave-functions. We compute the energy and correlations of the projected wave-function through a Monte Carlo sampling of real space configurations which obey the strict Hilbert space constraint (that the fermionic spinons are at half-filling on each site). We extend previous variational numerics performed on the same model in \cite{bose_dirac} by including more variational ansatzes/parameters which cover the pure DSL, as well as DSL+weak magnetic ordering such as ferromagnet (FM), zig-zag (ZZ), double zig-zag (dZZ), and spiral ordering, as well as Jastrow factors.\par

\begin{figure}[b]
 \centering
  \includegraphics[width=0.48\textwidth]{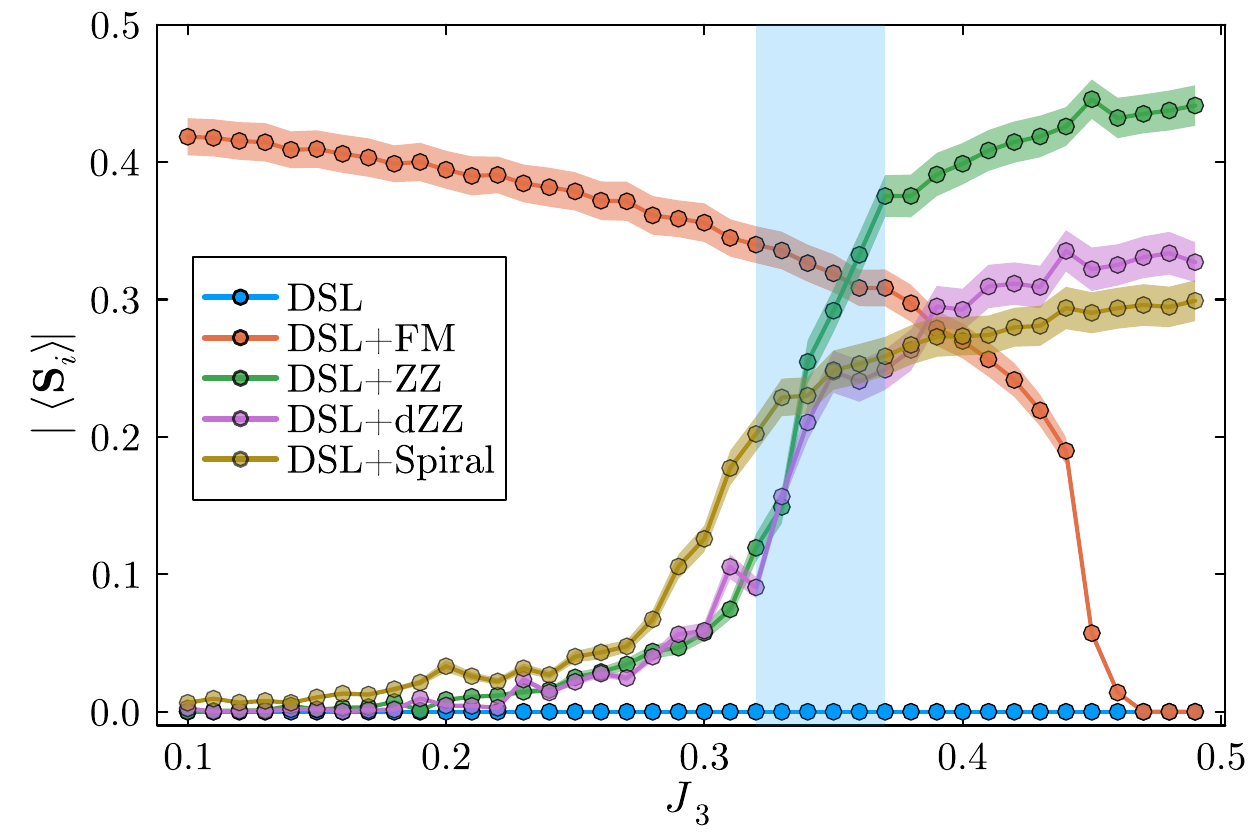}
 \caption{The ordered moment in the optimized VMC wave-functions for different symmetry breaking patterns as a function of \(J_3/J_1\). All the
 states correspond to in-plane magnetic orders. While the FM and ZZ orders at small and large $J_3/J_1$ are robust (ordered moment $\sim\! 0.45$) 
 the ordered moment in the intermediate regime (shaded in light blue) is weak $\sim\! 0.2$. }
\label{fig:VMC moments}
\end{figure}

The result of the optimum energies is shown in Fig.\ref{fig:VMC energies} where we find the ground state wave-function to be a projected DSL+FM state for \(J_3/J_1 \lesssim 0.32\), while for \(J_3 \gtrsim 0.37\) its a projected DSL+ZZ state. In the intermediate regime \(0.32 \lesssim J3 \lesssim 0.37 \), we find multiple competing ordered states, all within \(1-2\%\) in their optimum energies. This regime is also when the pure DSL is closest to the true ground state and consequently we hypothesize can be treated as the parent spin-liquid state from which all these weak orderings are arising from \cite{ASL_Hermele2005}. We 
also explored the ansatz corresponding to the DSL with additional Ising Neel order; this broken symmetry was reported in DMRG calculations \cite{Chernyshev2023} and pfFRG calculations \cite{j1j3_watanabe2022frustrated} but we do not find any parameter regime where its energy is lower than that of a pure DSL ansatz, so this instability is not found in our optimized VMC calculations.
The ordering moments are reported in Fig.\ref{fig:VMC moments} and are roughly $25$-$50\%$ of the full moment 
in the regime of interest (as compared to \(\approx 80\)-\(90\%\) in the FM and ZZ phases). Note that this moment will further be reduced upon incorporating magnon-like spin fluctuation of the internal Weiss fields, making it consistent with recent studies \cite{Chernyshev2023, Asim2024}, and allowing us to treat the state as weak ordering on top of a DSL.\par

\section{RPA Instabilities}
\begin{figure}[!ht]
 \centering
 \includegraphics[width=0.48\textwidth]{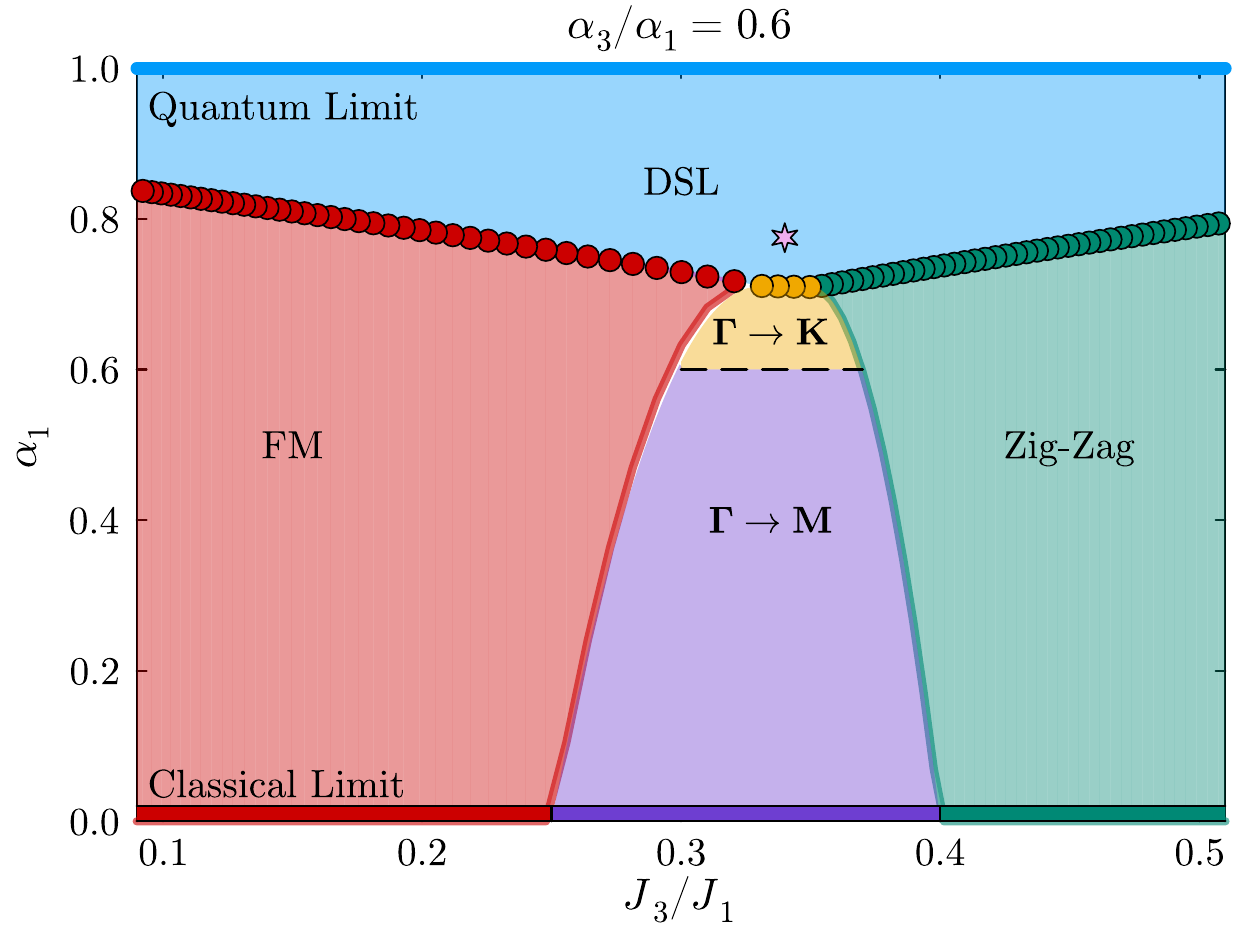}
 \caption{Phase diagram of the $J_1$-$J_3$ XY model made at a fixed ratio \(\a_3/\a_1=0.6\) and looking at the RPA instabilities of the DSL state as we tune $\alpha_1$ and $J_3/J_1$. For \(0.6 \lesssim \a_1 \lesssim 0.7\), the phase diagram matches roughly with previous DMRG numerics \cite{Chernyshev2023,Asim2024}.
 This phase diagram is plotted for fixed
 \(\a_3/\a_1=0.6\), but the phase diagram is qualitatively
 similar for a range of ratios
 \(\a_3/\a_1 \approx 0.5\)-\(0.8\). Star marks a representative parameter point 
 near the RPA instability in the regime of interest where all spin-response functions of the DSL are calculated in this paper.}
\label{fig:RPA instabilities 0.6}
\end{figure}

In this section, we want to study the RPA instabilities of the mean-field DSL state as we start tuning \(\a_{ij}\) from the purely quantum limit of \(\a_{ij}=1\) which introduces four-fermion interaction terms to be treated under RPA. The hopping scale of the fermions is \(t_1 \approx 0.13\a_1 J_1\), while the interaction for the RPA analysis scales like \((1-\a_1)J_1\) and \((1-\a_3)J_3\). The third neighbour hopping is roughly \(t_3/t_1\approx 0.1\)-\(0.2\) both in mean-field theory and in the VMC optimized variational parameter as detailed in Appendix.\ref{appendix:VMC}. For simplicity, we shall fix the ratio \(\a_3/\a_1\) in our analysis such that the obtained RPA phase diagram roughly matches the results of previous numerics, and tune \(\a_1\) along with the physical exchange parameter \(J_3/J_1\) to obtain Fig.\ref{fig:RPA instabilities 0.6}.\par

\begin{figure}[!t]
 \centering
 \includegraphics[width=0.45\textwidth]{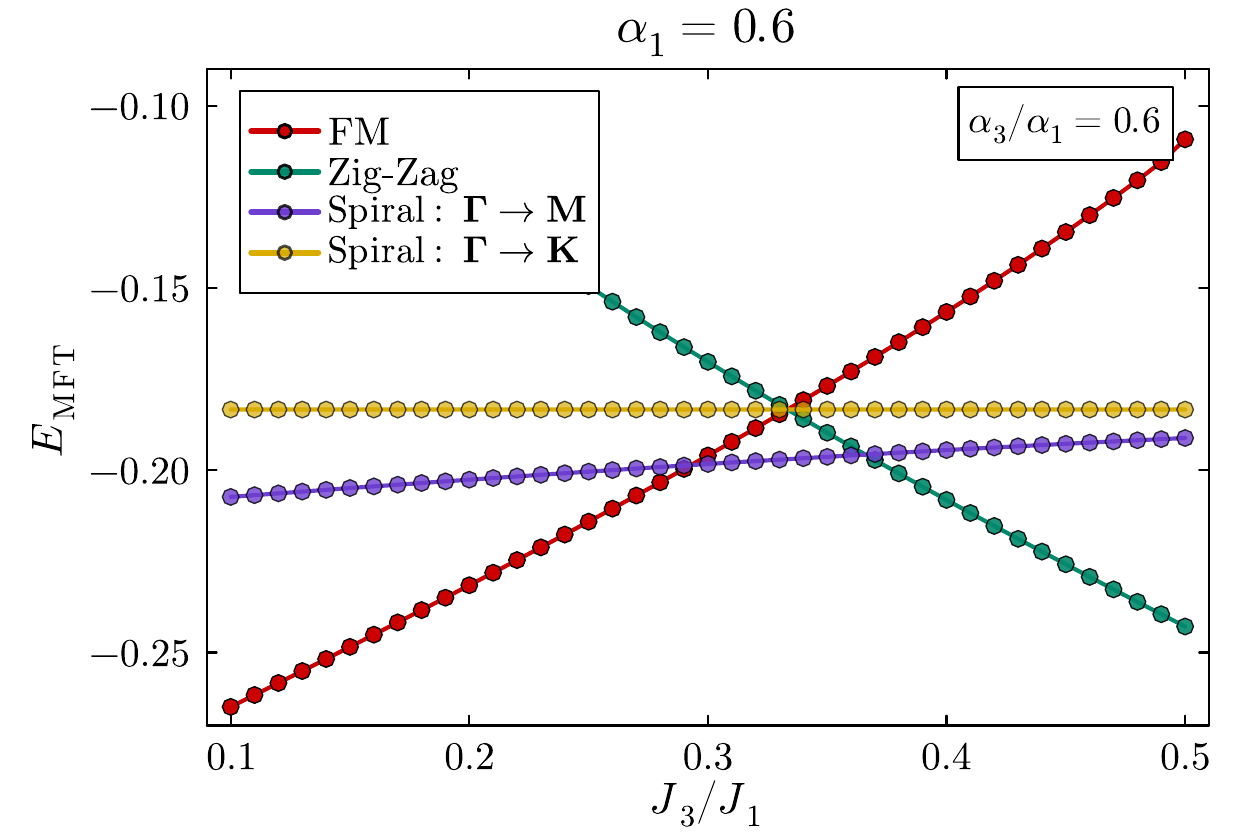}
 \caption{Self-consistent mean-field energies of symmetry broken DSL+ordering states computed in the regime where the RPA instability of the DSL leads
 to magnetic order. The calculation is done on large real-space lattices to accommodate commensurate collinear and incommensurate spiral orders. 
 We find that soon after the DSL transitions into an incommensurate order along \(\bGa\rightarrow\bK\) at \(\a_1^c\approx 0.7\), the system again transitions into an incommensurate order along \(\bGa\rightarrow\bM\) instead. This \(\bGa\rightarrow\bM\) is consistent with the classical limit and experiments
 on {\bcao} \cite{BCAO_Broholm2023}.}
 \label{fig:MFT PD}
\end{figure}

As we see in Fig.\ref{fig:RPA instabilities 0.6}, starting in the pure quantum limit at \(\a_1=0\), the DSL is stable as \(\a_1\) is decreased towards the classical limit until some critical value \(\a_1^c\). Beyond this, the system enters a symmetry broken magnetically ordered state with some wave-vector \(\bQ_c\). For \(J_3/J_1 \lesssim 0.32\), the RPA instability encountered is just a ferromagnet, while for \(J_3/J_1 \gtrsim 0.37\) the system transitions into the zig-zag ordered state. In the intermediate regime, \(0.32 \lesssim J_3/J_1 \lesssim 0.37\), the DSL first transitions into an incommensurate order with \(\bQ_c\) lying between \(\bGa = (0, 0)\) and \(\bK=(4\pi/3, 0)\). However, we find using a real-space mean-field theory that when \(\a_1\) is decreased further, the system undergoes one more transition into an incommensurate order with \(\bQ_c\) lying between \(\bGa\) and \(\bM=(\pi, \pi/\sqrt{3})\), as shown 
in Fig.\ref{fig:MFT PD}. This phase diagram with FM, ZZ, and and intermediate window of complex magnetic orders is broadly consistent with
DMRG numerical results \cite{Chernyshev2023,Asim2024}, and the type of spiral orders we find are consistent with one set of DMRG calculations \cite{Asim2024} 
and might be consistent with experimental observations on {\bcao} \cite{BCAO_Broholm2023}.

\section{Physical response functions}
In this section, we report the spin response functions of the DSL state after RPA correction near the phase boundary into a symmetry broken state in the intermediate \(J_3/J_1\) regime (closest to the exchange coupling ratio found for \bcao in \cite{Abinitio_Das2021, BCAO_Broholm2023}).  First, we calculate the sublattice-resolved bare mean-field spin response of the DSL spinons (appearing in Eq.\eqref{effective action h main}) \cite{Wen2002, Wen2015, Iqbal2020, YBK2003, BOcquet2001} using a simple bubble diagram 
\begin{equation}
   \label{chi bubble}
    \c_{0, ij}^{ab}(\bQ, i\Omega)\! =\!-\!\!\sum_{\bk,\;i\omega}  \!\! \text{Tr} [\frac{\s^{a}}{2}\mg_{ij}^{T}(\bk, i\w) \frac{\s^{b}}{2} \mg_{ji}^{T}(\bk+\bQ, i\w+i\Omega)]\,,
\end{equation}
where \(a\,,b\) represent spin-directions, and \(i\,,j\) mark sublattices, and \(\mg\) is the bare spinon Green's function. Following this, we can analytically continue Eq.\eqref{chi bubble} into real-frequency as \(\c_{0, ij}^{ab}(\bQ, \W+i\eta)\). Incorporating RPA corrections to the response, we get (Note that both the interaction, as well as the bare susceptibility depend on the chosen value of \(\{\a\}\) but we suppress that for now for simplicity in notation)
\begin{equation}
    \label{RPA bubble}
    \!\! \c_{ij}^{ab}(\bQ, \W) \!=\! \left[(\mathds{1}\!-\!\mj(\bQ)\!\cdot\!\c_0(\bQ, \W))^{-1}\!\cdot\! \c_0(\bQ, \W)\right]_{ij}^{ab}\,,
\end{equation}
where all algebraic operations like \(\mj\cdot \c_0\), or \(\c_0^{-1}\), are shorthands for \emph{matrix operations} in both spin-space, as well as sublattice space. Lastly, since we want to compare with experimental results, we Fourier transform over actual real-space position, getting rid of the sublattice index, and obtain \(\c_{ab}(\bQ, \W) = \sum_{i, j}\c_{ij}^{ab}(\bQ, \W) e^{-i \bQ\cdot(\br_i-\br_j)}\) with \(\br_1=(0, 0)\) and \(\br_2=(0, 1/\sqrt{3})\) being the two sublattice positions.

\subsection{THz}
\begin{figure}[!ht]
 \centering
 \includegraphics[width=0.45\textwidth]{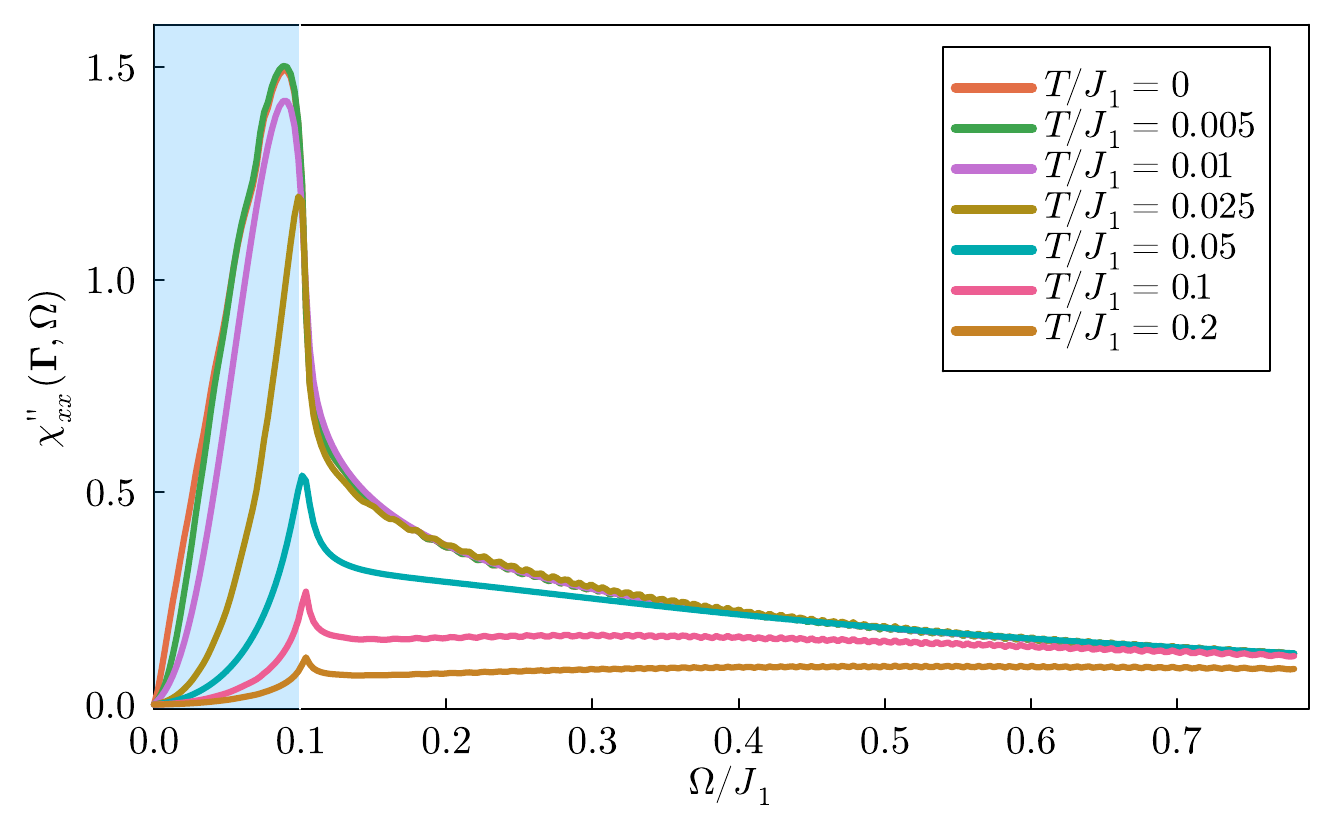}
 \caption{Temperature dependence of the THz response including RPA corrections
 near the instability towards incommensurate order (point marked with a star in Fig.\ref{fig:RPA instabilities 0.6}). The shaded region is the experimentally unexplored region in \cite{Armitage2022}. Compared to the bare mean-field response of the DSL \cite{bose_dirac}, the RPA corrections enhances the
 spectral weight near low energies at the \(\bGa\) point.}
\label{fig:THz with T}
\end{figure}
Fig.\ref{fig:THz with T} shows the real frequency zero wave-vector THz spin response, the imaginary
component $\c_{xx}^{''}(\bGa, \Omega)$, plotted as function of frequency $\Omega$, for various temperatures $T$ after incorporating RPA corrections. The calculation is done in the intermediate regime near \(J_3/J_1=0.35\), with the hopping scale \(t_1\approx 0.8\)-\(1\)\,meV assuming \(J_1\approx7\)-\(8\)\,meV \cite{Abinitio_Das2021, BCAO_Broholm2023},  \(t_3/t_1\approx 0.2\), and RPA corrections are calculated at \(\a_1\approx 0.7\). The response exhibits an enhanced low energy `peak' around $\Omega/J_1 \approx 0.1$ arising from the large low-energy density of states for spinon excitations around the entire Brillouin zone edge \cite{bose_dirac} (discussed in Appendix.\ref{appendix:MFT}). We note that the THz absorption data on {\bcao} \cite{Armitage2022} does not extend below $\lesssim 0.2$\,THz which corresponds to $\Omega/J_1 \!\lesssim\! 0.1$.

\subsection{Neutron scattering}

\begin{figure}[t]
 \centering
    \includegraphics[width=0.5\textwidth]{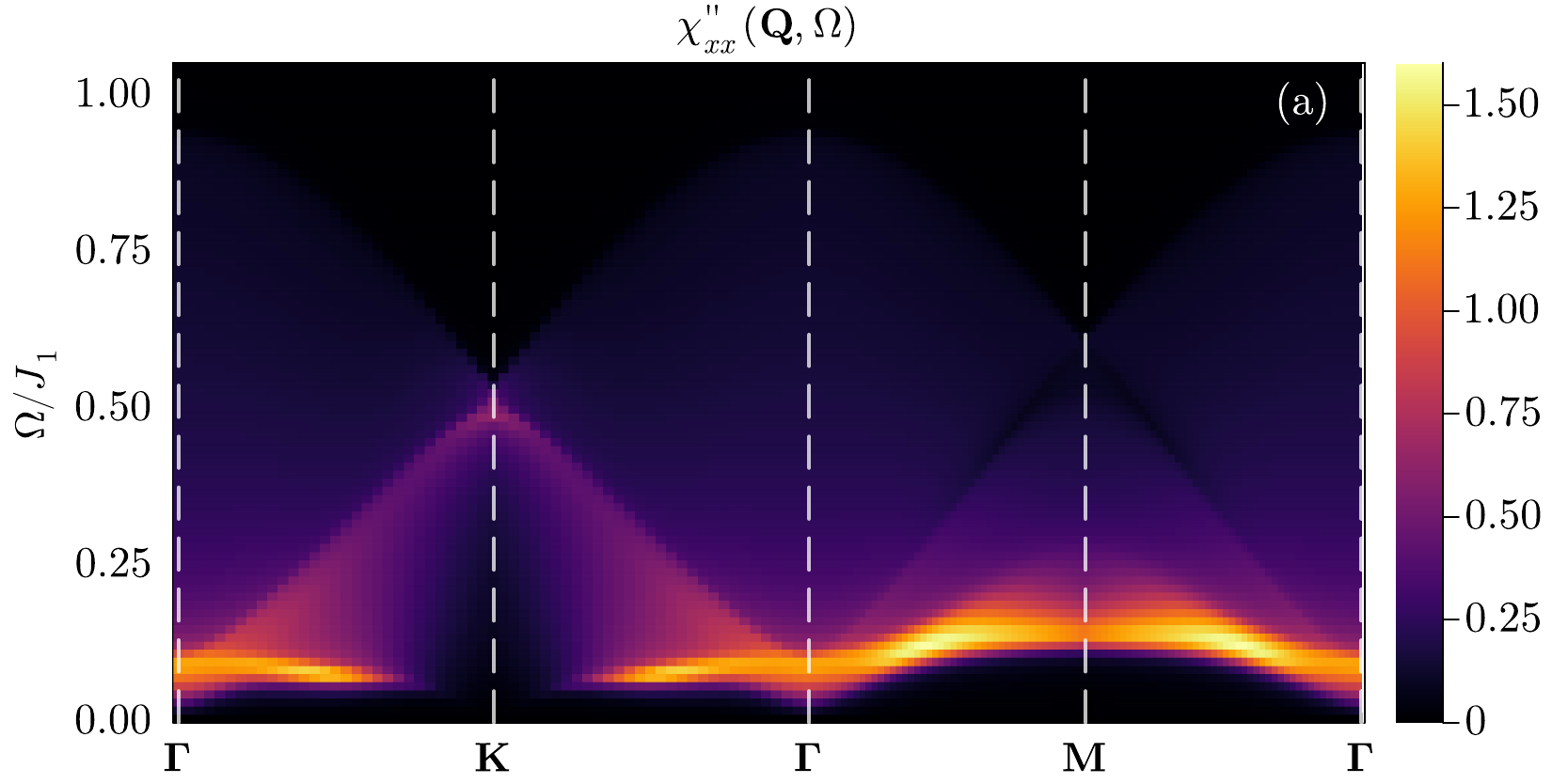}
    \includegraphics[width=0.5\textwidth]{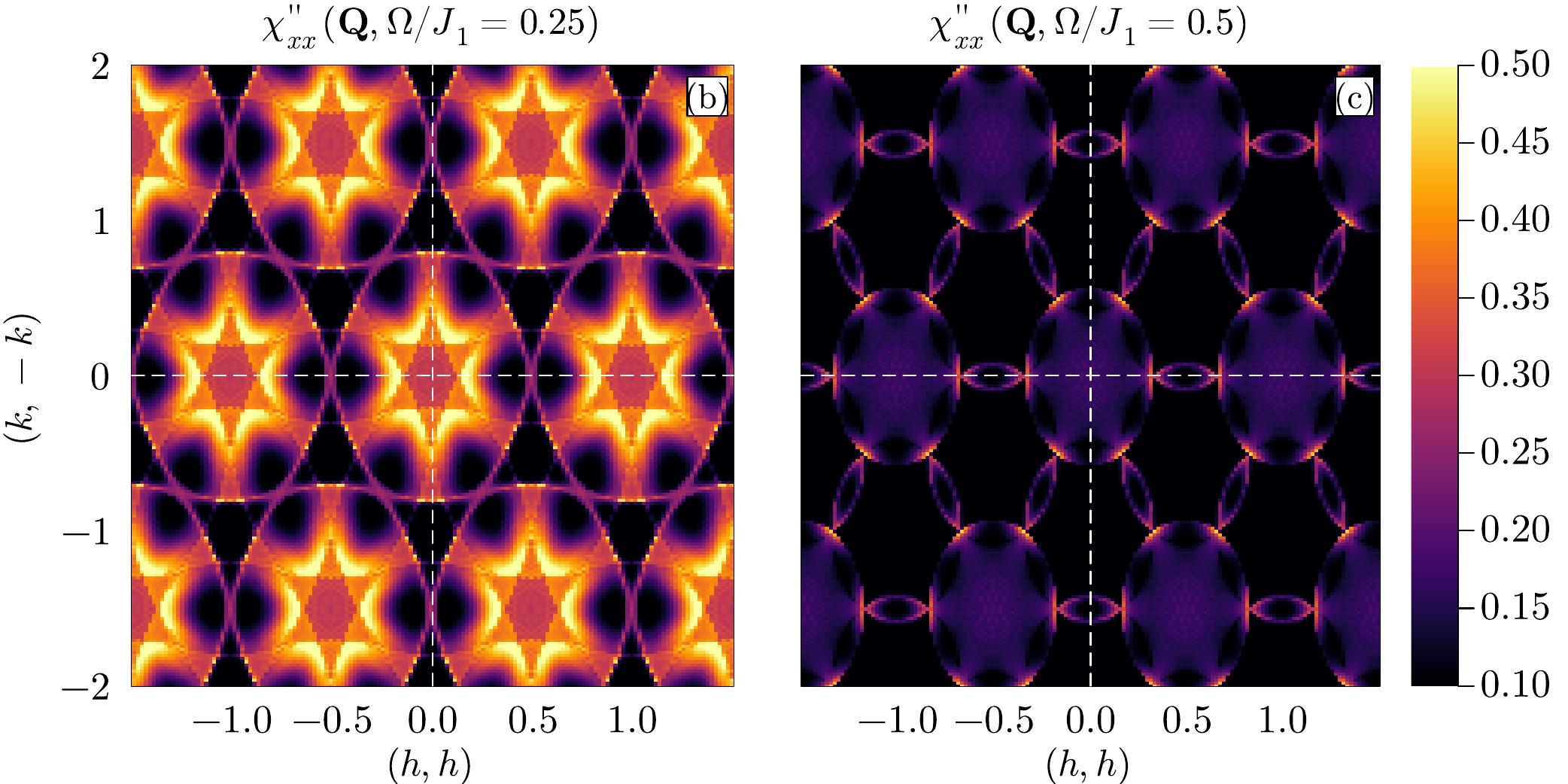}
 \caption{(a) The imaginary part of the magnetic susceptibility \(\c_{xx}^{''}(\bQ, \W)\) measured along a high-symmetry path in the Brillouin zone and throughout the bandwidth. We clearly see sharp modes running from \(\bGa\rightarrow\bM\) between \(\approx 0.05\)-\(0.2J_1\), and another one running from \(\bGa\rightarrow\bK\) between \(\approx 0.05\)-\(0.5J_1\). Furthermore, there is a continuum of response seen at both the \(\bGa\) and \(\bM\) points, all of which is consistent with experimental data reported in Fig.~4 of Ref.~\cite{BCAO_Broholm2023}. (b)-(c) \(\c_{xx}^{''}(\bQ, \W)\) 
 measured at constant energies \(\W=0.25J_1\) and \(\W=0.5J_1\) respectively, in the full Brillouin zone which is in qualitative agreement
 with experimental observations reported in Fig.~2 of Ref.\cite{BCAO_Broholm2023}. The momentum \(\bQ\) is expressed in the reciprocal basis  \(\bb_{1/2} = (2\pi, \pm 2\pi/\sqrt{3})\). The color-bar scales are in arbitrary units but consistent between the different plots. }
\label{fig:Inelastic Neutron plots}
\end{figure}
Fig.\ref{fig:Inelastic Neutron plots} shows the real frequency imaginary component of $\c_{xx}^{''}(\bQ, \Omega)$ for momenta along high-symmetry path and all energies, as well as fixed energy slices of the full Brillouin zone. We find clear sharp modes similar to magnon bands going along \(\bGa\rightarrow \bM\), as well as a higher energy mode going along \(\bGa\rightarrow\bK\). These sharp modes are seen with a continuum background, especially at \(\bGa\) and \(\bM\) points. The fixed energy slices also exhibit clear structure which match very well qualitatively with the experimental results seen in \cite{BCAO_Broholm2023},
albeit at energies which are slightly different from the energy slices chosen in the experimental
data \cite{BCAO_Broholm2023}; resolving this quantitative mismatch would go beyond RPA and incorporating spinon self-energy effects, as well as 
possible vertex corrections, all of which will potentially renormalize the band structure, and hence the precise energy scales.

\subsection{NMR}
\begin{figure}[!ht]
 \centering
 \includegraphics[width=0.48\textwidth]{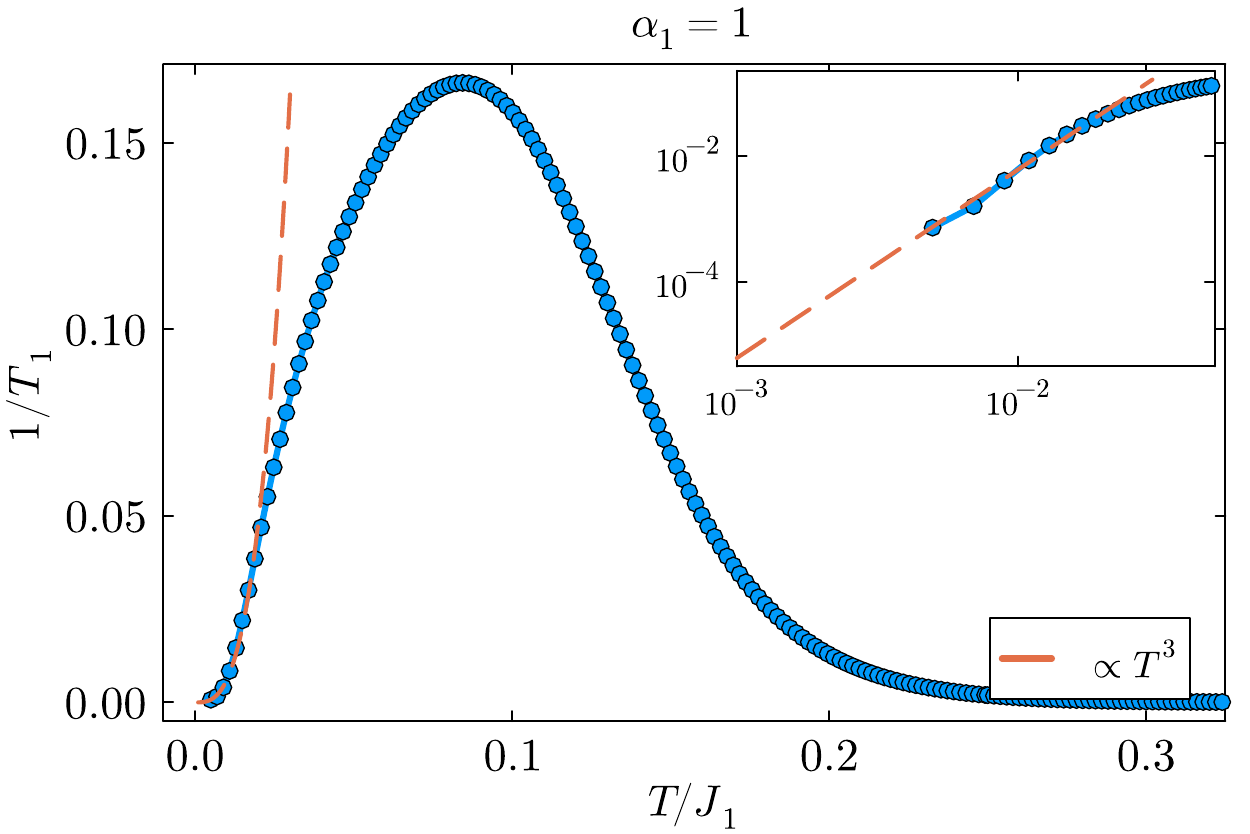}
 \caption{\(1/T_1\) (in arbitrary units) for NMR calculated at the mean-field level in the pure DSL. The bare response is also supplemented with temperature dependence of the hopping shown in Appendix.\ref{appendix:MFT}. In our calculations we assumed that the nuclear spin is placed at the center of, and coupled to all lattice spins on a hexagonal plaquette. The applied field is assumed to be in-plane, and subsequently the decay time depends on the local transverse magnetic susceptibility (weighted by appropriate form factors). The inset shows the low temperature power law fit which is found to be \(T^3\). }
\label{fig:NMR}
\end{figure}
The NMR decay time \(1/T_1\) is calculated in the mean-field level for the DSL state assuming that the nucleus is placed at the center of a hexagonal plaquette. The decay time \(1/T_1 \approx \lim_{\W\rightarrow 0} (T/2\W)\sum_{\bQ}[\c_{ij}^{+-}(\bQ, \W)F_{ij}(\bQ)]^{''}\) where \(F_{ij}(\bQ)\) are form factors from coupling the nuclear spin at the center of a hexagon too all the spins on the hexagonal plaquette.  The temperature dependence is shown in Fig.\ref{fig:NMR} where we find that the low energy limit follows a \(T^3\) power law as expected from a Dirac state, even after form-factor dependence. The decay time eventually dies off since the mean-field hopping vanishes as the temperature is increased (shown in Appendix.\ref{appendix:MFT}).

\section{In-Plane Zeeman field}
In this section, we introduce an in-plane Zeeman field \textit{s.t.} \(\mh \rightarrow \mh + B_x\sum_{i}S_i^x\) and study its effect on the phase diagram, magnetization, as well as all the previously calculated response functions.

\subsection{Magnetization}

\begin{figure}[!ht]
 \centering
 \includegraphics[width=0.45\textwidth]{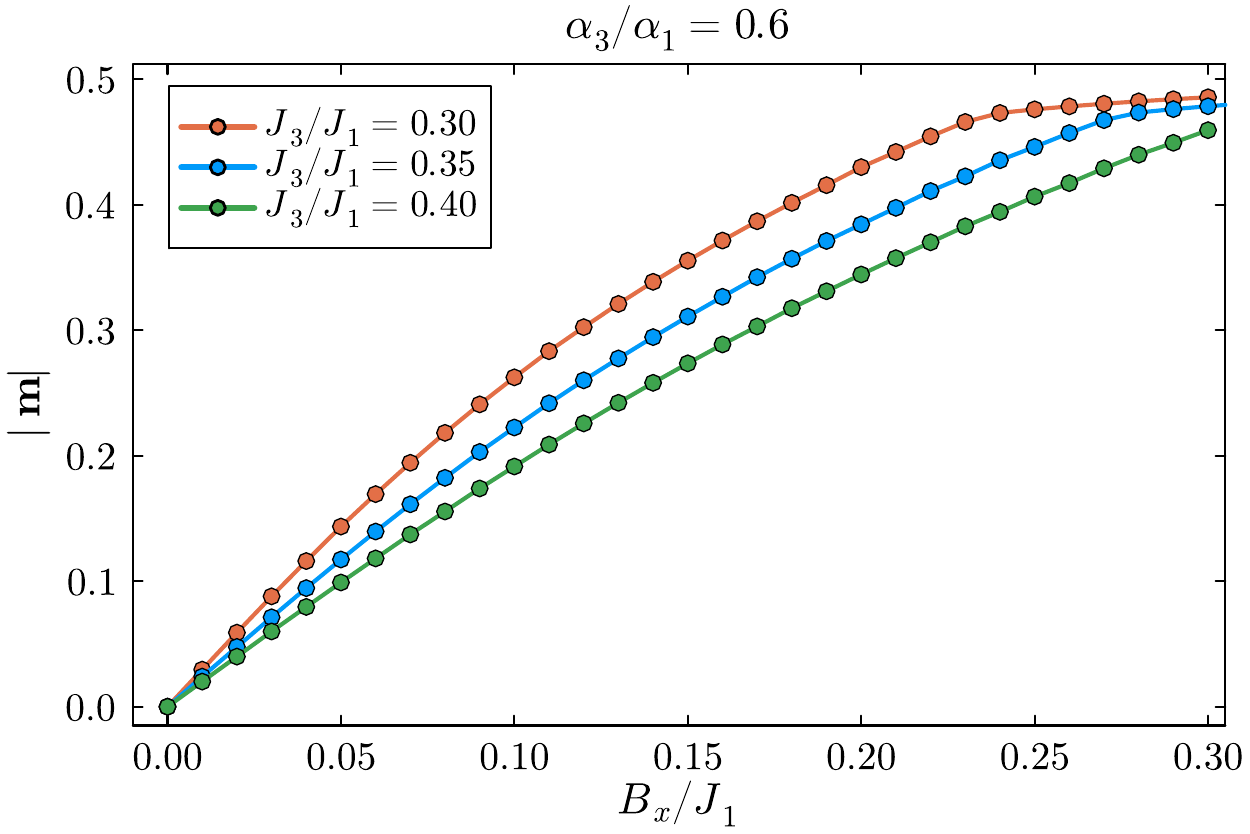}
 \caption{Mean-field in-plane magnetization \(|\mathbf{m}|\) as a function of applied Zeeman field \(B_x\) in the intermediate regime \(0.3\lesssim J_3/J_1 \lesssim 0.4\). The calculation is done assuming a fixed value \(\a_3/\a_1=0.6\), and \(\a_1\approx0.7\) near the RPA instability of the DSL marked in Fig.\ref{fig:RPA instabilities 0.6}.}
\label{fig:MFT mag with B field}
\end{figure}

In this section, we calculate the dependence of the net in-plane magnetization \(\mathbf{m} = (1/L)\sum_{i}\expval{\mathbf{S}_i}\) on the applied Zeeman field in the intermediate regime \(J_3/J_1 = 0.3\)-\(0.4\). The results from our modified mean-field theory are shown in Fig.\ref{fig:MFT mag with B field}. At the mean-field level we find a polarizing field which is roughly \(B_x/J_1\approx 0.1\)-\(0.15\). However, we expect mean-field theory to underestimate internal Weiss fields, and hence will magnetize much slower than the full Gutzwiller projected wave-function. We also find the dependence of the hoppings \(t_1\) and \(t_3\) in the presence of Zeeman field and plot the results in Appendix.\ref{appendix:MFT}.

\begin{figure}[!ht]
 \centering
 \includegraphics[width=0.45\textwidth]{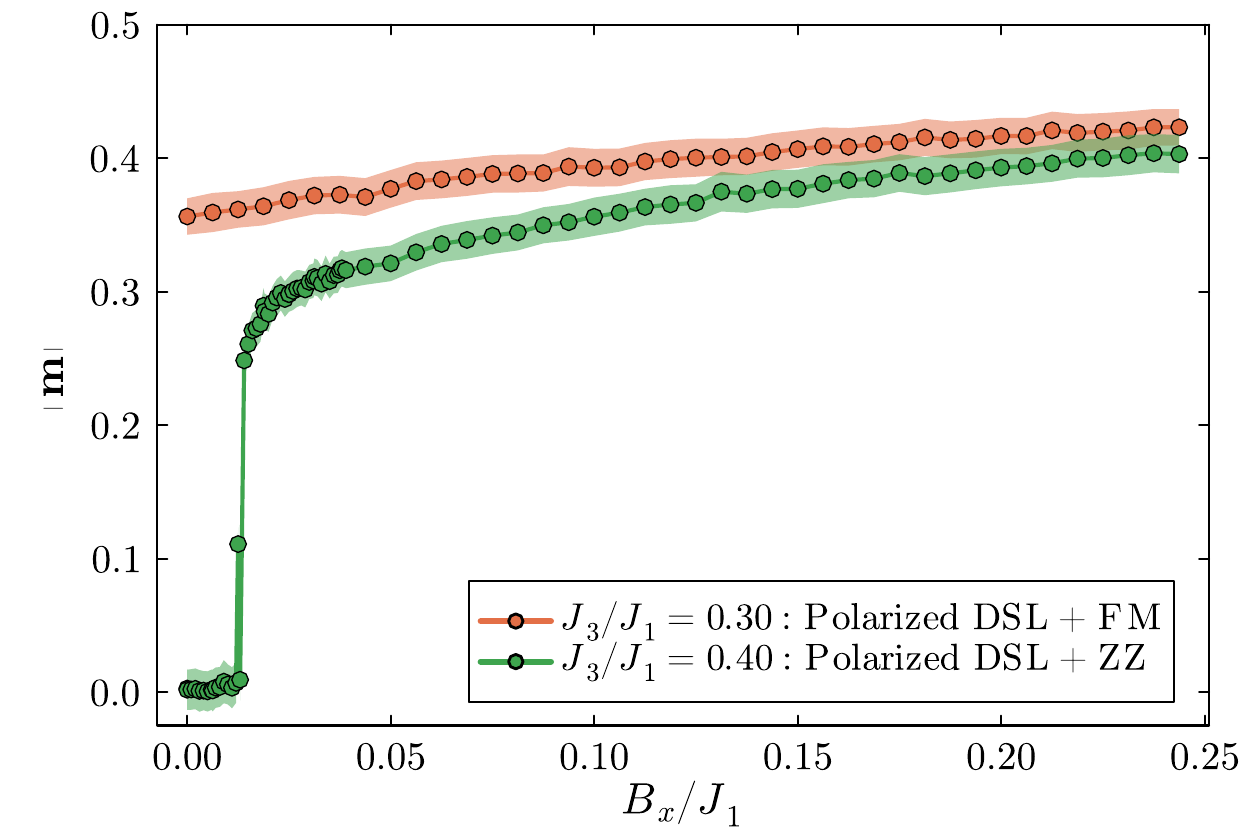}
 \caption{The net in-plane magnetization \(|\mathbf{m}|\) found in the optimum energy state found in VMC at fixed values of \( J_3/J_1=0.3\) and \(J_3/J_1=0.4\), in the presence of an applied in-plane magnetic field \(B_x\). The variational ansatz at the two points are taken to be the zero-field VMC ground states found in Fig.\ref{fig:VMC energies}, along with an additional uniform Weiss field to capture the effects of the applied Zeeman field.  }
\label{fig:VMC magnetization}
\end{figure}

 Subsequently, we repeat the same calculation using VMC where we optimize our variational states same as before, and report the net magnetization in the optimum state in Fig.\ref{fig:VMC magnetization}. Performing this analysis for DSL+Spiral or DSL+dZZ order in the intermediate regime is not feasible since they are never the actual ground states in our finite system VMC calculation, and hence will always magnetize even at zero magnetic field. However, from the behavior of the DSL+ZZ ansatz, we can infer that if those aforementioned states are somehow stabilized (either in the thermodynamic limit or through some perturbations of the pure XY model), we expect those state to also be magnetized very easily. For scale, the DSL+ZZ ansatz magnetizes at around \(B_x/J_1\approx 0.02\) ; assuming \(J_1\approx 7\,\)meV and a g-factor $g \!\approx\! 3$  \cite{Abinitio_Das2021, Armitage2022} yields a field scale for polarization
$\sim \! 0.5$\,T, which is in very good agreement with published data \cite{BCAO_Broholm2023, Asim2024}.\par

\subsection{RPA Instabilities}
\begin{figure}[!ht]
 \centering
 \includegraphics[width=0.45\textwidth]{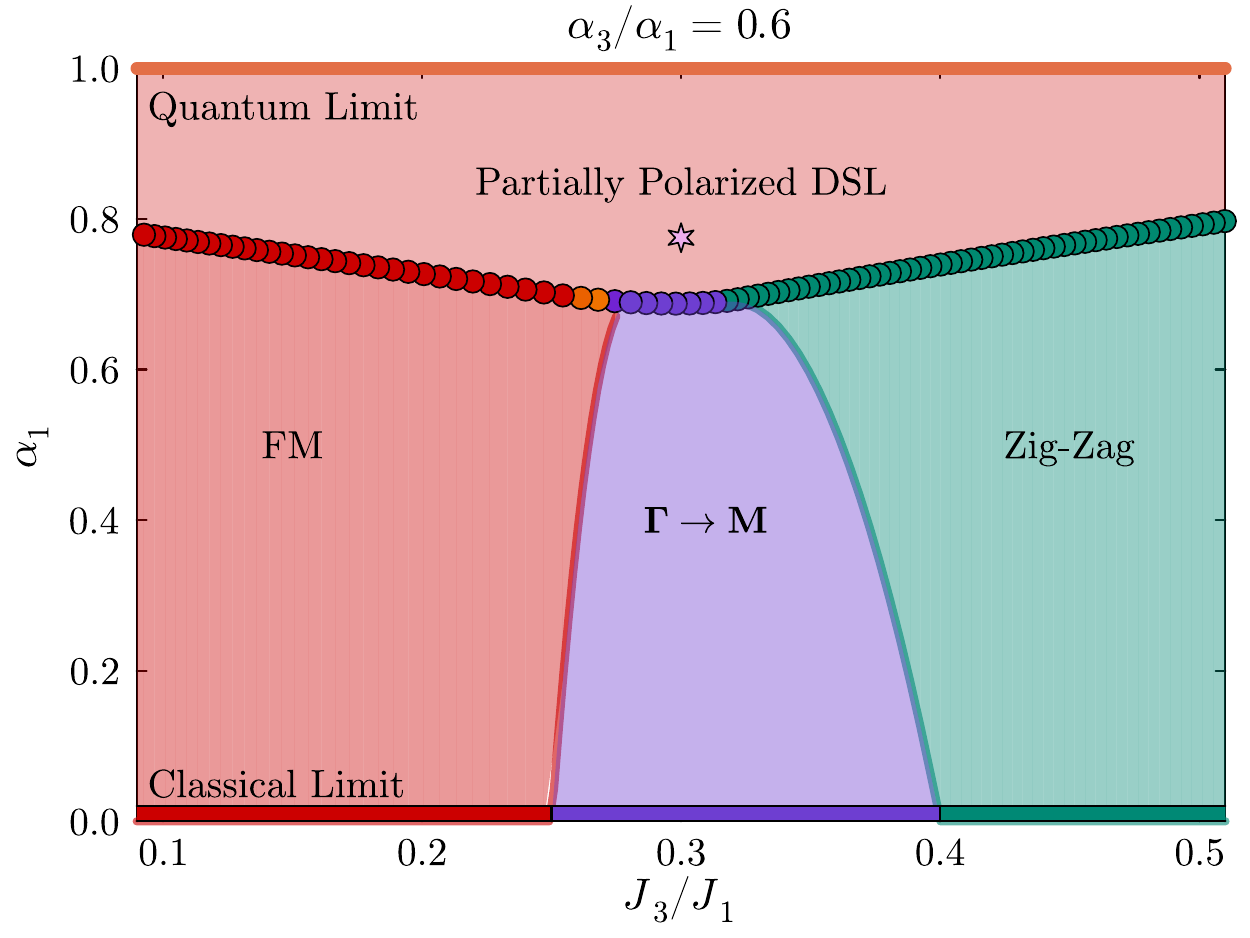}
 \caption{Phase diagram of the $J_1$-$J_3$ XY model made at a fixed ratio \(\a_3/\a_1=0.6\), in the presence of an applied Zeeman field \(B_x/J_1\approx 0.05\)-\(0.1\). Again, the phase diagram looks qualitatively similar for a range of \(\a_3/\a_1\approx0.5\)-\(0.8\). Star marks a representative parameter point near the RPA instability in the regime of interest where all spin-response functions of the DSL in the presence of a Zeeman field are calculated. }
\label{fig:RPA instabilities 0.6 with B}
\end{figure}
As we see in Fig.\ref{fig:RPA instabilities 0.6 with B}, starting in the pure quantum limit at \(\a_1=1\), the DSL is again stable as \(\a_1\) is decreased towards the classical limit until some critical value \(\a_1^c\). Beyond this, the system enters a symmetry broken magnetically ordered state with some wave-vector \(\bQ_c\) similar to the \(B_x=0\) case. The difference being that in the intermediate regime, the instability encountered is now directly an incommensurate ordering along \(\bGa\rightarrow\bM\). Note that since the ground state even in the pure quantum limit is a partially polarized DSL (which is gapped), there is no symmetry breaking transition in the small \(J_3/J_1\) limit into the ferromagnet. Instead it is a meta-magnetic type transition where the magnetization vs \(\a_1\) curve exhibits kinks at the transition. 

\subsection{Physical Response Function}
\begin{figure}[ht]
 \centering
 \includegraphics[width=0.45\textwidth]{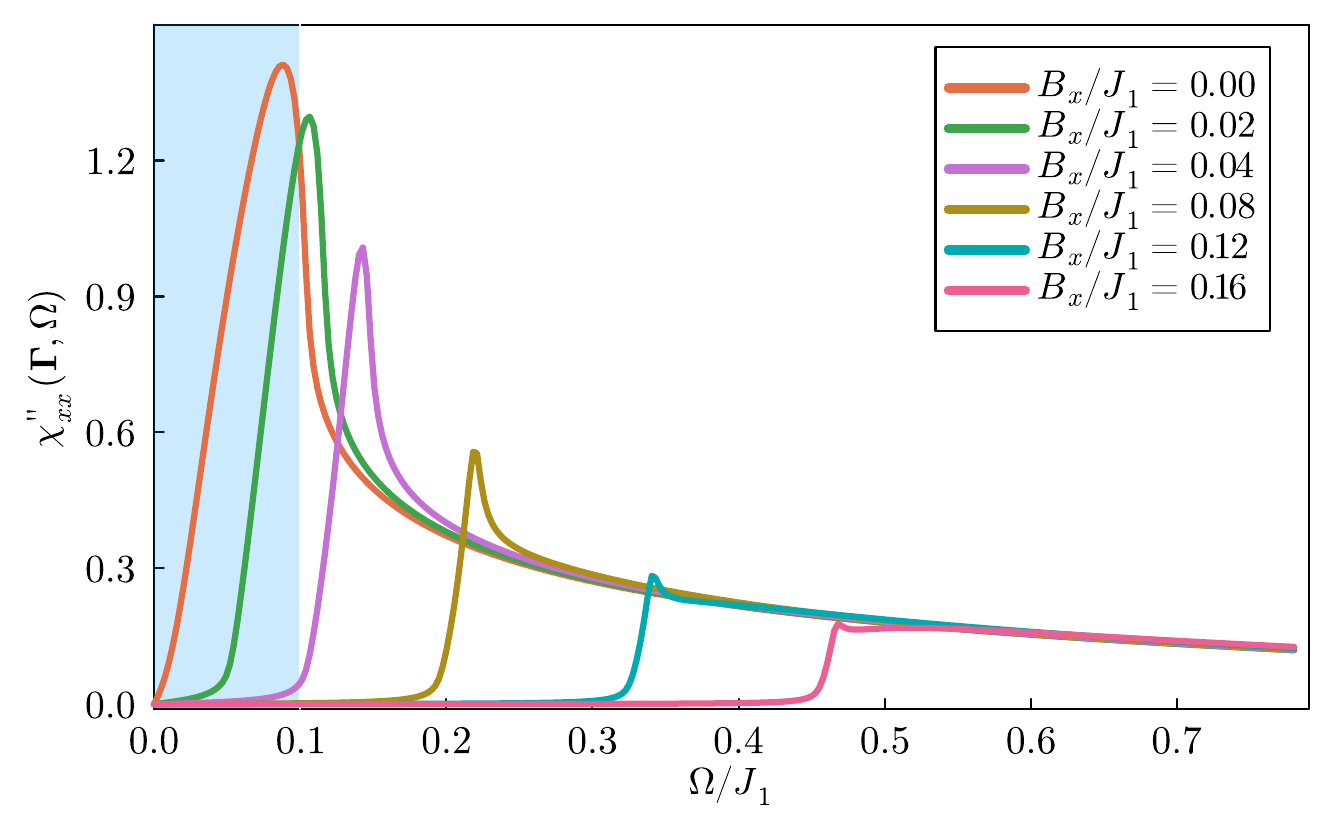}
 \caption{Zeeman field dependence of the THz response near the instability towards incommensurate order (point marked with a star in Fig.\ref{fig:RPA instabilities 0.6 with B}). Compared to the bare mean-field THz response of the DSL \cite{bose_dirac}, the RPA corrections replicate the \emph{overshooting} effect seen in experiments. The shaded region is the experimentally unexplored region in \cite{Armitage2022}.}
\label{fig:THz with B}
\end{figure}
\begin{figure}[!ht]
 \centering
    \includegraphics[width=0.5\textwidth]{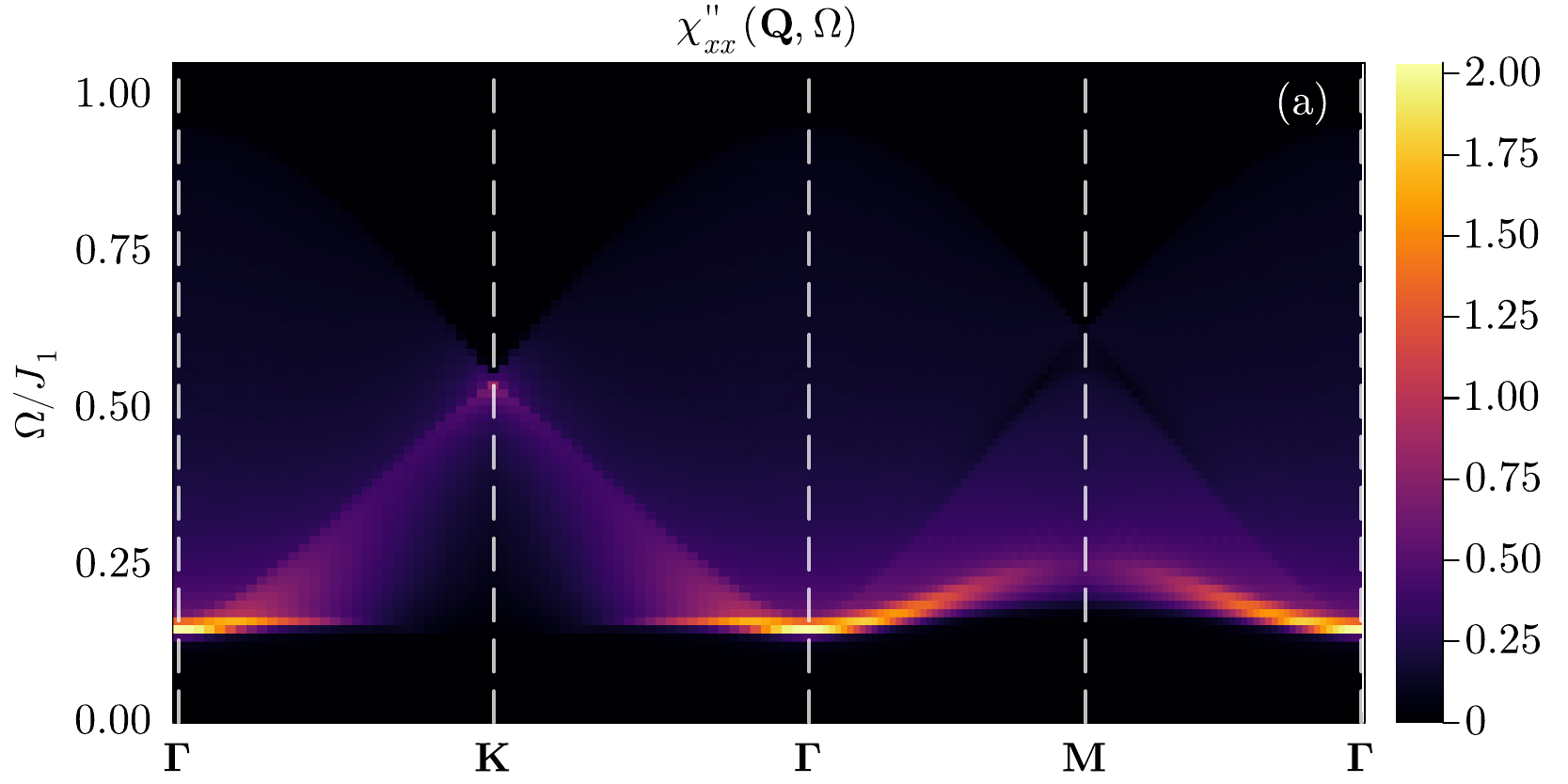}
    \includegraphics[width=0.5\textwidth]{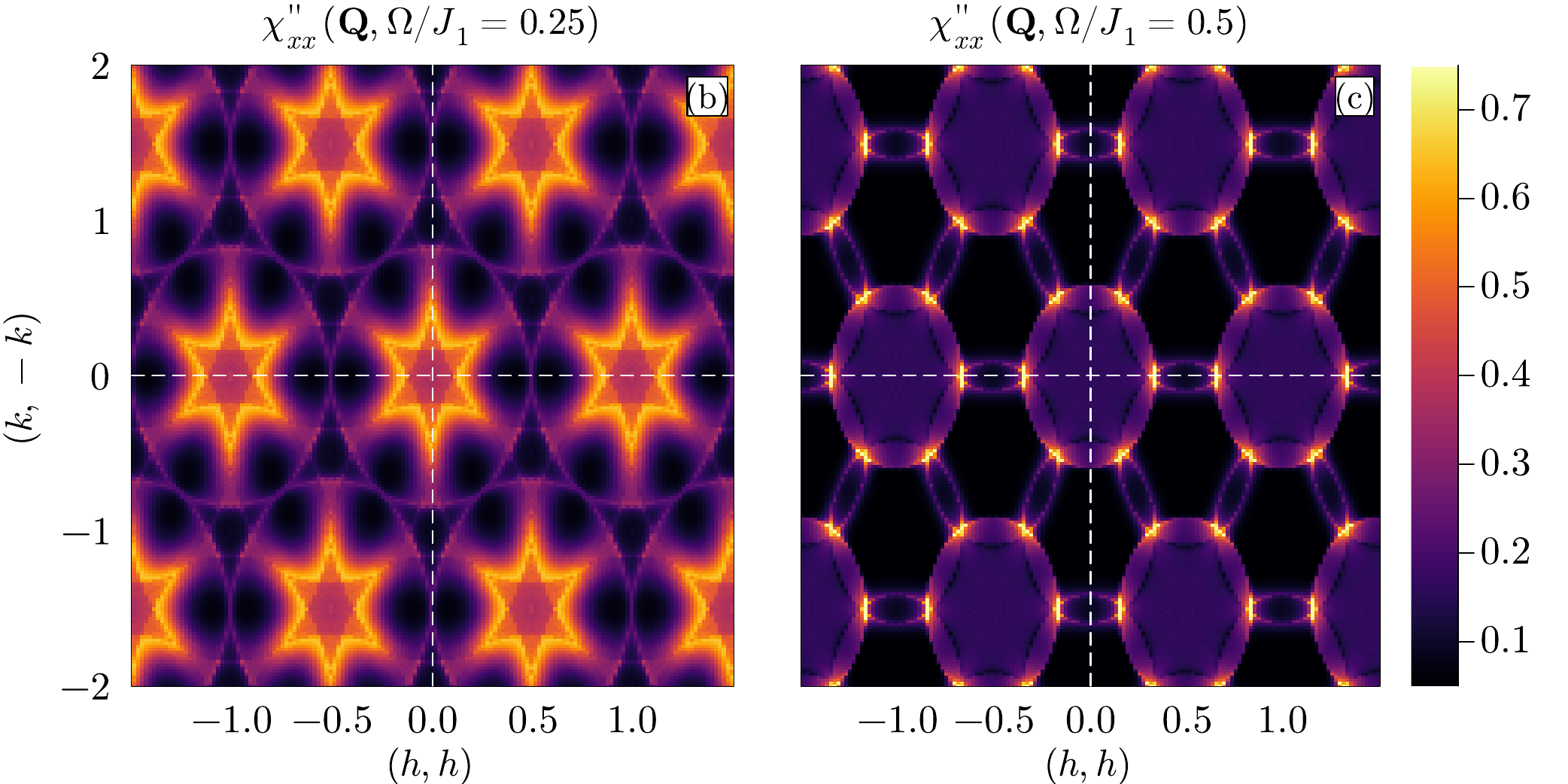}
 \caption{(a) The imaginary part of the magnetic susceptibility \(\c_{xx}^{''}(\bQ, \W)\) in the presence of an in-plane Zeeman field \(B_x/J_1\approx 0.05\)-\(0.1\) measured along a high-symmetry path in the Brillouin zone and throughout the bandwidth. The sharp mode between \(\bGa\rightarrow\bM\) is now flatter \(\approx0.15\)-\(0.25J_1\) because of the Zeeman field gapping out the Dirac nodes, while the one along \(\bGa\rightarrow\bK\) is weaker in intensity which is again in good agreement with experimental results reported in Fig.~4 of \cite{BCAO_Broholm2023}. (b)-(c) \(\c_{xx}^{''}(\bQ, \W)\) measured at constant energies \(\W=0.25J_1\) and \(\W=0.5J_1\) respectively in the full Brillouin zone. The momentum \(\bQ\) is expressed in the reciprocal basis  \(\bb_{1/2} = (2\pi, \pm 2\pi/\sqrt{3})\). The color-bar scales are in arbitrary units but consistent between plots. }
\label{fig:Inelastic Neutron plots B=1}
\end{figure}

At the mean-field level, the Zeeman field just gaps out the Dirac state (as shown in Appendix.\ref{appendix:MFT}), which is also seen in the RPA corrected response functions. The THz susceptibility is shown in Fig.\ref{fig:THz with B}. The high energy characteristics are similar to the \(B_x=0\) case, while the low energy shows a gap as expected. Meanwhile, the neutron scattering plots shown in Fig.\ref{fig:Inelastic Neutron plots B=1} exhibit flatter bandwidth of the sharp modes once the Zeeman field gaps out the Dirac nodes. The high energy response however looks similar to the \(B_x=0\) case.\par

\section{Discussion}
In this paper, we have shown that a modified phenomenological parton approach including RPA corrections is a useful route to understanding the phase diagram and spin dynamics in magnetic systems with weak ordering and a proximate spin-liquid state. In this approach, we split the spin Hamiltonian (written in terms of fermionic spinons) into two parts and decompose each using different HS fields : classical dynamic Weiss fields, as well as spinon hopping fields assumed to be condensed around their mean-field value. Subsequently we integrate out the spinons and are left with an effective action of the Weiss fields to leading order (equivalent to RPA). We then focus on a class of $J_1$-$J_3$ XY spin Hamiltonian on the honeycomb lattice relevant to the honeycomb cobaltates which showcase a plethora of magnetic orderings in a relatively narrow parameter regime. We show using VMC that there exists a regime where a pure DSL, and DSL+weak orderings like FM, ZZ, dZZ, and spiral order are all within \(1\)-\(2\%\) of each other in their optimized energies. Focusing on this regime, we employ our modified RPA method and reproduce the full many-body phase diagram. Furthermore, we illustrate how our RPA corrected magnetic susceptibility reproduces key features in THz spectroscopy \cite{Armitage2022} and inelastic neutron scattering experiments on \bcao \cite{BCAO_Broholm2023, Asim2024}. Following that, we study the system under an external in-plane Zeeman field and look at the phase diagram, magnetization curve, as well as physical response functions. We again find very good agreement with experimental results, especially for the polarizing field strength.\par
For future directions, it would be interesting to develop a controlled diagrammatic route to our methodology which can be formally justified for spin-\(1/2\) systems. Such a controlled setup can in principle be extended to include diagrams beyond RPA, and might perhaps lead to quantitative agreements with experimental results. Additionally, it would useful to explore the calculation of optimal parameters \(\{\a^{*}\}\) from the full free energy given in Eq.\eqref{total free energy}. It would also be interesting to study and connect our current results to the low energy theory which would consist of \(N_f=4\) flavour fermions coupled to a \(SU(2)\) gauge theory (not a \(U(1)\)) \cite{DiracMonopoles_PRX2020, Song2019, ASL_Hermele2005, DSL_Wen2004}. Specifically, one needs to understand what role do monopoles play in these possibly weakly ordered Dirac spin-liquids.\par
On the experimental side, there is a huge playground of materials which exhibit weak ordering but which may not be completely characterized by simple LSWT. There have been a series of work on triangular lattice anti-ferromagnet \ybmg where a similar RPA enhanced mean-field spinon fermi surface state has been successful in describing inelastic neutron scattering experiments \cite{triangleAFM2016, triangleAFM2017, triangleAFM_RPA}. Cobalt based materials on the triangular lattice are also described by some easy-plane spin-\(1/2\) Hamiltonian \textit{e.g.} \ncpo \cite{triangleSupersolid}, and \kcpo \cite{Zhu_2024}. Experiments on these materials show weak ordering at low temperatures, but exhibit spin-liquid like continuum in their INS spectrum, which makes them appropriate candidates for applying our methodology. In contrast, there is another triangular lattice material \ybzn which has recently garnered interest \cite{xu2023realizationu1diracquantum, J1J2_triangular, J1J2_dynamics}. It is a candidate material for hosting a Dirac spin liquid itself, described by a $J_1$-$J_2$ easy-axis model, and hence would be interesting to study using our approach as well. Investigating these extensions and candidate materials would improve our understanding of the interplay between spinons and magnons in the dynamics of frustrated magnets.

\bigskip

\acknowledgments
We acknowledge support from the Natural Sciences and Engineering Research Council (NSERC) of Canada. All numerical computations were performed on the Niagara supercomputer at the SciNet HPC Consortium and the Digital Research Alliance of Canada. The Julia
codes for tight-binding analysis and mean-field simulations are available online at \href{https://github.com/Anjishnubose/TightBindingToolkit.jl}{TightBindingToolkit.jl} and \href{https://github.com/Anjishnubose/MeanFieldToolkit.jl}{MeanFieldToolkit.jl} respectively. The bare susceptibility results were done in Julia as well as using TRIQS \cite{Parcollet_2015}

\bibliography{ref}

\begin{thebibliography}{75}%
\makeatletter
\providecommand \@ifxundefined [1]{%
 \@ifx{#1\undefined}
}%
\providecommand \@ifnum [1]{%
 \ifnum #1\expandafter \@firstoftwo
 \else \expandafter \@secondoftwo
 \fi
}%
\providecommand \@ifx [1]{%
 \ifx #1\expandafter \@firstoftwo
 \else \expandafter \@secondoftwo
 \fi
}%
\providecommand \natexlab [1]{#1}%
\providecommand \enquote  [1]{``#1''}%
\providecommand \bibnamefont  [1]{#1}%
\providecommand \bibfnamefont [1]{#1}%
\providecommand \citenamefont [1]{#1}%
\providecommand \href@noop [0]{\@secondoftwo}%
\providecommand \href [0]{\begingroup \@sanitize@url \@href}%
\providecommand \@href[1]{\@@startlink{#1}\@@href}%
\providecommand \@@href[1]{\endgroup#1\@@endlink}%
\providecommand \@sanitize@url [0]{\catcode `\\12\catcode `\$12\catcode `\&12\catcode `\#12\catcode `\^12\catcode `\_12\catcode `\%12\relax}%
\providecommand \@@startlink[1]{}%
\providecommand \@@endlink[0]{}%
\providecommand \url  [0]{\begingroup\@sanitize@url \@url }%
\providecommand \@url [1]{\endgroup\@href {#1}{\urlprefix }}%
\providecommand \urlprefix  [0]{URL }%
\providecommand \Eprint [0]{\href }%
\providecommand \doibase [0]{https://doi.org/}%
\providecommand \selectlanguage [0]{\@gobble}%
\providecommand \bibinfo  [0]{\@secondoftwo}%
\providecommand \bibfield  [0]{\@secondoftwo}%
\providecommand \translation [1]{[#1]}%
\providecommand \BibitemOpen [0]{}%
\providecommand \bibitemStop [0]{}%
\providecommand \bibitemNoStop [0]{.\EOS\space}%
\providecommand \EOS [0]{\spacefactor3000\relax}%
\providecommand \BibitemShut  [1]{\csname bibitem#1\endcsname}%
\let\auto@bib@innerbib\@empty
\bibitem [{\citenamefont {Knolle}\ and\ \citenamefont {Moessner}(2019)}]{QSL_Moessner2019}%
  \BibitemOpen
  \bibfield  {author} {\bibinfo {author} {\bibfnamefont {J.}~\bibnamefont {Knolle}}\ and\ \bibinfo {author} {\bibfnamefont {R.}~\bibnamefont {Moessner}},\ }\bibfield  {title} {\bibinfo {title} {A field guide to spin liquids},\ }\href {https://doi.org/https://doi.org/10.1146/annurev-conmatphys-031218-013401} {\bibfield  {journal} {\bibinfo  {journal} {Annual Review of Condensed Matter Physics}\ }\textbf {\bibinfo {volume} {10}},\ \bibinfo {pages} {451} (\bibinfo {year} {2019})}\BibitemShut {NoStop}%
\bibitem [{\citenamefont {Broholm}\ \emph {et~al.}(2020)\citenamefont {Broholm}, \citenamefont {Cava}, \citenamefont {Kivelson}, \citenamefont {Nocera}, \citenamefont {Norman},\ and\ \citenamefont {Senthil}}]{QSLreview_Broholm2020}%
  \BibitemOpen
  \bibfield  {author} {\bibinfo {author} {\bibfnamefont {C.}~\bibnamefont {Broholm}}, \bibinfo {author} {\bibfnamefont {R.~J.}\ \bibnamefont {Cava}}, \bibinfo {author} {\bibfnamefont {S.~A.}\ \bibnamefont {Kivelson}}, \bibinfo {author} {\bibfnamefont {D.~G.}\ \bibnamefont {Nocera}}, \bibinfo {author} {\bibfnamefont {M.~R.}\ \bibnamefont {Norman}},\ and\ \bibinfo {author} {\bibfnamefont {T.}~\bibnamefont {Senthil}},\ }\bibfield  {title} {\bibinfo {title} {Quantum spin liquids},\ }\href {https://doi.org/10.1126/science.aay0668} {\bibfield  {journal} {\bibinfo  {journal} {Science}\ }\textbf {\bibinfo {volume} {367}},\ \bibinfo {pages} {eaay0668} (\bibinfo {year} {2020})}\BibitemShut {NoStop}%
\bibitem [{\citenamefont {Chamorro}\ \emph {et~al.}(2021)\citenamefont {Chamorro}, \citenamefont {McQueen},\ and\ \citenamefont {Tran}}]{QSLreview_McQueen2021}%
  \BibitemOpen
  \bibfield  {author} {\bibinfo {author} {\bibfnamefont {J.~R.}\ \bibnamefont {Chamorro}}, \bibinfo {author} {\bibfnamefont {T.~M.}\ \bibnamefont {McQueen}},\ and\ \bibinfo {author} {\bibfnamefont {T.~T.}\ \bibnamefont {Tran}},\ }\bibfield  {title} {\bibinfo {title} {Chemistry of quantum spin liquids},\ }\bibfield  {booktitle} {\emph {\bibinfo {booktitle} {Chemical Reviews}},\ }\href {https://doi.org/10.1021/acs.chemrev.0c00641} {\bibfield  {journal} {\bibinfo  {journal} {Chemical Reviews}\ }\textbf {\bibinfo {volume} {121}},\ \bibinfo {pages} {2898} (\bibinfo {year} {2021})}\BibitemShut {NoStop}%
\bibitem [{\citenamefont {Savary}\ and\ \citenamefont {Balents}(2016)}]{QSLreview_Savary_2017}%
  \BibitemOpen
  \bibfield  {author} {\bibinfo {author} {\bibfnamefont {L.}~\bibnamefont {Savary}}\ and\ \bibinfo {author} {\bibfnamefont {L.}~\bibnamefont {Balents}},\ }\bibfield  {title} {\bibinfo {title} {Quantum spin liquids: a review},\ }\href {https://doi.org/10.1088/0034-4885/80/1/016502} {\bibfield  {journal} {\bibinfo  {journal} {Reports on Progress in Physics}\ }\textbf {\bibinfo {volume} {80}},\ \bibinfo {pages} {016502} (\bibinfo {year} {2016})}\BibitemShut {NoStop}%
\bibitem [{\citenamefont {Garanin}\ \emph {et~al.}(1992)\citenamefont {Garanin}, \citenamefont {Lutovinov},\ and\ \citenamefont {Panina}}]{magnon1992}%
  \BibitemOpen
  \bibfield  {author} {\bibinfo {author} {\bibfnamefont {D.}~\bibnamefont {Garanin}}, \bibinfo {author} {\bibfnamefont {V.}~\bibnamefont {Lutovinov}},\ and\ \bibinfo {author} {\bibfnamefont {L.}~\bibnamefont {Panina}},\ }\bibfield  {title} {\bibinfo {title} {The magnon-magnon interactions in easy-plane antiferromagnets},\ }\href {https://doi.org/https://doi.org/10.1016/0378-4371(92)90321-G} {\bibfield  {journal} {\bibinfo  {journal} {Physica A: Statistical Mechanics and its Applications}\ }\textbf {\bibinfo {volume} {184}},\ \bibinfo {pages} {523} (\bibinfo {year} {1992})}\BibitemShut {NoStop}%
\bibitem [{\citenamefont {Sourounis}\ and\ \citenamefont {Manchon}(2024)}]{magnon2024}%
  \BibitemOpen
  \bibfield  {author} {\bibinfo {author} {\bibfnamefont {K.}~\bibnamefont {Sourounis}}\ and\ \bibinfo {author} {\bibfnamefont {A.}~\bibnamefont {Manchon}},\ }\bibfield  {title} {\bibinfo {title} {Impact of magnon interactions on transport in honeycomb antiferromagnets},\ }\href {https://doi.org/10.1103/PhysRevB.110.054429} {\bibfield  {journal} {\bibinfo  {journal} {Phys. Rev. B}\ }\textbf {\bibinfo {volume} {110}},\ \bibinfo {pages} {054429} (\bibinfo {year} {2024})}\BibitemShut {NoStop}%
\bibitem [{\citenamefont {Wang}\ \emph {et~al.}(2024)\citenamefont {Wang}, \citenamefont {Mustafi}, \citenamefont {Fogh}, \citenamefont {Astrakhantsev}, \citenamefont {He}, \citenamefont {Bialo}, \citenamefont {Chan}, \citenamefont {Martinelli}, \citenamefont {Horio}, \citenamefont {Ivashko}, \citenamefont {Shaik}, \citenamefont {von Arx}, \citenamefont {Sassa}, \citenamefont {Paris}, \citenamefont {Fischer}, \citenamefont {Tseng}, \citenamefont {Christensen}, \citenamefont {Galdi}, \citenamefont {Schlom}, \citenamefont {Shen}, \citenamefont {Schmitt}, \citenamefont {Ronnow},\ and\ \citenamefont {Chang}}]{Wang2024}%
  \BibitemOpen
  \bibfield  {author} {\bibinfo {author} {\bibfnamefont {Q.}~\bibnamefont {Wang}}, \bibinfo {author} {\bibfnamefont {S.}~\bibnamefont {Mustafi}}, \bibinfo {author} {\bibfnamefont {E.}~\bibnamefont {Fogh}}, \bibinfo {author} {\bibfnamefont {N.}~\bibnamefont {Astrakhantsev}}, \bibinfo {author} {\bibfnamefont {Z.}~\bibnamefont {He}}, \bibinfo {author} {\bibfnamefont {I.}~\bibnamefont {Bialo}}, \bibinfo {author} {\bibfnamefont {Y.}~\bibnamefont {Chan}}, \bibinfo {author} {\bibfnamefont {L.}~\bibnamefont {Martinelli}}, \bibinfo {author} {\bibfnamefont {M.}~\bibnamefont {Horio}}, \bibinfo {author} {\bibfnamefont {O.}~\bibnamefont {Ivashko}}, \bibinfo {author} {\bibfnamefont {N.~E.}\ \bibnamefont {Shaik}}, \bibinfo {author} {\bibfnamefont {K.}~\bibnamefont {von Arx}}, \bibinfo {author} {\bibfnamefont {Y.}~\bibnamefont {Sassa}}, \bibinfo {author} {\bibfnamefont {E.}~\bibnamefont {Paris}}, \bibinfo {author} {\bibfnamefont {M.~H.}\ \bibnamefont {Fischer}}, \bibinfo {author} {\bibfnamefont {Y.}~\bibnamefont {Tseng}},
  \bibinfo {author} {\bibfnamefont {N.~B.}\ \bibnamefont {Christensen}}, \bibinfo {author} {\bibfnamefont {A.}~\bibnamefont {Galdi}}, \bibinfo {author} {\bibfnamefont {D.~G.}\ \bibnamefont {Schlom}}, \bibinfo {author} {\bibfnamefont {K.~M.}\ \bibnamefont {Shen}}, \bibinfo {author} {\bibfnamefont {T.}~\bibnamefont {Schmitt}}, \bibinfo {author} {\bibfnamefont {H.~M.}\ \bibnamefont {Ronnow}},\ and\ \bibinfo {author} {\bibfnamefont {J.}~\bibnamefont {Chang}},\ }\bibfield  {title} {\bibinfo {title} {{Magnon interactions in a moderately correlated Mott insulator}},\ }\href {https://doi.org/10.1038/s41467-024-49714-y} {\bibfield  {journal} {\bibinfo  {journal} {Nature Communications}\ }\textbf {\bibinfo {volume} {15}},\ \bibinfo {pages} {5348} (\bibinfo {year} {2024})}\BibitemShut {NoStop}%
\bibitem [{\citenamefont {Chubukov}\ \emph {et~al.}(1994)\citenamefont {Chubukov}, \citenamefont {Sachdev},\ and\ \citenamefont {Senthil}}]{Chubukov_largeS1994}%
  \BibitemOpen
  \bibfield  {author} {\bibinfo {author} {\bibfnamefont {A.~V.}\ \bibnamefont {Chubukov}}, \bibinfo {author} {\bibfnamefont {S.}~\bibnamefont {Sachdev}},\ and\ \bibinfo {author} {\bibfnamefont {T.}~\bibnamefont {Senthil}},\ }\bibfield  {title} {\bibinfo {title} {Large-s expansion for quantum antiferromagnets on a triangular lattice},\ }\href {https://doi.org/10.1088/0953-8984/6/42/019} {\bibfield  {journal} {\bibinfo  {journal} {Journal of Physics: Condensed Matter}\ }\textbf {\bibinfo {volume} {6}},\ \bibinfo {pages} {8891} (\bibinfo {year} {1994})}\BibitemShut {NoStop}%
\bibitem [{\citenamefont {Chubukov}\ and\ \citenamefont {Musaelian}(1994)}]{Chubokov_largeSHubbard}%
  \BibitemOpen
  \bibfield  {author} {\bibinfo {author} {\bibfnamefont {A.~V.}\ \bibnamefont {Chubukov}}\ and\ \bibinfo {author} {\bibfnamefont {K.~A.}\ \bibnamefont {Musaelian}},\ }\bibfield  {title} {\bibinfo {title} {Systematic 1/s study of the two-dimensional hubbard model at half-filling},\ }\href {https://doi.org/10.1103/PhysRevB.50.6238} {\bibfield  {journal} {\bibinfo  {journal} {Phys. Rev. B}\ }\textbf {\bibinfo {volume} {50}},\ \bibinfo {pages} {6238} (\bibinfo {year} {1994})}\BibitemShut {NoStop}%
\bibitem [{\citenamefont {Chubukov}\ and\ \citenamefont {Jolicoeur}(1992)}]{Chubokov_largeS1992}%
  \BibitemOpen
  \bibfield  {author} {\bibinfo {author} {\bibfnamefont {A.~V.}\ \bibnamefont {Chubukov}}\ and\ \bibinfo {author} {\bibfnamefont {T.}~\bibnamefont {Jolicoeur}},\ }\bibfield  {title} {\bibinfo {title} {Order-from-disorder phenomena in heisenberg antiferromagnets on a triangular lattice},\ }\href {https://doi.org/10.1103/PhysRevB.46.11137} {\bibfield  {journal} {\bibinfo  {journal} {Phys. Rev. B}\ }\textbf {\bibinfo {volume} {46}},\ \bibinfo {pages} {11137} (\bibinfo {year} {1992})}\BibitemShut {NoStop}%
\bibitem [{\citenamefont {Winter}\ \emph {et~al.}(2017)\citenamefont {Winter}, \citenamefont {Riedl}, \citenamefont {Maksimov}, \citenamefont {Chernyshev}, \citenamefont {Honecker},\ and\ \citenamefont {Valenti}}]{Chernyshev2017}%
  \BibitemOpen
  \bibfield  {author} {\bibinfo {author} {\bibfnamefont {S.~M.}\ \bibnamefont {Winter}}, \bibinfo {author} {\bibfnamefont {K.}~\bibnamefont {Riedl}}, \bibinfo {author} {\bibfnamefont {P.~A.}\ \bibnamefont {Maksimov}}, \bibinfo {author} {\bibfnamefont {A.~L.}\ \bibnamefont {Chernyshev}}, \bibinfo {author} {\bibfnamefont {A.}~\bibnamefont {Honecker}},\ and\ \bibinfo {author} {\bibfnamefont {R.}~\bibnamefont {Valenti}},\ }\bibfield  {title} {\bibinfo {title} {{Breakdown of magnons in a strongly spin-orbital coupled magnet}},\ }\href {https://doi.org/10.1038/s41467-017-01177-0} {\bibfield  {journal} {\bibinfo  {journal} {Nature Communications}\ }\textbf {\bibinfo {volume} {8}},\ \bibinfo {pages} {1152} (\bibinfo {year} {2017})}\BibitemShut {NoStop}%
\bibitem [{\citenamefont {Lakshmanan}(2011)}]{Lakshmanan_2011}%
  \BibitemOpen
  \bibfield  {author} {\bibinfo {author} {\bibfnamefont {M.}~\bibnamefont {Lakshmanan}},\ }\bibfield  {title} {\bibinfo {title} {The fascinating world of the {Landau$-$Lifshitz$-$Gilbert} equation: an overview},\ }\href {https://doi.org/10.1098/rsta.2010.0319} {\bibfield  {journal} {\bibinfo  {journal} {Philosophical Transactions of the Royal Society A: Mathematical, Physical and Engineering Sciences}\ }\textbf {\bibinfo {volume} {369}},\ \bibinfo {pages} {1280–1300} (\bibinfo {year} {2011})}\BibitemShut {NoStop}%
\bibitem [{\citenamefont {Zhang}\ \emph {et~al.}(2023{\natexlab{a}})\citenamefont {Zhang}, \citenamefont {Wilke},\ and\ \citenamefont {Kim}}]{YBK_magnon2023}%
  \BibitemOpen
  \bibfield  {author} {\bibinfo {author} {\bibfnamefont {E.~Z.}\ \bibnamefont {Zhang}}, \bibinfo {author} {\bibfnamefont {R.~H.}\ \bibnamefont {Wilke}},\ and\ \bibinfo {author} {\bibfnamefont {Y.~B.}\ \bibnamefont {Kim}},\ }\bibfield  {title} {\bibinfo {title} {Spin excitation continuum to topological magnon crossover and thermal hall conductivity in kitaev magnets},\ }\href {https://doi.org/10.1103/PhysRevB.107.184418} {\bibfield  {journal} {\bibinfo  {journal} {Phys. Rev. B}\ }\textbf {\bibinfo {volume} {107}},\ \bibinfo {pages} {184418} (\bibinfo {year} {2023}{\natexlab{a}})}\BibitemShut {NoStop}%
\bibitem [{\citenamefont {Halloran}\ \emph {et~al.}(2023)\citenamefont {Halloran}, \citenamefont {Desrochers}, \citenamefont {Zhang}, \citenamefont {Chen}, \citenamefont {Chern}, \citenamefont {Xu}, \citenamefont {Winn}, \citenamefont {Graves-Brook}, \citenamefont {Stone}, \citenamefont {Kolesnikov}, \citenamefont {Qiu}, \citenamefont {Zhong}, \citenamefont {Cava}, \citenamefont {Kim},\ and\ \citenamefont {Broholm}}]{BCAO_Broholm2023}%
  \BibitemOpen
  \bibfield  {author} {\bibinfo {author} {\bibfnamefont {T.}~\bibnamefont {Halloran}}, \bibinfo {author} {\bibfnamefont {F.}~\bibnamefont {Desrochers}}, \bibinfo {author} {\bibfnamefont {E.~Z.}\ \bibnamefont {Zhang}}, \bibinfo {author} {\bibfnamefont {T.}~\bibnamefont {Chen}}, \bibinfo {author} {\bibfnamefont {L.~E.}\ \bibnamefont {Chern}}, \bibinfo {author} {\bibfnamefont {Z.}~\bibnamefont {Xu}}, \bibinfo {author} {\bibfnamefont {B.}~\bibnamefont {Winn}}, \bibinfo {author} {\bibfnamefont {M.}~\bibnamefont {Graves-Brook}}, \bibinfo {author} {\bibfnamefont {M.~B.}\ \bibnamefont {Stone}}, \bibinfo {author} {\bibfnamefont {A.~I.}\ \bibnamefont {Kolesnikov}}, \bibinfo {author} {\bibfnamefont {Y.}~\bibnamefont {Qiu}}, \bibinfo {author} {\bibfnamefont {R.}~\bibnamefont {Zhong}}, \bibinfo {author} {\bibfnamefont {R.}~\bibnamefont {Cava}}, \bibinfo {author} {\bibfnamefont {Y.~B.}\ \bibnamefont {Kim}},\ and\ \bibinfo {author} {\bibfnamefont {C.}~\bibnamefont {Broholm}},\ }\bibfield  {title} {\bibinfo {title}
  {Geometrical frustration versus {Kitaev} interactions in {BaCo$_2$(AsO$_4$)$_2$}},\ }\href {https://doi.org/10.1073/pnas.2215509119} {\bibfield  {journal} {\bibinfo  {journal} {Proceedings of the National Academy of Sciences}\ }\textbf {\bibinfo {volume} {120}},\ \bibinfo {pages} {e2215509119} (\bibinfo {year} {2023})}\BibitemShut {NoStop}%
\bibitem [{\citenamefont {Hosoi}\ \emph {et~al.}(2022)\citenamefont {Hosoi}, \citenamefont {Zhang}, \citenamefont {Patri},\ and\ \citenamefont {Kim}}]{YBK_spinice2022}%
  \BibitemOpen
  \bibfield  {author} {\bibinfo {author} {\bibfnamefont {M.}~\bibnamefont {Hosoi}}, \bibinfo {author} {\bibfnamefont {E.~Z.}\ \bibnamefont {Zhang}}, \bibinfo {author} {\bibfnamefont {A.~S.}\ \bibnamefont {Patri}},\ and\ \bibinfo {author} {\bibfnamefont {Y.~B.}\ \bibnamefont {Kim}},\ }\bibfield  {title} {\bibinfo {title} {Uncovering footprints of dipolar-octupolar quantum spin ice from neutron scattering signatures},\ }\href {https://doi.org/10.1103/PhysRevLett.129.097202} {\bibfield  {journal} {\bibinfo  {journal} {Phys. Rev. Lett.}\ }\textbf {\bibinfo {volume} {129}},\ \bibinfo {pages} {097202} (\bibinfo {year} {2022})}\BibitemShut {NoStop}%
\bibitem [{\citenamefont {Schollw\"ock}(2005)}]{dmrg_schwol}%
  \BibitemOpen
  \bibfield  {author} {\bibinfo {author} {\bibfnamefont {U.}~\bibnamefont {Schollw\"ock}},\ }\bibfield  {title} {\bibinfo {title} {The density-matrix renormalization group},\ }\href {https://doi.org/10.1103/RevModPhys.77.259} {\bibfield  {journal} {\bibinfo  {journal} {Rev. Mod. Phys.}\ }\textbf {\bibinfo {volume} {77}},\ \bibinfo {pages} {259} (\bibinfo {year} {2005})}\BibitemShut {NoStop}%
\bibitem [{\citenamefont {Catarina}\ and\ \citenamefont {Murta}(2023)}]{dmrg_Catarina_2023}%
  \BibitemOpen
  \bibfield  {author} {\bibinfo {author} {\bibfnamefont {G.}~\bibnamefont {Catarina}}\ and\ \bibinfo {author} {\bibfnamefont {B.}~\bibnamefont {Murta}},\ }\bibfield  {title} {\bibinfo {title} {Density-matrix renormalization group: a pedagogical introduction},\ }\bibfield  {journal} {\bibinfo  {journal} {The European Physical Journal B}\ }\textbf {\bibinfo {volume} {96}},\ \href {https://doi.org/10.1140/epjb/s10051-023-00575-2} {10.1140/epjb/s10051-023-00575-2} (\bibinfo {year} {2023})\BibitemShut {NoStop}%
\bibitem [{\citenamefont {Becca}\ and\ \citenamefont {Sorella}(2017)}]{becca_sorella_2017}%
  \BibitemOpen
  \bibfield  {author} {\bibinfo {author} {\bibfnamefont {F.}~\bibnamefont {Becca}}\ and\ \bibinfo {author} {\bibfnamefont {S.}~\bibnamefont {Sorella}},\ }\href {https://doi.org/10.1017/9781316417041} {\emph {\bibinfo {title} {Quantum Monte Carlo Approaches for Correlated Systems}}}\ (\bibinfo  {publisher} {Cambridge University Press},\ \bibinfo {year} {2017})\BibitemShut {NoStop}%
\bibitem [{\citenamefont {Song}(2024)}]{song2024neuralquantumstatesvariational}%
  \BibitemOpen
  \bibfield  {author} {\bibinfo {author} {\bibfnamefont {Y.}~\bibnamefont {Song}},\ }\href {https://arxiv.org/abs/2406.01017} {\bibinfo {title} {Neural quantum states in variational monte carlo method: A brief summary}} (\bibinfo {year} {2024}),\ \Eprint {https://arxiv.org/abs/2406.01017} {arXiv:2406.01017 [cond-mat.str-el]} \BibitemShut {NoStop}%
\bibitem [{\citenamefont {Noculak}\ and\ \citenamefont {Reuther}(2024)}]{pffrg_Bfield}%
  \BibitemOpen
  \bibfield  {author} {\bibinfo {author} {\bibfnamefont {V.}~\bibnamefont {Noculak}}\ and\ \bibinfo {author} {\bibfnamefont {J.}~\bibnamefont {Reuther}},\ }\bibfield  {title} {\bibinfo {title} {Pseudo-fermion functional renormalization group with magnetic fields},\ }\href {https://doi.org/10.1103/PhysRevB.109.174414} {\bibfield  {journal} {\bibinfo  {journal} {Phys. Rev. B}\ }\textbf {\bibinfo {volume} {109}},\ \bibinfo {pages} {174414} (\bibinfo {year} {2024})}\BibitemShut {NoStop}%
\bibitem [{\citenamefont {Müller}\ \emph {et~al.}(2024)\citenamefont {Müller}, \citenamefont {Kiese}, \citenamefont {Niggemann}, \citenamefont {Sbierski}, \citenamefont {Reuther}, \citenamefont {Trebst}, \citenamefont {Thomale},\ and\ \citenamefont {Iqbal}}]{pffrg_rev}%
  \BibitemOpen
  \bibfield  {author} {\bibinfo {author} {\bibfnamefont {T.}~\bibnamefont {Müller}}, \bibinfo {author} {\bibfnamefont {D.}~\bibnamefont {Kiese}}, \bibinfo {author} {\bibfnamefont {N.}~\bibnamefont {Niggemann}}, \bibinfo {author} {\bibfnamefont {B.}~\bibnamefont {Sbierski}}, \bibinfo {author} {\bibfnamefont {J.}~\bibnamefont {Reuther}}, \bibinfo {author} {\bibfnamefont {S.}~\bibnamefont {Trebst}}, \bibinfo {author} {\bibfnamefont {R.}~\bibnamefont {Thomale}},\ and\ \bibinfo {author} {\bibfnamefont {Y.}~\bibnamefont {Iqbal}},\ }\bibfield  {title} {\bibinfo {title} {Pseudo-fermion functional renormalization group for spin models},\ }\href {https://doi.org/10.1088/1361-6633/ad208c} {\bibfield  {journal} {\bibinfo  {journal} {Reports on Progress in Physics}\ }\textbf {\bibinfo {volume} {87}},\ \bibinfo {pages} {036501} (\bibinfo {year} {2024})}\BibitemShut {NoStop}%
\bibitem [{\citenamefont {Chern}\ \emph {et~al.}(2024)\citenamefont {Chern}, \citenamefont {Desrochers}, \citenamefont {Kim},\ and\ \citenamefont {Castelnovo}}]{pffrg_pyrocholore}%
  \BibitemOpen
  \bibfield  {author} {\bibinfo {author} {\bibfnamefont {L.~E.}\ \bibnamefont {Chern}}, \bibinfo {author} {\bibfnamefont {F.}~\bibnamefont {Desrochers}}, \bibinfo {author} {\bibfnamefont {Y.~B.}\ \bibnamefont {Kim}},\ and\ \bibinfo {author} {\bibfnamefont {C.}~\bibnamefont {Castelnovo}},\ }\bibfield  {title} {\bibinfo {title} {Pseudofermion functional renormalization group study of dipolar-octupolar pyrochlore magnets},\ }\href {https://doi.org/10.1103/PhysRevB.109.184421} {\bibfield  {journal} {\bibinfo  {journal} {Phys. Rev. B}\ }\textbf {\bibinfo {volume} {109}},\ \bibinfo {pages} {184421} (\bibinfo {year} {2024})}\BibitemShut {NoStop}%
\bibitem [{\citenamefont {Coleman}(2015)}]{Coleman_2015}%
  \BibitemOpen
  \bibfield  {author} {\bibinfo {author} {\bibfnamefont {P.}~\bibnamefont {Coleman}},\ }\href@noop {} {\emph {\bibinfo {title} {Introduction to Many-Body Physics}}}\ (\bibinfo  {publisher} {Cambridge University Press},\ \bibinfo {year} {2015})\BibitemShut {NoStop}%
\bibitem [{\citenamefont {Bardeen}\ \emph {et~al.}(1957)\citenamefont {Bardeen}, \citenamefont {Cooper},\ and\ \citenamefont {Schrieffer}}]{BCS1957}%
  \BibitemOpen
  \bibfield  {author} {\bibinfo {author} {\bibfnamefont {J.}~\bibnamefont {Bardeen}}, \bibinfo {author} {\bibfnamefont {L.~N.}\ \bibnamefont {Cooper}},\ and\ \bibinfo {author} {\bibfnamefont {J.~R.}\ \bibnamefont {Schrieffer}},\ }\bibfield  {title} {\bibinfo {title} {Theory of superconductivity},\ }\href {https://doi.org/10.1103/PhysRev.108.1175} {\bibfield  {journal} {\bibinfo  {journal} {Phys. Rev.}\ }\textbf {\bibinfo {volume} {108}},\ \bibinfo {pages} {1175} (\bibinfo {year} {1957})}\BibitemShut {NoStop}%
\bibitem [{\citenamefont {Anderson}(1966)}]{Anderson1966}%
  \BibitemOpen
  \bibfield  {author} {\bibinfo {author} {\bibfnamefont {P.~W.}\ \bibnamefont {Anderson}},\ }\bibfield  {title} {\bibinfo {title} {Considerations on the flow of superfluid helium},\ }\href {https://doi.org/10.1103/RevModPhys.38.298} {\bibfield  {journal} {\bibinfo  {journal} {Rev. Mod. Phys.}\ }\textbf {\bibinfo {volume} {38}},\ \bibinfo {pages} {298} (\bibinfo {year} {1966})}\BibitemShut {NoStop}%
\bibitem [{\citenamefont {Hermele}\ \emph {et~al.}(2004)\citenamefont {Hermele}, \citenamefont {Senthil}, \citenamefont {Fisher}, \citenamefont {Lee}, \citenamefont {Nagaosa},\ and\ \citenamefont {Wen}}]{DSL_Wen2004}%
  \BibitemOpen
  \bibfield  {author} {\bibinfo {author} {\bibfnamefont {M.}~\bibnamefont {Hermele}}, \bibinfo {author} {\bibfnamefont {T.}~\bibnamefont {Senthil}}, \bibinfo {author} {\bibfnamefont {M.~P.~A.}\ \bibnamefont {Fisher}}, \bibinfo {author} {\bibfnamefont {P.~A.}\ \bibnamefont {Lee}}, \bibinfo {author} {\bibfnamefont {N.}~\bibnamefont {Nagaosa}},\ and\ \bibinfo {author} {\bibfnamefont {X.-G.}\ \bibnamefont {Wen}},\ }\bibfield  {title} {\bibinfo {title} {Stability of $u(1)$ spin liquids in two dimensions},\ }\href {https://doi.org/10.1103/PhysRevB.70.214437} {\bibfield  {journal} {\bibinfo  {journal} {Phys. Rev. B}\ }\textbf {\bibinfo {volume} {70}},\ \bibinfo {pages} {214437} (\bibinfo {year} {2004})}\BibitemShut {NoStop}%
\bibitem [{\citenamefont {Hermele}\ \emph {et~al.}(2005)\citenamefont {Hermele}, \citenamefont {Senthil},\ and\ \citenamefont {Fisher}}]{ASL_Hermele2005}%
  \BibitemOpen
  \bibfield  {author} {\bibinfo {author} {\bibfnamefont {M.}~\bibnamefont {Hermele}}, \bibinfo {author} {\bibfnamefont {T.}~\bibnamefont {Senthil}},\ and\ \bibinfo {author} {\bibfnamefont {M.~P.~A.}\ \bibnamefont {Fisher}},\ }\bibfield  {title} {\bibinfo {title} {Algebraic spin liquid as the mother of many competing orders},\ }\href {https://doi.org/10.1103/PhysRevB.72.104404} {\bibfield  {journal} {\bibinfo  {journal} {Phys. Rev. B}\ }\textbf {\bibinfo {volume} {72}},\ \bibinfo {pages} {104404} (\bibinfo {year} {2005})}\BibitemShut {NoStop}%
\bibitem [{\citenamefont {Song}\ \emph {et~al.}(2020)\citenamefont {Song}, \citenamefont {He}, \citenamefont {Vishwanath},\ and\ \citenamefont {Wang}}]{DiracMonopoles_PRX2020}%
  \BibitemOpen
  \bibfield  {author} {\bibinfo {author} {\bibfnamefont {X.-Y.}\ \bibnamefont {Song}}, \bibinfo {author} {\bibfnamefont {Y.-C.}\ \bibnamefont {He}}, \bibinfo {author} {\bibfnamefont {A.}~\bibnamefont {Vishwanath}},\ and\ \bibinfo {author} {\bibfnamefont {C.}~\bibnamefont {Wang}},\ }\bibfield  {title} {\bibinfo {title} {From spinon band topology to the symmetry quantum numbers of monopoles in dirac spin liquids},\ }\href {https://doi.org/10.1103/PhysRevX.10.011033} {\bibfield  {journal} {\bibinfo  {journal} {Phys. Rev. X}\ }\textbf {\bibinfo {volume} {10}},\ \bibinfo {pages} {011033} (\bibinfo {year} {2020})}\BibitemShut {NoStop}%
\bibitem [{\citenamefont {Song}\ \emph {et~al.}(2019)\citenamefont {Song}, \citenamefont {Wang}, \citenamefont {Vishwanath},\ and\ \citenamefont {He}}]{Song2019}%
  \BibitemOpen
  \bibfield  {author} {\bibinfo {author} {\bibfnamefont {X.-Y.}\ \bibnamefont {Song}}, \bibinfo {author} {\bibfnamefont {C.}~\bibnamefont {Wang}}, \bibinfo {author} {\bibfnamefont {A.}~\bibnamefont {Vishwanath}},\ and\ \bibinfo {author} {\bibfnamefont {Y.-C.}\ \bibnamefont {He}},\ }\bibfield  {title} {\bibinfo {title} {{Unifying description of competing orders in two-dimensional quantum magnets}},\ }\href {https://doi.org/10.1038/s41467-019-11727-3} {\bibfield  {journal} {\bibinfo  {journal} {Nature Communications}\ }\textbf {\bibinfo {volume} {10}},\ \bibinfo {pages} {4254} (\bibinfo {year} {2019})}\BibitemShut {NoStop}%
\bibitem [{\citenamefont {Wen}(2007)}]{QSLgauge_XGWenBook}%
  \BibitemOpen
  \bibfield  {author} {\bibinfo {author} {\bibfnamefont {X.~G.}\ \bibnamefont {Wen}},\ }\href {https://doi.org/10.1093/acprof:oso/9780199227259.001.0001} {\emph {\bibinfo {title} {{Quantum field theory of many-body systems: from the origin of sound to an origin of light and electrons}}}}\ (\bibinfo  {publisher} {Oxford University Press},\ \bibinfo {address} {Oxford},\ \bibinfo {year} {2007})\BibitemShut {NoStop}%
\bibitem [{\citenamefont {Lee}(2014)}]{QSLgauge_Lee2014}%
  \BibitemOpen
  \bibfield  {author} {\bibinfo {author} {\bibfnamefont {P.~A.}\ \bibnamefont {Lee}},\ }\bibfield  {title} {\bibinfo {title} {Quantum spin liquid: a tale of emergence from frustration},\ }\href {https://doi.org/10.1088/1742-6596/529/1/012001} {\bibfield  {journal} {\bibinfo  {journal} {Journal of Physics: Conference Series}\ }\textbf {\bibinfo {volume} {529}},\ \bibinfo {pages} {012001} (\bibinfo {year} {2014})}\BibitemShut {NoStop}%
\bibitem [{\citenamefont {Brinckmann}\ and\ \citenamefont {Lee}(1997)}]{Lee1997}%
  \BibitemOpen
  \bibfield  {author} {\bibinfo {author} {\bibfnamefont {J.}~\bibnamefont {Brinckmann}}\ and\ \bibinfo {author} {\bibfnamefont {P.~A.}\ \bibnamefont {Lee}},\ }\bibfield  {title} {\bibinfo {title} {Spin susceptibility and the $\pi$-excitation in underdoped cuprates},\ }\href {https://api.semanticscholar.org/CorpusID:98622029} {\bibfield  {journal} {\bibinfo  {journal} {Journal of Physics and Chemistry of Solids}\ }\textbf {\bibinfo {volume} {59}},\ \bibinfo {pages} {1811} (\bibinfo {year} {1997})}\BibitemShut {NoStop}%
\bibitem [{\citenamefont {Brinckmann}\ and\ \citenamefont {Lee}(2001)}]{Lee2001}%
  \BibitemOpen
  \bibfield  {author} {\bibinfo {author} {\bibfnamefont {J.}~\bibnamefont {Brinckmann}}\ and\ \bibinfo {author} {\bibfnamefont {P.~A.}\ \bibnamefont {Lee}},\ }\bibfield  {title} {\bibinfo {title} {Renormalized mean-field theory of neutron scattering in cuprate superconductors},\ }\href {https://doi.org/10.1103/PhysRevB.65.014502} {\bibfield  {journal} {\bibinfo  {journal} {Phys. Rev. B}\ }\textbf {\bibinfo {volume} {65}},\ \bibinfo {pages} {014502} (\bibinfo {year} {2001})}\BibitemShut {NoStop}%
\bibitem [{\citenamefont {Brinckmann}\ and\ \citenamefont {Lee}(1999)}]{Lee1999_boson}%
  \BibitemOpen
  \bibfield  {author} {\bibinfo {author} {\bibfnamefont {J.}~\bibnamefont {Brinckmann}}\ and\ \bibinfo {author} {\bibfnamefont {P.~A.}\ \bibnamefont {Lee}},\ }\bibfield  {title} {\bibinfo {title} {Slave boson approach to neutron scattering in ${{\mathrm{YBa}}_{2}{\mathrm{Cu}}_{3}O}_{6+\mathit{y}}$ superconductors},\ }\href {https://doi.org/10.1103/PhysRevLett.82.2915} {\bibfield  {journal} {\bibinfo  {journal} {Phys. Rev. Lett.}\ }\textbf {\bibinfo {volume} {82}},\ \bibinfo {pages} {2915} (\bibinfo {year} {1999})}\BibitemShut {NoStop}%
\bibitem [{\citenamefont {Abanov}\ and\ \citenamefont {Chubukov}(2000)}]{Chubokov_spinFermion2000}%
  \BibitemOpen
  \bibfield  {author} {\bibinfo {author} {\bibfnamefont {A.}~\bibnamefont {Abanov}}\ and\ \bibinfo {author} {\bibfnamefont {A.~V.}\ \bibnamefont {Chubukov}},\ }\bibfield  {title} {\bibinfo {title} {Spin-fermion model near the quantum critical point: One-loop renormalization group results},\ }\href {https://doi.org/10.1103/PhysRevLett.84.5608} {\bibfield  {journal} {\bibinfo  {journal} {Phys. Rev. Lett.}\ }\textbf {\bibinfo {volume} {84}},\ \bibinfo {pages} {5608} (\bibinfo {year} {2000})}\BibitemShut {NoStop}%
\bibitem [{\citenamefont {Ar.~Abanov}\ and\ \citenamefont {Schmalian}(2003)}]{Chubokov_spinFermion2003}%
  \BibitemOpen
  \bibfield  {author} {\bibinfo {author} {\bibfnamefont {A.~V.~C.}\ \bibnamefont {Ar.~Abanov}}\ and\ \bibinfo {author} {\bibfnamefont {J.}~\bibnamefont {Schmalian}},\ }\bibfield  {title} {\bibinfo {title} {Quantum-critical theory of the spin-fermion model and its application to cuprates: Normal state analysis},\ }\href {https://doi.org/10.1080/0001873021000057123} {\bibfield  {journal} {\bibinfo  {journal} {Advances in Physics}\ }\textbf {\bibinfo {volume} {52}},\ \bibinfo {pages} {119} (\bibinfo {year} {2003})},\ \Eprint {https://arxiv.org/abs/https://doi.org/10.1080/0001873021000057123} {https://doi.org/10.1080/0001873021000057123} \BibitemShut {NoStop}%
\bibitem [{\citenamefont {Balents}\ and\ \citenamefont {Starykh}(2020)}]{Starykh_2020}%
  \BibitemOpen
  \bibfield  {author} {\bibinfo {author} {\bibfnamefont {L.}~\bibnamefont {Balents}}\ and\ \bibinfo {author} {\bibfnamefont {O.~A.}\ \bibnamefont {Starykh}},\ }\bibfield  {title} {\bibinfo {title} {Collective spinon spin wave in a magnetized u(1) spin liquid},\ }\href {https://doi.org/10.1103/PhysRevB.101.020401} {\bibfield  {journal} {\bibinfo  {journal} {Phys. Rev. B}\ }\textbf {\bibinfo {volume} {101}},\ \bibinfo {pages} {020401} (\bibinfo {year} {2020})}\BibitemShut {NoStop}%
\bibitem [{\citenamefont {Regnault}\ \emph {et~al.}(1977)\citenamefont {Regnault}, \citenamefont {Burlet},\ and\ \citenamefont {Rossat-Mignod}}]{BCAO_regnault1977}%
  \BibitemOpen
  \bibfield  {author} {\bibinfo {author} {\bibfnamefont {L.}~\bibnamefont {Regnault}}, \bibinfo {author} {\bibfnamefont {P.}~\bibnamefont {Burlet}},\ and\ \bibinfo {author} {\bibfnamefont {J.}~\bibnamefont {Rossat-Mignod}},\ }\bibfield  {title} {\bibinfo {title} {Magnetic ordering in a planar {XY} model: {BaCo$_2$(AsO$_4$)$_2$}},\ }\href {https://doi.org/https://doi.org/10.1016/0378-4363(77)90635-0} {\bibfield  {journal} {\bibinfo  {journal} {Physica B+C}\ }\textbf {\bibinfo {volume} {86-88}},\ \bibinfo {pages} {660} (\bibinfo {year} {1977})}\BibitemShut {NoStop}%
\bibitem [{\citenamefont {Regnault}\ \emph {et~al.}(2018)\citenamefont {Regnault}, \citenamefont {Boullier},\ and\ \citenamefont {Lorenzo}}]{BCAO_Regnault2018}%
  \BibitemOpen
  \bibfield  {author} {\bibinfo {author} {\bibfnamefont {L.~P.}\ \bibnamefont {Regnault}}, \bibinfo {author} {\bibfnamefont {C.}~\bibnamefont {Boullier}},\ and\ \bibinfo {author} {\bibfnamefont {J.~E.}\ \bibnamefont {Lorenzo}},\ }\bibfield  {title} {\bibinfo {title} {Polarized-neutron investigation of magnetic ordering and spin dynamics in {BaCo$_{2}$(AsO$_{4}$)$_{2}$} frustrated honeycomb-lattice magnet},\ }\href {https://doi.org/10.1016/j.heliyon.2018.e00507} {\bibfield  {journal} {\bibinfo  {journal} {Heliyon}\ }\textbf {\bibinfo {volume} {4}},\ \bibinfo {pages} {E00507} (\bibinfo {year} {2018})}\BibitemShut {NoStop}%
\bibitem [{\citenamefont {Zhong}\ \emph {et~al.}(2019)\citenamefont {Zhong}, \citenamefont {Gao}, \citenamefont {Ong},\ and\ \citenamefont {Cava}}]{BCAO_Zhong2019}%
  \BibitemOpen
  \bibfield  {author} {\bibinfo {author} {\bibfnamefont {R.}~\bibnamefont {Zhong}}, \bibinfo {author} {\bibfnamefont {T.}~\bibnamefont {Gao}}, \bibinfo {author} {\bibfnamefont {N.~P.}\ \bibnamefont {Ong}},\ and\ \bibinfo {author} {\bibfnamefont {R.~J.}\ \bibnamefont {Cava}},\ }\bibfield  {title} {\bibinfo {title} {Weak-field induced nonmagnetic state in a {Co}-based honeycomb},\ }\href {https://doi.org/10.1126/sciadv.aay6953} {\bibfield  {journal} {\bibinfo  {journal} {Science Advances}\ }\textbf {\bibinfo {volume} {6 (4)}},\ \bibinfo {pages} {aay6953} (\bibinfo {year} {2019})}\BibitemShut {NoStop}%
\bibitem [{\citenamefont {Zhang}\ \emph {et~al.}(2023{\natexlab{b}})\citenamefont {Zhang}, \citenamefont {Xu}, \citenamefont {Halloran}, \citenamefont {Zhong}, \citenamefont {Broholm}, \citenamefont {Cava}, \citenamefont {Drichko},\ and\ \citenamefont {Armitage}}]{Armitage2022}%
  \BibitemOpen
  \bibfield  {author} {\bibinfo {author} {\bibfnamefont {X.}~\bibnamefont {Zhang}}, \bibinfo {author} {\bibfnamefont {Y.}~\bibnamefont {Xu}}, \bibinfo {author} {\bibfnamefont {T.}~\bibnamefont {Halloran}}, \bibinfo {author} {\bibfnamefont {R.}~\bibnamefont {Zhong}}, \bibinfo {author} {\bibfnamefont {C.}~\bibnamefont {Broholm}}, \bibinfo {author} {\bibfnamefont {R.~J.}\ \bibnamefont {Cava}}, \bibinfo {author} {\bibfnamefont {N.}~\bibnamefont {Drichko}},\ and\ \bibinfo {author} {\bibfnamefont {N.~P.}\ \bibnamefont {Armitage}},\ }\bibfield  {title} {\bibinfo {title} {A magnetic continuum in the cobalt-based honeycomb magnet {BaCo$_2$(AsO$_4$)$_2$}},\ }\href {https://doi.org/10.1038/s41563-022-01403-1} {\bibfield  {journal} {\bibinfo  {journal} {Nature Materials}\ }\textbf {\bibinfo {volume} {22}},\ \bibinfo {pages} {58} (\bibinfo {year} {2023}{\natexlab{b}})}\BibitemShut {NoStop}%
\bibitem [{\citenamefont {{Tu}}\ \emph {et~al.}(2022)\citenamefont {{Tu}}, \citenamefont {{Dai}}, \citenamefont {{Zhang}}, \citenamefont {{Zhao}}, \citenamefont {{Jin}}, \citenamefont {{Gao}}, \citenamefont {{Dai}},\ and\ \citenamefont {{Li}}}]{BCAO_thermalcond_Li2022}%
  \BibitemOpen
  \bibfield  {author} {\bibinfo {author} {\bibfnamefont {C.}~\bibnamefont {{Tu}}}, \bibinfo {author} {\bibfnamefont {D.}~\bibnamefont {{Dai}}}, \bibinfo {author} {\bibfnamefont {X.}~\bibnamefont {{Zhang}}}, \bibinfo {author} {\bibfnamefont {C.}~\bibnamefont {{Zhao}}}, \bibinfo {author} {\bibfnamefont {X.}~\bibnamefont {{Jin}}}, \bibinfo {author} {\bibfnamefont {B.}~\bibnamefont {{Gao}}}, \bibinfo {author} {\bibfnamefont {P.}~\bibnamefont {{Dai}}},\ and\ \bibinfo {author} {\bibfnamefont {S.}~\bibnamefont {{Li}}},\ }\bibfield  {title} {\bibinfo {title} {{Evidence for gapless quantum spin liquid in a honeycomb lattice}},\ }\href@noop {} {\bibfield  {journal} {\bibinfo  {journal} {arXiv e-prints}\ ,\ \bibinfo {eid} {arXiv:2212.07322}} (\bibinfo {year} {2022})},\ \Eprint {https://arxiv.org/abs/2212.07322} {arXiv:2212.07322 [cond-mat.str-el]} \BibitemShut {NoStop}%
\bibitem [{\citenamefont {Nair}\ \emph {et~al.}(2018)\citenamefont {Nair}, \citenamefont {Brown}, \citenamefont {Coldren}, \citenamefont {Hester}, \citenamefont {Gelfand}, \citenamefont {Podlesnyak}, \citenamefont {Huang},\ and\ \citenamefont {Ross}}]{BCPO_Nair2018}%
  \BibitemOpen
  \bibfield  {author} {\bibinfo {author} {\bibfnamefont {H.~S.}\ \bibnamefont {Nair}}, \bibinfo {author} {\bibfnamefont {J.~M.}\ \bibnamefont {Brown}}, \bibinfo {author} {\bibfnamefont {E.}~\bibnamefont {Coldren}}, \bibinfo {author} {\bibfnamefont {G.}~\bibnamefont {Hester}}, \bibinfo {author} {\bibfnamefont {M.~P.}\ \bibnamefont {Gelfand}}, \bibinfo {author} {\bibfnamefont {A.}~\bibnamefont {Podlesnyak}}, \bibinfo {author} {\bibfnamefont {Q.}~\bibnamefont {Huang}},\ and\ \bibinfo {author} {\bibfnamefont {K.~A.}\ \bibnamefont {Ross}},\ }\bibfield  {title} {\bibinfo {title} {Short-range order in the quantum {XXZ} honeycomb lattice material {${\mathrm{BaCo}}_{2}{({\mathrm{PO}}_{4})}_{2}$}},\ }\href {https://doi.org/10.1103/PhysRevB.97.134409} {\bibfield  {journal} {\bibinfo  {journal} {Phys. Rev. B}\ }\textbf {\bibinfo {volume} {97}},\ \bibinfo {pages} {134409} (\bibinfo {year} {2018})}\BibitemShut {NoStop}%
\bibitem [{\citenamefont {Lefrancois}\ \emph {et~al.}(2016)\citenamefont {Lefrancois}, \citenamefont {Songvilay}, \citenamefont {Robert}, \citenamefont {Nataf}, \citenamefont {Jordan}, \citenamefont {Chaix}, \citenamefont {Colin}, \citenamefont {Lejay}, \citenamefont {Hadj-Azzem}, \citenamefont {Ballou},\ and\ \citenamefont {Simonet}}]{ncto_simonet2016}%
  \BibitemOpen
  \bibfield  {author} {\bibinfo {author} {\bibfnamefont {E.}~\bibnamefont {Lefrancois}}, \bibinfo {author} {\bibfnamefont {M.}~\bibnamefont {Songvilay}}, \bibinfo {author} {\bibfnamefont {J.}~\bibnamefont {Robert}}, \bibinfo {author} {\bibfnamefont {G.}~\bibnamefont {Nataf}}, \bibinfo {author} {\bibfnamefont {E.}~\bibnamefont {Jordan}}, \bibinfo {author} {\bibfnamefont {L.}~\bibnamefont {Chaix}}, \bibinfo {author} {\bibfnamefont {C.~V.}\ \bibnamefont {Colin}}, \bibinfo {author} {\bibfnamefont {P.}~\bibnamefont {Lejay}}, \bibinfo {author} {\bibfnamefont {A.}~\bibnamefont {Hadj-Azzem}}, \bibinfo {author} {\bibfnamefont {R.}~\bibnamefont {Ballou}},\ and\ \bibinfo {author} {\bibfnamefont {V.}~\bibnamefont {Simonet}},\ }\bibfield  {title} {\bibinfo {title} {Magnetic properties of the honeycomb oxide {Na$_3$Co$_2$TeO$_6$}},\ }\href {https://doi.org/10.1103/PhysRevB.94.214416} {\bibfield  {journal} {\bibinfo  {journal} {Phys. Rev. B}\ }\textbf {\bibinfo {volume} {94}},\ \bibinfo {pages} {214416} (\bibinfo {year}
  {2016})}\BibitemShut {NoStop}%
\bibitem [{\citenamefont {Songvilay}\ \emph {et~al.}(2020)\citenamefont {Songvilay}, \citenamefont {Robert}, \citenamefont {Petit}, \citenamefont {Rodriguez-Rivera}, \citenamefont {Ratcliff}, \citenamefont {Damay}, \citenamefont {Bal\'edent}, \citenamefont {Jim\'enez-Ruiz}, \citenamefont {Lejay}, \citenamefont {Pachoud}, \citenamefont {Hadj-Azzem}, \citenamefont {Simonet},\ and\ \citenamefont {Stock}}]{ncto_ncso_stock2020}%
  \BibitemOpen
  \bibfield  {author} {\bibinfo {author} {\bibfnamefont {M.}~\bibnamefont {Songvilay}}, \bibinfo {author} {\bibfnamefont {J.}~\bibnamefont {Robert}}, \bibinfo {author} {\bibfnamefont {S.}~\bibnamefont {Petit}}, \bibinfo {author} {\bibfnamefont {J.~A.}\ \bibnamefont {Rodriguez-Rivera}}, \bibinfo {author} {\bibfnamefont {W.~D.}\ \bibnamefont {Ratcliff}}, \bibinfo {author} {\bibfnamefont {F.}~\bibnamefont {Damay}}, \bibinfo {author} {\bibfnamefont {V.}~\bibnamefont {Bal\'edent}}, \bibinfo {author} {\bibfnamefont {M.}~\bibnamefont {Jim\'enez-Ruiz}}, \bibinfo {author} {\bibfnamefont {P.}~\bibnamefont {Lejay}}, \bibinfo {author} {\bibfnamefont {E.}~\bibnamefont {Pachoud}}, \bibinfo {author} {\bibfnamefont {A.}~\bibnamefont {Hadj-Azzem}}, \bibinfo {author} {\bibfnamefont {V.}~\bibnamefont {Simonet}},\ and\ \bibinfo {author} {\bibfnamefont {C.}~\bibnamefont {Stock}},\ }\bibfield  {title} {\bibinfo {title} {Kitaev interactions in the {Co} honeycomb antiferromagnets {Na$_3$Co$_2$SbO$_6$} and {Na$_3$Co$_2$TeO$_6$}},\
  }\href {https://doi.org/10.1103/PhysRevB.102.224429} {\bibfield  {journal} {\bibinfo  {journal} {Phys. Rev. B}\ }\textbf {\bibinfo {volume} {102}},\ \bibinfo {pages} {224429} (\bibinfo {year} {2020})}\BibitemShut {NoStop}%
\bibitem [{\citenamefont {Yang}\ \emph {et~al.}(2022)\citenamefont {Yang}, \citenamefont {Kim}, \citenamefont {Choi}, \citenamefont {Lee}, \citenamefont {Lin}, \citenamefont {Ma}, \citenamefont {Kratochvilova}, \citenamefont {Proschek}, \citenamefont {Moon}, \citenamefont {Lee}, \citenamefont {Oh},\ and\ \citenamefont {Park}}]{Kappaxy_NCoTeO_Park2022}%
  \BibitemOpen
  \bibfield  {author} {\bibinfo {author} {\bibfnamefont {H.}~\bibnamefont {Yang}}, \bibinfo {author} {\bibfnamefont {C.}~\bibnamefont {Kim}}, \bibinfo {author} {\bibfnamefont {Y.}~\bibnamefont {Choi}}, \bibinfo {author} {\bibfnamefont {J.~H.}\ \bibnamefont {Lee}}, \bibinfo {author} {\bibfnamefont {G.}~\bibnamefont {Lin}}, \bibinfo {author} {\bibfnamefont {J.}~\bibnamefont {Ma}}, \bibinfo {author} {\bibfnamefont {M.}~\bibnamefont {Kratochvilova}}, \bibinfo {author} {\bibfnamefont {P.}~\bibnamefont {Proschek}}, \bibinfo {author} {\bibfnamefont {E.-G.}\ \bibnamefont {Moon}}, \bibinfo {author} {\bibfnamefont {K.~H.}\ \bibnamefont {Lee}}, \bibinfo {author} {\bibfnamefont {Y.~S.}\ \bibnamefont {Oh}},\ and\ \bibinfo {author} {\bibfnamefont {J.-G.}\ \bibnamefont {Park}},\ }\bibfield  {title} {\bibinfo {title} {Significant thermal {Hall} effect in the $3d$ cobalt {Kitaev} system {Na$_2$Co$_2$TeO$_6$}},\ }\href {https://doi.org/10.1103/PhysRevB.106.L081116} {\bibfield  {journal} {\bibinfo  {journal} {Phys. Rev. B}\
  }\textbf {\bibinfo {volume} {106}},\ \bibinfo {pages} {L081116} (\bibinfo {year} {2022})}\BibitemShut {NoStop}%
\bibitem [{\citenamefont {Zhang}\ \emph {et~al.}(2023{\natexlab{c}})\citenamefont {Zhang}, \citenamefont {Lee}, \citenamefont {Woods}, \citenamefont {Thomas}, \citenamefont {Movshovich}, \citenamefont {Brosha}, \citenamefont {Huang}, \citenamefont {Zhou}, \citenamefont {Zapf},\ and\ \citenamefont {Lee}}]{LANL_ncto2022}%
  \BibitemOpen
  \bibfield  {author} {\bibinfo {author} {\bibfnamefont {S.}~\bibnamefont {Zhang}}, \bibinfo {author} {\bibfnamefont {S.}~\bibnamefont {Lee}}, \bibinfo {author} {\bibfnamefont {A.~J.}\ \bibnamefont {Woods}}, \bibinfo {author} {\bibfnamefont {S.~M.}\ \bibnamefont {Thomas}}, \bibinfo {author} {\bibfnamefont {R.}~\bibnamefont {Movshovich}}, \bibinfo {author} {\bibfnamefont {E.}~\bibnamefont {Brosha}}, \bibinfo {author} {\bibfnamefont {Q.}~\bibnamefont {Huang}}, \bibinfo {author} {\bibfnamefont {H.}~\bibnamefont {Zhou}}, \bibinfo {author} {\bibfnamefont {V.~S.}\ \bibnamefont {Zapf}},\ and\ \bibinfo {author} {\bibfnamefont {M.}~\bibnamefont {Lee}},\ }\href@noop {} {\bibinfo {title} {Electronic and magnetic phase diagrams of {Kitaev} quantum spin liquid candidate {Na$_2$Co$_2$TeO$_6$}}} (\bibinfo {year} {2023}{\natexlab{c}}),\ \Eprint {https://arxiv.org/abs/2212.03849} {arXiv:2212.03849 [cond-mat.str-el]} \BibitemShut {NoStop}%
\bibitem [{\citenamefont {Guang}\ \emph {et~al.}(2023)\citenamefont {Guang}, \citenamefont {Li}, \citenamefont {Luo}, \citenamefont {Huang}, \citenamefont {Wang}, \citenamefont {Yue}, \citenamefont {Xia}, \citenamefont {Li}, \citenamefont {Zhao}, \citenamefont {Chen}, \citenamefont {Zhou},\ and\ \citenamefont {Sun}}]{kappa_NCTO_Sun_PRB_2023}%
  \BibitemOpen
  \bibfield  {author} {\bibinfo {author} {\bibfnamefont {S.}~\bibnamefont {Guang}}, \bibinfo {author} {\bibfnamefont {N.}~\bibnamefont {Li}}, \bibinfo {author} {\bibfnamefont {R.~L.}\ \bibnamefont {Luo}}, \bibinfo {author} {\bibfnamefont {Q.}~\bibnamefont {Huang}}, \bibinfo {author} {\bibfnamefont {Y.}~\bibnamefont {Wang}}, \bibinfo {author} {\bibfnamefont {X.}~\bibnamefont {Yue}}, \bibinfo {author} {\bibfnamefont {K.}~\bibnamefont {Xia}}, \bibinfo {author} {\bibfnamefont {Q.}~\bibnamefont {Li}}, \bibinfo {author} {\bibfnamefont {X.}~\bibnamefont {Zhao}}, \bibinfo {author} {\bibfnamefont {G.}~\bibnamefont {Chen}}, \bibinfo {author} {\bibfnamefont {H.}~\bibnamefont {Zhou}},\ and\ \bibinfo {author} {\bibfnamefont {X.}~\bibnamefont {Sun}},\ }\bibfield  {title} {\bibinfo {title} {Thermal transport of fractionalized antiferromagnetic and field-induced states in the {Kitaev material} {${\mathrm{Na}}_{2}{\mathrm{Co}}_{2}{\mathrm{TeO}}_{6}$}},\ }\href {https://doi.org/10.1103/PhysRevB.107.184423} {\bibfield
  {journal} {\bibinfo  {journal} {Phys. Rev. B}\ }\textbf {\bibinfo {volume} {107}},\ \bibinfo {pages} {184423} (\bibinfo {year} {2023})}\BibitemShut {NoStop}%
\bibitem [{\citenamefont {Yan}\ \emph {et~al.}(2019)\citenamefont {Yan}, \citenamefont {Okamoto}, \citenamefont {Wu}, \citenamefont {Zheng}, \citenamefont {Zhou}, \citenamefont {Cao},\ and\ \citenamefont {McGuire}}]{ncso_mcguire2019}%
  \BibitemOpen
  \bibfield  {author} {\bibinfo {author} {\bibfnamefont {J.-Q.}\ \bibnamefont {Yan}}, \bibinfo {author} {\bibfnamefont {S.}~\bibnamefont {Okamoto}}, \bibinfo {author} {\bibfnamefont {Y.}~\bibnamefont {Wu}}, \bibinfo {author} {\bibfnamefont {Q.}~\bibnamefont {Zheng}}, \bibinfo {author} {\bibfnamefont {H.~D.}\ \bibnamefont {Zhou}}, \bibinfo {author} {\bibfnamefont {H.~B.}\ \bibnamefont {Cao}},\ and\ \bibinfo {author} {\bibfnamefont {M.~A.}\ \bibnamefont {McGuire}},\ }\bibfield  {title} {\bibinfo {title} {Magnetic order in single crystals of {${\mathrm{Na}}_{3}{\mathrm{Co}}_{2}{\mathrm{SbO}}_{6}$} with a honeycomb arrangement of {${3\mathrm{d}}^{7}\phantom{\rule{0.28em}{0ex}}{\mathrm{Co}}^{2+}$} ions},\ }\href {https://doi.org/10.1103/PhysRevMaterials.3.074405} {\bibfield  {journal} {\bibinfo  {journal} {Phys. Rev. Materials}\ }\textbf {\bibinfo {volume} {3}},\ \bibinfo {pages} {074405} (\bibinfo {year} {2019})}\BibitemShut {NoStop}%
\bibitem [{\citenamefont {Liu}\ and\ \citenamefont {Khaliullin}(2018)}]{CoKitaev_Liu2018}%
  \BibitemOpen
  \bibfield  {author} {\bibinfo {author} {\bibfnamefont {H.}~\bibnamefont {Liu}}\ and\ \bibinfo {author} {\bibfnamefont {G.}~\bibnamefont {Khaliullin}},\ }\bibfield  {title} {\bibinfo {title} {Pseudospin exchange interactions in ${d}^{7}$ cobalt compounds: Possible realization of the {Kitaev} model},\ }\href {https://doi.org/10.1103/PhysRevB.97.014407} {\bibfield  {journal} {\bibinfo  {journal} {Phys. Rev. B}\ }\textbf {\bibinfo {volume} {97}},\ \bibinfo {pages} {014407} (\bibinfo {year} {2018})}\BibitemShut {NoStop}%
\bibitem [{\citenamefont {Liu}\ \emph {et~al.}(2020)\citenamefont {Liu}, \citenamefont {Chaloupka},\ and\ \citenamefont {Khaliullin}}]{CoKitaev_Liu2020}%
  \BibitemOpen
  \bibfield  {author} {\bibinfo {author} {\bibfnamefont {H.}~\bibnamefont {Liu}}, \bibinfo {author} {\bibfnamefont {J.~c.~v.}\ \bibnamefont {Chaloupka}},\ and\ \bibinfo {author} {\bibfnamefont {G.}~\bibnamefont {Khaliullin}},\ }\bibfield  {title} {\bibinfo {title} {Kitaev spin liquid in $3d$ transition metal compounds},\ }\href {https://doi.org/10.1103/PhysRevLett.125.047201} {\bibfield  {journal} {\bibinfo  {journal} {Phys. Rev. Lett.}\ }\textbf {\bibinfo {volume} {125}},\ \bibinfo {pages} {047201} (\bibinfo {year} {2020})}\BibitemShut {NoStop}%
\bibitem [{\citenamefont {Das}\ \emph {et~al.}(2021)\citenamefont {Das}, \citenamefont {Voleti}, \citenamefont {Saha-Dasgupta},\ and\ \citenamefont {Paramekanti}}]{Abinitio_Das2021}%
  \BibitemOpen
  \bibfield  {author} {\bibinfo {author} {\bibfnamefont {S.}~\bibnamefont {Das}}, \bibinfo {author} {\bibfnamefont {S.}~\bibnamefont {Voleti}}, \bibinfo {author} {\bibfnamefont {T.}~\bibnamefont {Saha-Dasgupta}},\ and\ \bibinfo {author} {\bibfnamefont {A.}~\bibnamefont {Paramekanti}},\ }\bibfield  {title} {\bibinfo {title} {Xy magnetism, {Kitaev} exchange, and long-range frustration in the ${J}_{\mathrm{eff}}=\frac{1}{2}$ honeycomb cobaltates},\ }\href {https://doi.org/10.1103/PhysRevB.104.134425} {\bibfield  {journal} {\bibinfo  {journal} {Phys. Rev. B}\ }\textbf {\bibinfo {volume} {104}},\ \bibinfo {pages} {134425} (\bibinfo {year} {2021})}\BibitemShut {NoStop}%
\bibitem [{\citenamefont {Samanta}\ \emph {et~al.}(2022)\citenamefont {Samanta}, \citenamefont {Detrattanawichai}, \citenamefont {Na-Phattalung},\ and\ \citenamefont {Kim}}]{Abinitio_HSKim2022}%
  \BibitemOpen
  \bibfield  {author} {\bibinfo {author} {\bibfnamefont {S.}~\bibnamefont {Samanta}}, \bibinfo {author} {\bibfnamefont {P.}~\bibnamefont {Detrattanawichai}}, \bibinfo {author} {\bibfnamefont {S.}~\bibnamefont {Na-Phattalung}},\ and\ \bibinfo {author} {\bibfnamefont {H.-S.}\ \bibnamefont {Kim}},\ }\bibfield  {title} {\bibinfo {title} {Active orbital degree of freedom and potential spin-orbit-entangled moments in the {Kitaev} magnet candidate {${\mathrm{BaCo}}_{2}{({\mathrm{AsO}}_{4})}_{2}$}},\ }\href {https://doi.org/10.1103/PhysRevB.106.195136} {\bibfield  {journal} {\bibinfo  {journal} {Phys. Rev. B}\ }\textbf {\bibinfo {volume} {106}},\ \bibinfo {pages} {195136} (\bibinfo {year} {2022})}\BibitemShut {NoStop}%
\bibitem [{\citenamefont {Maksimov}\ \emph {et~al.}(2022)\citenamefont {Maksimov}, \citenamefont {Ushakov}, \citenamefont {Pchelkina}, \citenamefont {Li}, \citenamefont {Winter},\ and\ \citenamefont {Streltsov}}]{Abinitio_Streltsov2022}%
  \BibitemOpen
  \bibfield  {author} {\bibinfo {author} {\bibfnamefont {P.~A.}\ \bibnamefont {Maksimov}}, \bibinfo {author} {\bibfnamefont {A.~V.}\ \bibnamefont {Ushakov}}, \bibinfo {author} {\bibfnamefont {Z.~V.}\ \bibnamefont {Pchelkina}}, \bibinfo {author} {\bibfnamefont {Y.}~\bibnamefont {Li}}, \bibinfo {author} {\bibfnamefont {S.~M.}\ \bibnamefont {Winter}},\ and\ \bibinfo {author} {\bibfnamefont {S.~V.}\ \bibnamefont {Streltsov}},\ }\bibfield  {title} {\bibinfo {title} {Ab initio guided minimal model for the ``{Kitaev}'' material {${\mathrm{BaCo}}_{2}$(${\mathrm{AsO}}_{4}{)}_{2}$}: {Importance} of direct hopping, third-neighbor exchange, and quantum fluctuations},\ }\href {https://doi.org/10.1103/PhysRevB.106.165131} {\bibfield  {journal} {\bibinfo  {journal} {Phys. Rev. B}\ }\textbf {\bibinfo {volume} {106}},\ \bibinfo {pages} {165131} (\bibinfo {year} {2022})}\BibitemShut {NoStop}%
\bibitem [{\citenamefont {Jiang}\ \emph {et~al.}(2023)\citenamefont {Jiang}, \citenamefont {White},\ and\ \citenamefont {Chernyshev}}]{Chernyshev2023}%
  \BibitemOpen
  \bibfield  {author} {\bibinfo {author} {\bibfnamefont {S.}~\bibnamefont {Jiang}}, \bibinfo {author} {\bibfnamefont {S.~R.}\ \bibnamefont {White}},\ and\ \bibinfo {author} {\bibfnamefont {A.~L.}\ \bibnamefont {Chernyshev}},\ }\bibfield  {title} {\bibinfo {title} {Quantum phases in the honeycomb-lattice ${J}_{1}$--${J}_{3}$ ferro-antiferromagnetic model},\ }\href {https://doi.org/10.1103/PhysRevB.108.L180406} {\bibfield  {journal} {\bibinfo  {journal} {Phys. Rev. B}\ }\textbf {\bibinfo {volume} {108}},\ \bibinfo {pages} {L180406} (\bibinfo {year} {2023})}\BibitemShut {NoStop}%
\bibitem [{\citenamefont {Maksimov}\ and\ \citenamefont {Chernyshev}(2022)}]{spinwaveXXZ_chernyshev_PRB2022}%
  \BibitemOpen
  \bibfield  {author} {\bibinfo {author} {\bibfnamefont {P.~A.}\ \bibnamefont {Maksimov}}\ and\ \bibinfo {author} {\bibfnamefont {A.~L.}\ \bibnamefont {Chernyshev}},\ }\bibfield  {title} {\bibinfo {title} {Easy-plane anisotropic-exchange magnets on a honeycomb lattice: Quantum effects and dealing with them},\ }\href {https://doi.org/10.1103/PhysRevB.106.214411} {\bibfield  {journal} {\bibinfo  {journal} {Phys. Rev. B}\ }\textbf {\bibinfo {volume} {106}},\ \bibinfo {pages} {214411} (\bibinfo {year} {2022})}\BibitemShut {NoStop}%
\bibitem [{\citenamefont {Watanabe}\ \emph {et~al.}(2022)\citenamefont {Watanabe}, \citenamefont {Trebst},\ and\ \citenamefont {Hickey}}]{j1j3_watanabe2022frustrated}%
  \BibitemOpen
  \bibfield  {author} {\bibinfo {author} {\bibfnamefont {Y.}~\bibnamefont {Watanabe}}, \bibinfo {author} {\bibfnamefont {S.}~\bibnamefont {Trebst}},\ and\ \bibinfo {author} {\bibfnamefont {C.}~\bibnamefont {Hickey}},\ }\href@noop {} {\bibinfo {title} {Frustrated ferromagnetism of honeycomb cobaltates: Incommensurate spirals, quantum disordered phases, and out-of-plane {Ising} order}} (\bibinfo {year} {2022}),\ \Eprint {https://arxiv.org/abs/2212.14053} {arXiv:2212.14053 [cond-mat.str-el]} \BibitemShut {NoStop}%
\bibitem [{\citenamefont {Fouet}\ \emph {et~al.}(2001)\citenamefont {Fouet}, \citenamefont {Sindzingre},\ and\ \citenamefont {Lhuillier}}]{j1j2j3_Fouet2001}%
  \BibitemOpen
  \bibfield  {author} {\bibinfo {author} {\bibfnamefont {J.~B.}\ \bibnamefont {Fouet}}, \bibinfo {author} {\bibfnamefont {P.}~\bibnamefont {Sindzingre}},\ and\ \bibinfo {author} {\bibfnamefont {C.}~\bibnamefont {Lhuillier}},\ }\bibfield  {title} {\bibinfo {title} {An investigation of the quantum {J1-J2-J3} model on the honeycomb lattice},\ }\href {https://doi.org/10.1007/s100510170273} {\bibfield  {journal} {\bibinfo  {journal} {The European Physical Journal B - Condensed Matter and Complex Systems}\ }\textbf {\bibinfo {volume} {20}},\ \bibinfo {pages} {241} (\bibinfo {year} {2001})}\BibitemShut {NoStop}%
\bibitem [{\citenamefont {Safari}\ \emph {et~al.}(2024)\citenamefont {Safari}, \citenamefont {Bateman-Hemphill}, \citenamefont {Mitra}, \citenamefont {Desrochers}, \citenamefont {Zhang}, \citenamefont {Shafeek}, \citenamefont {Ferrenti}, \citenamefont {McQueen}, \citenamefont {Shekhter}, \citenamefont {Köllö}, \citenamefont {Kim}, \citenamefont {Ramshaw},\ and\ \citenamefont {Modic}}]{Asim2024}%
  \BibitemOpen
  \bibfield  {author} {\bibinfo {author} {\bibfnamefont {S.}~\bibnamefont {Safari}}, \bibinfo {author} {\bibfnamefont {W.}~\bibnamefont {Bateman-Hemphill}}, \bibinfo {author} {\bibfnamefont {A.}~\bibnamefont {Mitra}}, \bibinfo {author} {\bibfnamefont {F.}~\bibnamefont {Desrochers}}, \bibinfo {author} {\bibfnamefont {E.~Z.}\ \bibnamefont {Zhang}}, \bibinfo {author} {\bibfnamefont {L.}~\bibnamefont {Shafeek}}, \bibinfo {author} {\bibfnamefont {A.}~\bibnamefont {Ferrenti}}, \bibinfo {author} {\bibfnamefont {T.~M.}\ \bibnamefont {McQueen}}, \bibinfo {author} {\bibfnamefont {A.}~\bibnamefont {Shekhter}}, \bibinfo {author} {\bibfnamefont {Z.}~\bibnamefont {Köllö}}, \bibinfo {author} {\bibfnamefont {Y.~B.}\ \bibnamefont {Kim}}, \bibinfo {author} {\bibfnamefont {B.~J.}\ \bibnamefont {Ramshaw}},\ and\ \bibinfo {author} {\bibfnamefont {K.~A.}\ \bibnamefont {Modic}},\ }\href {https://arxiv.org/abs/2403.15315} {\bibinfo {title} {Quantum fluctuations suppress the critical fields in baco$_2$(aso$_4$)$_2$}} (\bibinfo
  {year} {2024}),\ \Eprint {https://arxiv.org/abs/2403.15315} {arXiv:2403.15315 [cond-mat.str-el]} \BibitemShut {NoStop}%
\bibitem [{\citenamefont {Bose}\ \emph {et~al.}(2023)\citenamefont {Bose}, \citenamefont {Routh}, \citenamefont {Voleti}, \citenamefont {Saha}, \citenamefont {Kumar}, \citenamefont {Saha-Dasgupta},\ and\ \citenamefont {Paramekanti}}]{bose_dirac}%
  \BibitemOpen
  \bibfield  {author} {\bibinfo {author} {\bibfnamefont {A.}~\bibnamefont {Bose}}, \bibinfo {author} {\bibfnamefont {M.}~\bibnamefont {Routh}}, \bibinfo {author} {\bibfnamefont {S.}~\bibnamefont {Voleti}}, \bibinfo {author} {\bibfnamefont {S.~K.}\ \bibnamefont {Saha}}, \bibinfo {author} {\bibfnamefont {M.}~\bibnamefont {Kumar}}, \bibinfo {author} {\bibfnamefont {T.}~\bibnamefont {Saha-Dasgupta}},\ and\ \bibinfo {author} {\bibfnamefont {A.}~\bibnamefont {Paramekanti}},\ }\bibfield  {title} {\bibinfo {title} {Proximate dirac spin liquid in the honeycomb lattice ${J}_{1}\text{\ensuremath{-}}{J}_{3}$ xxz model: Numerical study and application to cobaltates},\ }\href {https://doi.org/10.1103/PhysRevB.108.174422} {\bibfield  {journal} {\bibinfo  {journal} {Phys. Rev. B}\ }\textbf {\bibinfo {volume} {108}},\ \bibinfo {pages} {174422} (\bibinfo {year} {2023})}\BibitemShut {NoStop}%
\bibitem [{\citenamefont {Polyakov}(1974)}]{polyakov}%
  \BibitemOpen
  \bibfield  {author} {\bibinfo {author} {\bibfnamefont {A.~M.}\ \bibnamefont {Polyakov}},\ }\bibfield  {title} {\bibinfo {title} {{Particle Spectrum in Quantum Field Theory}},\ }\href@noop {} {\bibfield  {journal} {\bibinfo  {journal} {JETP Lett.}\ }\textbf {\bibinfo {volume} {20}},\ \bibinfo {pages} {194} (\bibinfo {year} {1974})}\BibitemShut {NoStop}%
\bibitem [{\citenamefont {Zhou}\ and\ \citenamefont {Wen}(2003)}]{Wen2002}%
  \BibitemOpen
  \bibfield  {author} {\bibinfo {author} {\bibfnamefont {Y.}~\bibnamefont {Zhou}}\ and\ \bibinfo {author} {\bibfnamefont {X.-G.}\ \bibnamefont {Wen}},\ }\href {https://arxiv.org/abs/cond-mat/0210662} {\bibinfo {title} {Quantum orders and spin liquids in cs$_2$cucl$_4$}} (\bibinfo {year} {2003}),\ \Eprint {https://arxiv.org/abs/cond-mat/0210662} {arXiv:cond-mat/0210662 [cond-mat.str-el]} \BibitemShut {NoStop}%
\bibitem [{\citenamefont {Mei}\ and\ \citenamefont {Wen}(2015)}]{Wen2015}%
  \BibitemOpen
  \bibfield  {author} {\bibinfo {author} {\bibfnamefont {J.-W.}\ \bibnamefont {Mei}}\ and\ \bibinfo {author} {\bibfnamefont {X.-G.}\ \bibnamefont {Wen}},\ }\href {https://arxiv.org/abs/1507.03007} {\bibinfo {title} {Fractionalized spin-wave continuum in spin liquid states on the kagome lattice}} (\bibinfo {year} {2015}),\ \Eprint {https://arxiv.org/abs/1507.03007} {arXiv:1507.03007 [cond-mat.str-el]} \BibitemShut {NoStop}%
\bibitem [{\citenamefont {Sonnenschein}\ \emph {et~al.}(2020)\citenamefont {Sonnenschein}, \citenamefont {Chauhan}, \citenamefont {Iqbal},\ and\ \citenamefont {Reuther}}]{Iqbal2020}%
  \BibitemOpen
  \bibfield  {author} {\bibinfo {author} {\bibfnamefont {J.}~\bibnamefont {Sonnenschein}}, \bibinfo {author} {\bibfnamefont {A.}~\bibnamefont {Chauhan}}, \bibinfo {author} {\bibfnamefont {Y.}~\bibnamefont {Iqbal}},\ and\ \bibinfo {author} {\bibfnamefont {J.}~\bibnamefont {Reuther}},\ }\bibfield  {title} {\bibinfo {title} {Projective symmetry group classifications of quantum spin liquids on the simple cubic, body centered cubic, and face centered cubic lattices},\ }\href {https://doi.org/10.1103/PhysRevB.102.125140} {\bibfield  {journal} {\bibinfo  {journal} {Phys. Rev. B}\ }\textbf {\bibinfo {volume} {102}},\ \bibinfo {pages} {125140} (\bibinfo {year} {2020})}\BibitemShut {NoStop}%
\bibitem [{\citenamefont {Chung}\ \emph {et~al.}(2003)\citenamefont {Chung}, \citenamefont {Voelker},\ and\ \citenamefont {Kim}}]{YBK2003}%
  \BibitemOpen
  \bibfield  {author} {\bibinfo {author} {\bibfnamefont {C.-H.}\ \bibnamefont {Chung}}, \bibinfo {author} {\bibfnamefont {K.}~\bibnamefont {Voelker}},\ and\ \bibinfo {author} {\bibfnamefont {Y.~B.}\ \bibnamefont {Kim}},\ }\bibfield  {title} {\bibinfo {title} {Statistics of spinons in the spin-liquid phase of ${\mathrm{cs}}_{2}{\mathrm{cucl}}_{4}$},\ }\href {https://doi.org/10.1103/PhysRevB.68.094412} {\bibfield  {journal} {\bibinfo  {journal} {Phys. Rev. B}\ }\textbf {\bibinfo {volume} {68}},\ \bibinfo {pages} {094412} (\bibinfo {year} {2003})}\BibitemShut {NoStop}%
\bibitem [{\citenamefont {Bocquet}\ \emph {et~al.}(2001)\citenamefont {Bocquet}, \citenamefont {Essler}, \citenamefont {Tsvelik},\ and\ \citenamefont {Gogolin}}]{BOcquet2001}%
  \BibitemOpen
  \bibfield  {author} {\bibinfo {author} {\bibfnamefont {M.}~\bibnamefont {Bocquet}}, \bibinfo {author} {\bibfnamefont {F.~H.~L.}\ \bibnamefont {Essler}}, \bibinfo {author} {\bibfnamefont {A.~M.}\ \bibnamefont {Tsvelik}},\ and\ \bibinfo {author} {\bibfnamefont {A.~O.}\ \bibnamefont {Gogolin}},\ }\bibfield  {title} {\bibinfo {title} {Finite-temperature dynamical magnetic susceptibility of quasi-one-dimensional frustrated spin-$\frac{1}{2}$ heisenberg antiferromagnets},\ }\href {https://doi.org/10.1103/PhysRevB.64.094425} {\bibfield  {journal} {\bibinfo  {journal} {Phys. Rev. B}\ }\textbf {\bibinfo {volume} {64}},\ \bibinfo {pages} {094425} (\bibinfo {year} {2001})}\BibitemShut {NoStop}%
\bibitem [{\citenamefont {Li}\ \emph {et~al.}(2016)\citenamefont {Li}, \citenamefont {Wang},\ and\ \citenamefont {Chen}}]{triangleAFM2016}%
  \BibitemOpen
  \bibfield  {author} {\bibinfo {author} {\bibfnamefont {Y.-D.}\ \bibnamefont {Li}}, \bibinfo {author} {\bibfnamefont {X.}~\bibnamefont {Wang}},\ and\ \bibinfo {author} {\bibfnamefont {G.}~\bibnamefont {Chen}},\ }\bibfield  {title} {\bibinfo {title} {Anisotropic spin model of strong spin-orbit-coupled triangular antiferromagnets},\ }\href {https://doi.org/10.1103/PhysRevB.94.035107} {\bibfield  {journal} {\bibinfo  {journal} {Phys. Rev. B}\ }\textbf {\bibinfo {volume} {94}},\ \bibinfo {pages} {035107} (\bibinfo {year} {2016})}\BibitemShut {NoStop}%
\bibitem [{\citenamefont {Li}\ \emph {et~al.}(2017)\citenamefont {Li}, \citenamefont {Lu},\ and\ \citenamefont {Chen}}]{triangleAFM2017}%
  \BibitemOpen
  \bibfield  {author} {\bibinfo {author} {\bibfnamefont {Y.-D.}\ \bibnamefont {Li}}, \bibinfo {author} {\bibfnamefont {Y.-M.}\ \bibnamefont {Lu}},\ and\ \bibinfo {author} {\bibfnamefont {G.}~\bibnamefont {Chen}},\ }\bibfield  {title} {\bibinfo {title} {Spinon fermi surface $u(1)$ spin liquid in the spin-orbit-coupled triangular-lattice mott insulator ${\mathrm{ybmggao}}_{4}$},\ }\href {https://doi.org/10.1103/PhysRevB.96.054445} {\bibfield  {journal} {\bibinfo  {journal} {Phys. Rev. B}\ }\textbf {\bibinfo {volume} {96}},\ \bibinfo {pages} {054445} (\bibinfo {year} {2017})}\BibitemShut {NoStop}%
\bibitem [{\citenamefont {Li}\ and\ \citenamefont {Chen}(2017)}]{triangleAFM_RPA}%
  \BibitemOpen
  \bibfield  {author} {\bibinfo {author} {\bibfnamefont {Y.-D.}\ \bibnamefont {Li}}\ and\ \bibinfo {author} {\bibfnamefont {G.}~\bibnamefont {Chen}},\ }\bibfield  {title} {\bibinfo {title} {Detecting spin fractionalization in a spinon fermi surface spin liquid},\ }\href {https://doi.org/10.1103/PhysRevB.96.075105} {\bibfield  {journal} {\bibinfo  {journal} {Phys. Rev. B}\ }\textbf {\bibinfo {volume} {96}},\ \bibinfo {pages} {075105} (\bibinfo {year} {2017})}\BibitemShut {NoStop}%
\bibitem [{\citenamefont {Jia}\ \emph {et~al.}(2024)\citenamefont {Jia}, \citenamefont {Ma}, \citenamefont {Wang},\ and\ \citenamefont {Chen}}]{triangleSupersolid}%
  \BibitemOpen
  \bibfield  {author} {\bibinfo {author} {\bibfnamefont {H.}~\bibnamefont {Jia}}, \bibinfo {author} {\bibfnamefont {B.}~\bibnamefont {Ma}}, \bibinfo {author} {\bibfnamefont {Z.~D.}\ \bibnamefont {Wang}},\ and\ \bibinfo {author} {\bibfnamefont {G.}~\bibnamefont {Chen}},\ }\bibfield  {title} {\bibinfo {title} {Quantum spin supersolid as a precursory dirac spin liquid in a triangular lattice antiferromagnet},\ }\href {https://doi.org/10.1103/PhysRevResearch.6.033031} {\bibfield  {journal} {\bibinfo  {journal} {Phys. Rev. Res.}\ }\textbf {\bibinfo {volume} {6}},\ \bibinfo {pages} {033031} (\bibinfo {year} {2024})}\BibitemShut {NoStop}%
\bibitem [{\citenamefont {Zhu}\ \emph {et~al.}(2024)\citenamefont {Zhu}, \citenamefont {Romerio}, \citenamefont {Steiger}, \citenamefont {Nabi}, \citenamefont {Murai}, \citenamefont {Ohira-Kawamura}, \citenamefont {Povarov}, \citenamefont {Skourski}, \citenamefont {Sibille}, \citenamefont {Keller}, \citenamefont {Yan}, \citenamefont {Gvasaliya},\ and\ \citenamefont {Zheludev}}]{Zhu_2024}%
  \BibitemOpen
  \bibfield  {author} {\bibinfo {author} {\bibfnamefont {M.}~\bibnamefont {Zhu}}, \bibinfo {author} {\bibfnamefont {V.}~\bibnamefont {Romerio}}, \bibinfo {author} {\bibfnamefont {N.}~\bibnamefont {Steiger}}, \bibinfo {author} {\bibfnamefont {S.}~\bibnamefont {Nabi}}, \bibinfo {author} {\bibfnamefont {N.}~\bibnamefont {Murai}}, \bibinfo {author} {\bibfnamefont {S.}~\bibnamefont {Ohira-Kawamura}}, \bibinfo {author} {\bibfnamefont {K.}~\bibnamefont {Povarov}}, \bibinfo {author} {\bibfnamefont {Y.}~\bibnamefont {Skourski}}, \bibinfo {author} {\bibfnamefont {R.}~\bibnamefont {Sibille}}, \bibinfo {author} {\bibfnamefont {L.}~\bibnamefont {Keller}}, \bibinfo {author} {\bibfnamefont {Z.}~\bibnamefont {Yan}}, \bibinfo {author} {\bibfnamefont {S.}~\bibnamefont {Gvasaliya}},\ and\ \bibinfo {author} {\bibfnamefont {A.}~\bibnamefont {Zheludev}},\ }\bibfield  {title} {\bibinfo {title} {Continuum excitations in a spin supersolid on a triangular lattice},\ }\bibfield  {journal} {\bibinfo  {journal} {Physical Review Letters}\
  }\textbf {\bibinfo {volume} {133}},\ \href {https://doi.org/10.1103/physrevlett.133.186704} {10.1103/physrevlett.133.186704} (\bibinfo {year} {2024})\BibitemShut {NoStop}%
\bibitem [{\citenamefont {Xu}\ \emph {et~al.}(2023)\citenamefont {Xu}, \citenamefont {Bag}, \citenamefont {Sherman}, \citenamefont {Yadav}, \citenamefont {Kolesnikov}, \citenamefont {Podlesnyak}, \citenamefont {Moore},\ and\ \citenamefont {Haravifard}}]{xu2023realizationu1diracquantum}%
  \BibitemOpen
  \bibfield  {author} {\bibinfo {author} {\bibfnamefont {S.}~\bibnamefont {Xu}}, \bibinfo {author} {\bibfnamefont {R.}~\bibnamefont {Bag}}, \bibinfo {author} {\bibfnamefont {N.~E.}\ \bibnamefont {Sherman}}, \bibinfo {author} {\bibfnamefont {L.}~\bibnamefont {Yadav}}, \bibinfo {author} {\bibfnamefont {A.~I.}\ \bibnamefont {Kolesnikov}}, \bibinfo {author} {\bibfnamefont {A.~A.}\ \bibnamefont {Podlesnyak}}, \bibinfo {author} {\bibfnamefont {J.~E.}\ \bibnamefont {Moore}},\ and\ \bibinfo {author} {\bibfnamefont {S.}~\bibnamefont {Haravifard}},\ }\href {https://arxiv.org/abs/2305.20040} {\bibinfo {title} {Realization of u(1) dirac quantum spin liquid in ybzn2gao5}} (\bibinfo {year} {2023}),\ \Eprint {https://arxiv.org/abs/2305.20040} {arXiv:2305.20040 [cond-mat.str-el]} \BibitemShut {NoStop}%
\bibitem [{\citenamefont {Iqbal}\ \emph {et~al.}(2016)\citenamefont {Iqbal}, \citenamefont {Hu}, \citenamefont {Thomale}, \citenamefont {Poilblanc},\ and\ \citenamefont {Becca}}]{J1J2_triangular}%
  \BibitemOpen
  \bibfield  {author} {\bibinfo {author} {\bibfnamefont {Y.}~\bibnamefont {Iqbal}}, \bibinfo {author} {\bibfnamefont {W.-J.}\ \bibnamefont {Hu}}, \bibinfo {author} {\bibfnamefont {R.}~\bibnamefont {Thomale}}, \bibinfo {author} {\bibfnamefont {D.}~\bibnamefont {Poilblanc}},\ and\ \bibinfo {author} {\bibfnamefont {F.}~\bibnamefont {Becca}},\ }\bibfield  {title} {\bibinfo {title} {Spin liquid nature in the heisenberg ${J}_{1}\ensuremath{-}{J}_{2}$ triangular antiferromagnet},\ }\href {https://doi.org/10.1103/PhysRevB.93.144411} {\bibfield  {journal} {\bibinfo  {journal} {Phys. Rev. B}\ }\textbf {\bibinfo {volume} {93}},\ \bibinfo {pages} {144411} (\bibinfo {year} {2016})}\BibitemShut {NoStop}%
\bibitem [{\citenamefont {Ferrari}\ and\ \citenamefont {Becca}(2019)}]{J1J2_dynamics}%
  \BibitemOpen
  \bibfield  {author} {\bibinfo {author} {\bibfnamefont {F.}~\bibnamefont {Ferrari}}\ and\ \bibinfo {author} {\bibfnamefont {F.}~\bibnamefont {Becca}},\ }\bibfield  {title} {\bibinfo {title} {Dynamical structure factor of the ${J}_{1}\ensuremath{-}{J}_{2}$ heisenberg model on the triangular lattice: Magnons, spinons, and gauge fields},\ }\href {https://doi.org/10.1103/PhysRevX.9.031026} {\bibfield  {journal} {\bibinfo  {journal} {Phys. Rev. X}\ }\textbf {\bibinfo {volume} {9}},\ \bibinfo {pages} {031026} (\bibinfo {year} {2019})}\BibitemShut {NoStop}%
\bibitem [{\citenamefont {Parcollet}\ \emph {et~al.}(2015)\citenamefont {Parcollet}, \citenamefont {Ferrero}, \citenamefont {Ayral}, \citenamefont {Hafermann}, \citenamefont {Krivenko}, \citenamefont {Messio},\ and\ \citenamefont {Seth}}]{Parcollet_2015}%
  \BibitemOpen
  \bibfield  {author} {\bibinfo {author} {\bibfnamefont {O.}~\bibnamefont {Parcollet}}, \bibinfo {author} {\bibfnamefont {M.}~\bibnamefont {Ferrero}}, \bibinfo {author} {\bibfnamefont {T.}~\bibnamefont {Ayral}}, \bibinfo {author} {\bibfnamefont {H.}~\bibnamefont {Hafermann}}, \bibinfo {author} {\bibfnamefont {I.}~\bibnamefont {Krivenko}}, \bibinfo {author} {\bibfnamefont {L.}~\bibnamefont {Messio}},\ and\ \bibinfo {author} {\bibfnamefont {P.}~\bibnamefont {Seth}},\ }\bibfield  {title} {\bibinfo {title} {Triqs: A toolbox for research on interacting quantum systems},\ }\href {https://doi.org/10.1016/j.cpc.2015.04.023} {\bibfield  {journal} {\bibinfo  {journal} {Computer Physics Communications}\ }\textbf {\bibinfo {volume} {196}},\ \bibinfo {pages} {398–415} (\bibinfo {year} {2015})}\BibitemShut {NoStop}%
\end{thebibliography}%

\appendix 
\onecolumngrid
\clearpage
\section{Path Integral Formulation}\label{appendix:path integral}

Suppose we have a general spin Hamiltonian on a lattice
\begin{equation}
    \label{spin ham}
    \mh(\t) = \sum_{\bd}J_{ij}^{ab}(\bd)\sum_{\br}  S_i^a(\br, \t)S_j^b(\br+\bd, \t)\,,
\end{equation}
where \(\br\) marks the unit cell, \(i\,,j\) represent sublattice degree of freedom, and \(a\,,b\) represent spin directions. Now we decompose the spin operator using fermionic partons as
\begin{equation}
    \label{fermion parton}
    S_i^a(\br, \t) = \frac{1}{2}f^{\dagger}_{i, \a}(\br, \t)\cdot\s^{a}_{\a\b}\cdot f_{i, \b}(\br, \t)\,,
\end{equation}
under the constraint that 
\begin{equation}
    \label{parton number constraint}
    \sum_{\a}f^{\dagger}_{i, \a}(\br, \t)f_{i, \a}(\br, \t) = 1 \:\:\:\forall \:\:\:\br\,,i\,,\t\,.
\end{equation}
This constraint can be imposed using a Lagrange multiplier \(a_0(\br, \t)\). The total parton action hence looks like (at zero temperature)
\begin{equation}
    \label{parton action}
    \ms[\brf, f, a_0] = \int_{0}^{\infty}d\t \left\{\sum_{\br}\brf_{i, \a}(\br, \t)\left[\partial_{\t}-a_0(\br, \t)\right]f_{i, \a}(\br, \t) + \mh(\t)\right\}\,,
\end{equation}
and the path integral looks like
\begin{equation}
    \label{path integral}
    \mz = \int \md[\brf, f\,,a_0] \:e^{-\ms[\brf, f, a_0]}\,.
\end{equation}

\subsection{Hubbard Stratonovich Decoupling}
The action consists of a four-fermion interaction term. This can be decoupled using Hubbard Stratonovich (HS) fields. However, this can be done in two channels : (a) when the fermions are hopping, (b) when there is magnetic ordering. Furthermore, both these channels can exist simultaneously with some relative ratio which can depend on the bond being decoupled as well.\par
In the ordering channel, we have that
\begin{equation}
    \label{magnetic HS}
    \exp\left(-S_i^a J_{ij}^{ab} S_j^b\right) = \int \prod_{a} d m_i^a \exp\left(m_i^a J_{ij}^{ab} m_j^b - 2S_i^a J_{ij}^{ab} m_j^b\right)\,,
\end{equation}
where \(m_i^a(\br, \t)\) can be interpreted as the dynamic magnetization on the site labeled by \((\br, i)\) (we will be using the magnetization instead of the usual Weiss fields to simplify some steps, but at the end will switch back to Weiss fields). For the full Hamiltonian, we get
\begin{equation}
    \label{magnetic HS full ham}
    \begin{split}
        \exp\left(-\int d\t\mh(\t)\right) = \int \md[m] \exp\left(\sum_{\bd}\sum_{\br}J_{ij}^{ab}(\bd)  \left[m_i^a(\br, \t)m_j^b(\br+\bd, \t)-2 S_i^a(\br, \t)m_j^b (\br+\bd, \t)\right]\right)\,.
    \end{split}
\end{equation}
If this HS field condenses \textit{i.e.} \(\expval{m_i^a(\br, \t)} = \expval{S_i^a(\br)} = m_i^a(\br)\) and gets some finite expectaion value, the system will undergo symmetry breaking and exhibit magnetic ordering. The lowest energy condensate can be found through self-consistent mean-field theory. However, one can also consider excitations around such a condensate \(m_i^a(\br, \t) = m_i^a(\br)+\d m(\br, \t)\). If we consider excitations around the mean-field condensate to leading order which will be quadratic, it is equivalent to performing random phase approximation (RPA). Meanwhile, in the hopping channel, we have that
\begin{equation}
    \label{hopping HS}
    \exp\left(-\frac{1}{2}S_i^a J_{ij}^{ab} S_j^b\right) = \int \prod_{\a, \b} d (w_{ij}^{ab})^{*} d w_{ij}^{ab} \exp\left(\frac{1}{4} J_{ij}^{ab} \s^a_{\a\b}\s^{b}_{\g\d} \left[\brf_{i, \a}f_{j, \d}(w_{ij}^{\b\g})^* + \brf_{j, \g}f_{i, \b}w_{ij}^{\a\d} - w_{ij}^{\a\d}(w_{ij}^{\b\g})^* \right]\right)\,,
\end{equation}
where the hopping HS field lives on the bonds, instead of on-site. For the full Hamiltonian, we get
\begin{equation}
    \label{hopping HS full}
    \begin{split}
        \exp\left(-\int d\t\mh(\t)\right) &= \int \md[\brw, w] \exp\bigg(\frac{1}{4} \sum_{\bd}\sum_{\br}J_{ij}^{ab}(\bd) \s^a_{\a\b}\s^{b}_{\g\d} \bigg[- w_{ij}^{\a\d}(\br+\bd/2, \t)\brw_{ij}^{\b\g}(\br+\bd/2, \t) \\
        & +\brf_{i, \a}(\br, \t)f_{j, \d}(\br+\bd, \t)\brw_{ij}^{\b\g}(\br+\bd/2, \t) + \brf_{j, \g}(\br+\bd, \t)f_{i, \b}(\br, \t)w_{ij}^{\a\d}(\br+\bd/2, \t) \bigg]\bigg)\,,
    \end{split}
\end{equation}
When \(w_{ij}^{\a\b}\) gains an expectation value \(\expval{w_{ij}^{\a\b}(\br+\bd/2, \t)} = \expval{f^{\dagger}_{i,\a}(\br)f_{j, \b}(\br+\bd)} = w_{ij}^{\a\b}(\br+\bd/2)\) , we get the fermions hopping on the lattice in some frozen background gauge flux determined by the phases of \(w_{ij}^{\a\b}\). This gauge is the same gauge redundancy which arises when writing Eq.\eqref{fermion parton}, which for simplicity's sake we are going to assume is just \(U(1)\) (can be \(SU(2)\) also) . 
This separation can be made explicit by writing \(w_{ij}^{\a\b} = |w_{ij}^{ab}|e^{-i a_{ij}}\), such that phase modulations are captured by gauge excitations around the background gauge flux which can be determined self-consistently.\par
Now, in general, we can imagine splitting the Hamiltonian on each bond as \(\mh_{ij} = \a_{ij} \mh_{ij}^{w}+(1-\a_{ij})\mh_{{ij}^m}\), where \(\mh_{ij}^{w}\) is decomposed using Eq.\eqref{hopping HS full}, while \(\mh_{ij}^{m}\) is decomposed using Eq.\eqref{magnetic HS full ham}. In total we can write down the full path integral as
\begin{equation}
    \label{full action}
    \mz = \int \md[\brf, f, \brw, w, m, a_0]e^{-\ms_0[\brf, f, a_0]}e^{-\ms[\brf, f, \brw, w ; \{\a\}]}e^{-\ms[\brf, f, m ; \{1-\a\}]}\,.
\end{equation}
If we calculate the path integral \emph{exactly}, then the partition function \emph{should not} depend on \(\{\a\}\). However, we are going to assume that the \(w_{ij}^{\a\b}\) fields are condensed around their uniform mean-field values \textit{i.e.} all the gauge fluctuations in the system are suppressed. Then, the resulting partition function can in general depend on \(\{\a\}\) as 
\begin{equation}
    \label{full action no gauge}
    \begin{split}
            \mz &= \int \md[a_0, \brw, w]e^{-\ms_{\mathrm{eff}}[a_0\,,\brw, w ; \{\a\}]}\,,\\
            e^{-\ms_{\mathrm{eff}}[a_0\,,\brw, w ; \{\a\}]} &= \exp\left(-\sum_{\br, \bd}\frac{\a_{ij}(\bd)J_{ij}^{ab}(\bd) }{4} \s^a_{\a\b}\s^{b}_{\g\d} w_{ij}^{\a\d}(\br+\bd/2, \t)\brw_{ij}^{\b\g}(\br+\bd/2, \t) \right)\int \md[\brf, f, m] e^{-\ms[\brf, f, m ; \{\a\}]}\,,
    \end{split}
\end{equation}
where
\begin{equation}
    \label{effective action}
    \begin{split}
        \ms[\brf, f, m ; \{\a\}] &= \int_{0}^{\infty}d\t \sum_{\br}\bigg\{\brf_{i, \a}(\br, \t)\left[\partial_{\t}-a_0\right]f_{i, \a}(\br, \t)\\
        &-\sum_{\br}\sum_{\bd}\left[\brf_{i, \a}(\br, \t)t_{ij}^{\a\b}(\bd ; \{\a\})f_{j, \b}(\br+\bd, \t) + \mathrm{h.c.}\right]\\
        &+\frac{1}{2}\sum_{\br}\sum_{\bd}(1-\a_{ij}(\bd))J_{ij}^{ab}(\bd)\left[\brf_{i, \a}(\br, \t)\s_{\a\b}^{a} m_j^b(\br+\bd, \t) f_{i, \b}(\br, \t) + (\mathrm{i\leftrightarrow j})\right]\\
        &-\sum_{\br}\sum_{\bd}(1-\a_{ij}(\bd))J_{ij}^{ab}(\bd) m_{i}^{a}(\br, \t)m_{j}^{b}(\br+\bd, \t)\,,\\
        &\equiv \ms_0[\brf, f ; \{\a\}] + \ms_{\mathrm{int}}[\brf, f, m ; \{\a\}] + \ms_0[m ; \{\a\}]\,,
    \end{split}
\end{equation}
where \(t_{ij}^{\a\b}(\bd ; \{\a\}) = (1/4)\a_{ij}(\bd)J_{ij}^{ab}(\bd)\s^{a}_{\a\d}\s^{b}_{\g\b}\brw_{ij}^{\d\g}\). 
\subsection{Integrating out spinons}
We switch to momentum space and write down each piece of the action as
\begin{equation}
    \label{momentum space}
    \begin{split}
      \ms_0[\brf, f ; \{\a\}] &= \sum_{i\w}\sum_{\bk}\brf_{i, \a}(\bk, i\w)\left[(i\w-a_0)\d_{ij}\d^{\a\b} + \mh_{ij}^{\a\b}(\bk ; \{\a\})\right]f_{j, \b}(\bk, i\w)\,,\\
      \ms_0[m ; \{\a\}] & = -\sum_{i\w}\sum_{\bk} m_i^a(-\bk, -i\w) \mj_{ij}^{ab}(\bk ; \{\a\}) m_b^j(\bk, i\w)\,,\\
      \ms_{\mathrm{int}}[\brf, f, m ; \{\a\}] & = \frac{1}{2}\sum_{i\W, i\w}\sum_{\bQ, \bk} m_i^a(-\bQ, -i\W)\mj_{ij}^{ab}(\bQ ; \{\a\}) \brf_{j, \a}(\bk+\bQ, i\w+i\W)\s^{b}_{\a\b}f_{j, \b}(\bk, i\w)\,,
    \end{split}
\end{equation}
where \(J_{ij}^{ab}(\bk) = \sum_{\bd}J_{ij}^{ab}(\bd)e^{-i\bk\cdot\bd}\), \(\tilde{J}_{ij}^{ab}(\bk ; \{\a\}) = \sum_{\bd}\a_{ij}(\bd)J_{ij}^{ab}(\bd)e^{-i\bk\cdot\bd}\) and \(\mj_{ij}^{ab}(\bk ; \{\a\}) = J_{ij}^{ab}(\bk) - \tilde{J}_{ij}^{ab}(\bk ; \{\a\})\). The free hopping Hamiltonian in momentum space looks like
\(\mh_{ij}^{\a\b}(\bk ; \{\a\}) = -\sum_{\bd}t_{ij}^{\a\b}(\bd ; \{\a\}) e^{-i\bk\cdot\bd}\). Combining everything we get
\begin{equation}
    \label{combined action}
    \begin{split}
        \ms[\brf, f, m ; \{\a\}] &= \sum_{i\w}\sum_{\bk} \left\{\brf_{i, \a}(\bk, i\w)\left[\mg_0^{-1}(\bk, i\w ; \{\a\})\right]_{ij}^{\a\b}f_{j, \b}(\bk, i\w) + m_i^a(-\bk, -i\w)\left[G_m^{-1}(\bk, i\w ; \{\a\})\right]_{ij}^{ab}m_j^b(\bk, i\w)\right\}\\
        &+\frac{1}{2}\sum_{i\W, i\w}\sum_{\bQ, \bk} \mj_{ij}^{ab}(\bQ ; \{\a\}) \s^{b}_{\a\b} m_i^a(-\bQ, -i\W) \brf_{j, \a}(\bk+\bQ, i\w+i\W)f_{j, \b}(\bk, i\w)\,.
    \end{split}
\end{equation}
We then want to integrate out the spinons and get an effective action in \(\ms_{\mathrm{eff}}[m ; \{\a\}]\) as 
\begin{equation}
    \label{integrating out fermions}
    e^{-\ms[m ; \{\a\}]} = \int\md[\brf, f] e^{-\ms_0[\brf, f ; \{\a\}]}e^{-\ms_{\mathrm{int}}[\brf, f, m ; \{\a\}]} = \mz_0 \expval{e^{-\ms_{\mathrm{int}}[\brf, f, m ; \{\a\}]}}_{0}\,.
\end{equation}
Now, we assume that the magnetization field is fluctuating weakly around its mean-field value \(m = \expval{m}+\d m\). However, if our mean-field solution is a spin liquid itself, \(\expval{m}=0\). Hence, the expectation value in Eq.\eqref{integrating out fermions} can be calculated in perturbation theory since the magnetization field \(m\) itself can be treated as weak \(\expval{e^{-\ms_{\mathrm{int}}}} = \sum_{n=0}^{\infty}(-\frac{1}{n!})\expval{\ms_{\mathrm{int}}^n}\). We can further drop the first order term which is linear in \(m\) since that term is just a disconnected diagram and will get canceled in the normalization. At second order, we get
\begin{equation}
    \label{second order expectation}
    \begin{split}
        \expval{\ms_{\mathrm{int}}^2} & = \sum_{i\W}\sum_{\bQ}m_i^a(-\bQ, -i\W)\cdot [\mj(\bQ ; \{\a\})\cdot\c_0(\bQ, i\W ; \{\a\})\cdot\mj(\bQ ; \{\a\})]_{ij}^{ab}\cdot  m_j^b(\bQ, i\W)\,,
    \end{split}
\end{equation}
where \(\c_0\) is the bare susceptibility bubble diagram
\begin{equation}
    \label{bare chi}
    \c_0(\bQ, i\W ; \{\a\})_{ij}^{ab} =  -\frac{1}{4}\sum_{i\w}\sum_{\bk}\mathrm{Tr}\left[\s^a\cdot \mg_{0, ji}^{T}(\bk, i\w ; \{\a\})\cdot \s^b\cdot\mg_{0, ij}^{T}(\bk+\bQ, i\w+i\W ; \{\a\})\right]\,.
\end{equation}
Combining with \(\ms_0[m]\), we have
\begin{equation}
    \label{effective action m}
    \ms_{\mathrm{eff}}[m ; \{\a\}] = \sum_{i\W}\sum_{\bQ}m_i^a(-\bQ, -i\W)\cdot [\mj(\bQ ; \{\a\})-\mj(\bQ ; \{\a\})\cdot\c_0(\bQ, i\W ; \{\a\})\cdot\mj(\bQ ; \{\a\})]_{ij}^{ab}\cdot  m_j^b(\bQ, i\W)\,.
\end{equation}
This gives us that
\begin{equation}
    \label{expectation mag}
    \expval{m_i^a(-\bQ, -i\W)m_j^b(\bQ, i\W)} = \left(\left[\mj(\bQ ; \{\a\})-\mj(\bQ ; \{\a\})\cdot\c_0(\bQ, i\W ; \{\a\})\cdot\mj(\bQ ; \{\a\})\right]^{-1}\right)_{ij}^{ab}\,.
\end{equation}
We can now switch back to the usual weiss fields \(h_{i}^a(\bQ, i\W) = \mj_{ij}^{ab}(\bQ)\cdot m_{j}^{b}(\bQ, i\W)\) instead of the magnetization, and get
\begin{equation}
    \label{effective action h}
    \ms_{\mathrm{eff}}[h ; \{\a\}] = \sum_{i\W}\sum_{\bQ}h_i^a(-\bQ, -i\W)\cdot [\mj^{-1}(\bQ ; \{\a\})-\c_0(\bQ, i\W ; \{\a\})]_{ij}^{ab}\cdot  h_j^b(\bQ, i\W)\,.
\end{equation}
\begin{equation}
    \label{expectation weiss}
    \expval{h_i^a(-\bQ, -i\W)h_j^b(\bQ, i\W)} = [\mj^{-1}(\bQ ; \{\a\})-\c_0(\bQ, i\W ; \{\a\})]_{ij}^{ab}\,.
\end{equation}
In the effective action, if \(\det[\mj(\bQ ; \{\a\})^{-1}-\c_0(\bQ, \W+i\eta ; \{\a\})]<0\) for some \(\bQ_c\) at \(\W=0\), we have an instability. 

\subsection{RPA corrections to response functions}
To get the physical susceptibility of the system, we have to introduce an external field \(h_{\mathrm{ext}}\), and find the effective action in terms of these fields. Doing so will give us an action with the same form as Eq.\eqref{effective action h}, but with the RPA corrected susceptibility which will look like 
\begin{equation}
    \label{effective action h external}
    \ms_{\mathrm{eff}}[h_{\mathrm{ext}} ; \{\a\}] = \sum_{i\W}\sum_{\bQ}h_{\mathrm{ext}, i}^a(-\bQ, -i\W)\cdot [\c^{-1}(\bQ, i\W ; \{\a\})]_{ij}^{ab}\cdot  h_{\mathrm{ext}, j}^b(\bQ, i\W)\,,
\end{equation}
where \(\c^{-1}(\bQ, i\W ; \{\a\}) = \c_0^{-1}(\bQ, i\W ; \{\a\})-\mj(\bQ ; \{\a\})\).

\section{VMC} \label{appendix:VMC}

\begin{figure}[!ht]
 \centering
 \includegraphics[width=0.4\textwidth]{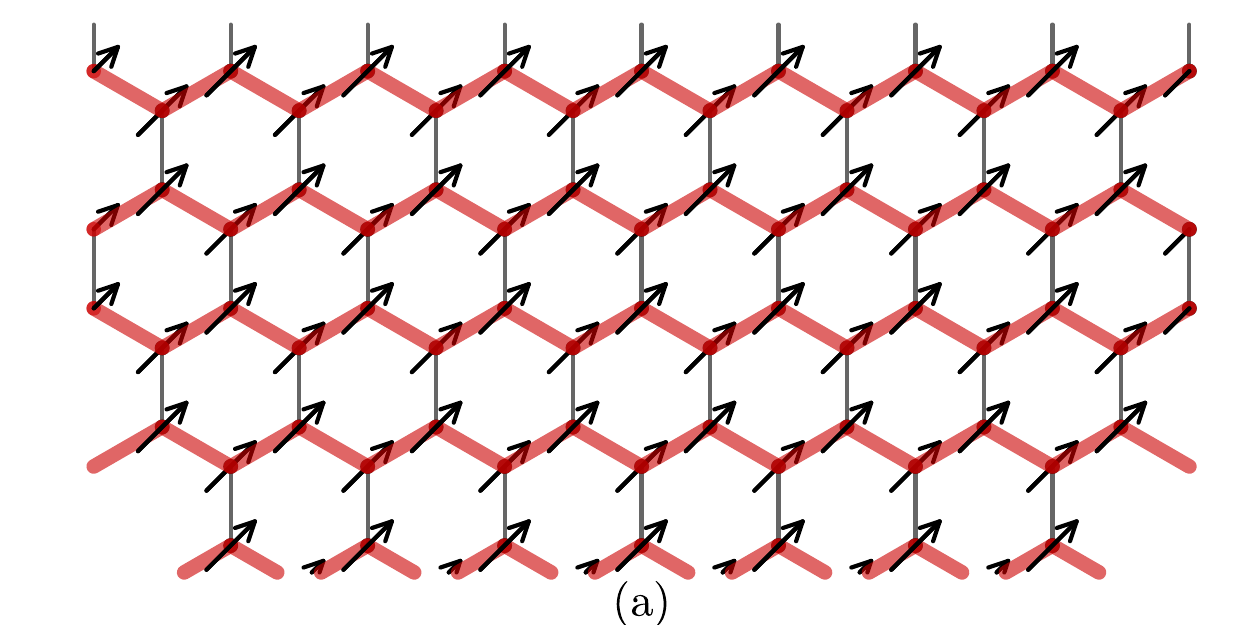}
 \includegraphics[width=0.4\textwidth]{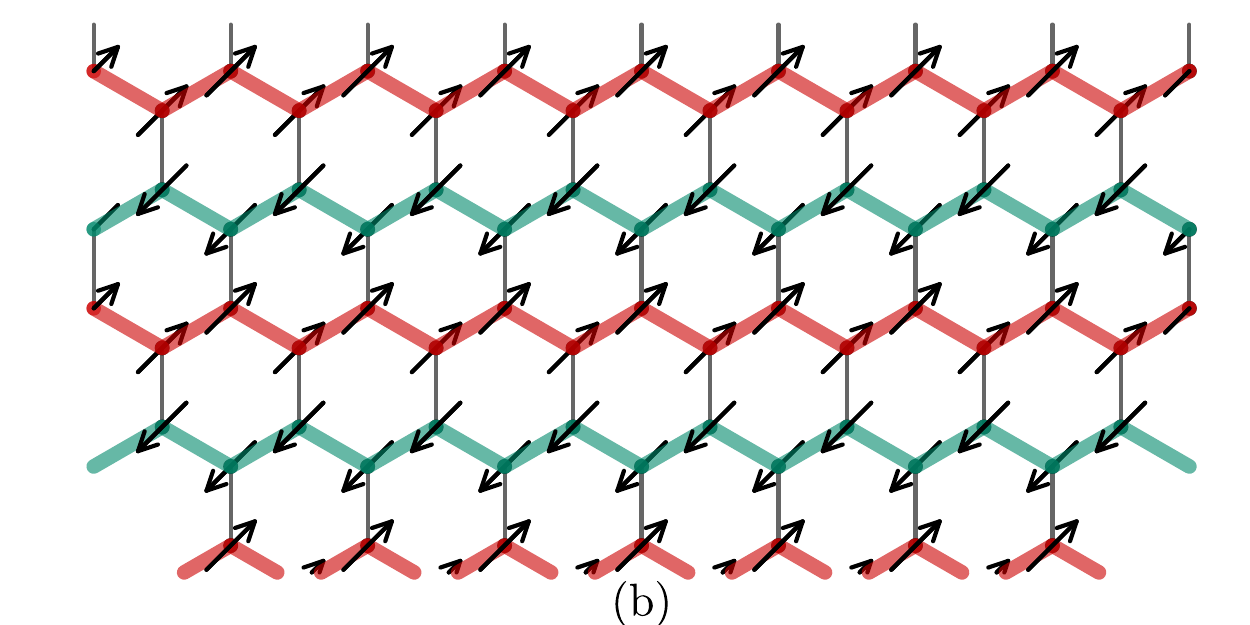}
 \includegraphics[width=0.4\textwidth]{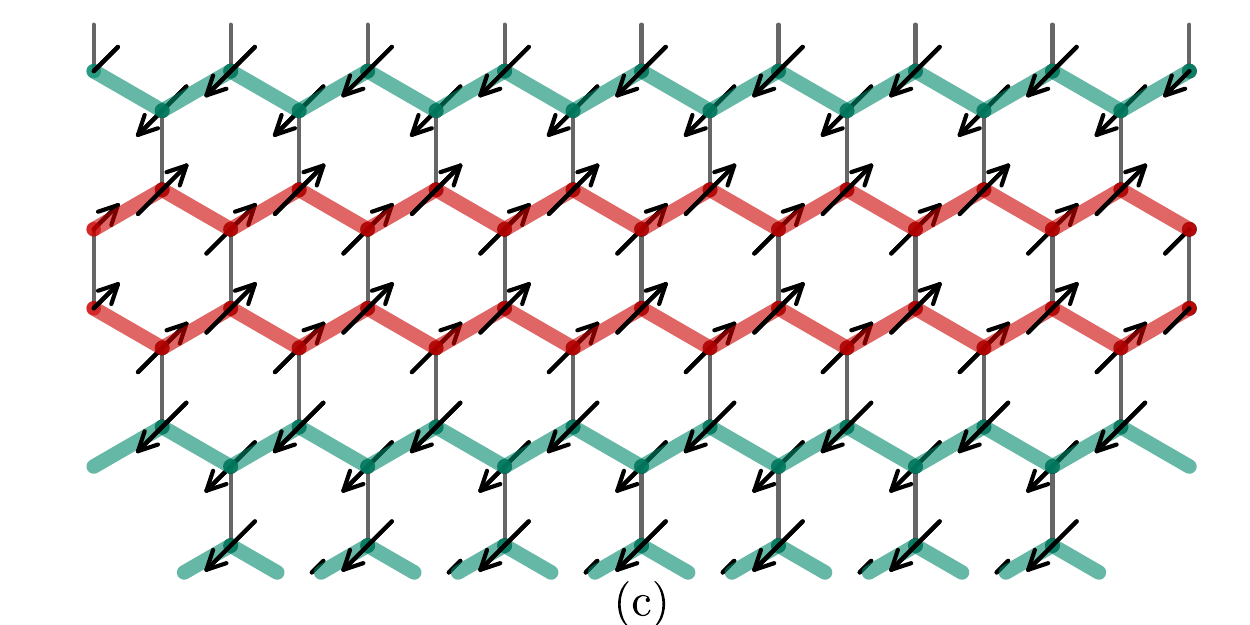}
 \includegraphics[width=0.4\textwidth]{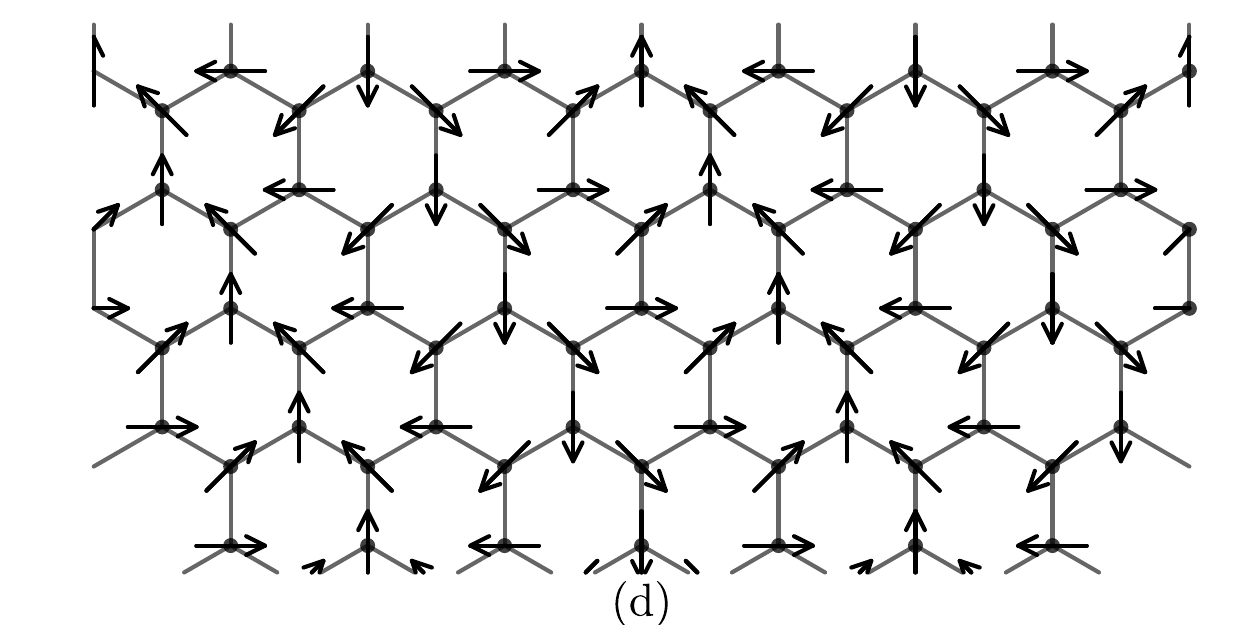}
 \caption{The various spin-rotation symmetry broken states considered in the VMC ansatzes. (a) A simple in-plane ferromagnet (FM) with \(\bQ=\bGa\), (b) an in-plane zig-zag (ZZ) order with alternating chains at \(\bQ=\bM\), (c) an in-plane double zig-zag (dZZ) order with alternating chains at \(\bQ=\bM/2\), and (d) a co-planar incommensurate (Spiral) order with a wavevector lying in \(\bQ\in \bGa\rightarrow \bM\).  }
\label{fig:VMC orderings}
\end{figure}

In our VMC implementation, the auxillary parton Hamiltonian we have considered looks like
\begin{equation}
    \label{vmc aux ham}
    \begin{split}
        \mh^{f} = -t_1\sum_{\expval{i, j}_1}\left(f^{\dagger}_{i, \a}\s^z_{\a\b}f_{j, \b}+\mathrm{h.c.}\right)-t_3\sum_{\expval{i, j}_3}\left(f^{\dagger}_{i, \a}\s^z_{\a\b}f_{j, \b}+\mathrm{h.c.}\right)-\frac{1}{2}\sum_{i}h_i^a f^{\dagger}_{i, \a}\s^a_{\a\b}f_{i, \b}\,,
    \end{split}
\end{equation}
where \(t_1=1\) can be held fixed. The site-dependent Weiss fields \(h_i^a\) depend on the specific ordering pattern being considered. In this paper we worked with a \(L=3\times12\times2\) honeycomb lattice with a quadrupled (along one direction) unit cell to accommodate the different ordering patterns as shown in Fig.\ref{fig:VMC orderings}. All the in-plane collinear orders can be captured by a single weiss field strength \(h\) whereas the co-planar spiral order requires three independent Weiss fields to keep track of.\par

\begin{figure}[!ht]
 \centering
 \includegraphics[width=0.5\textwidth]{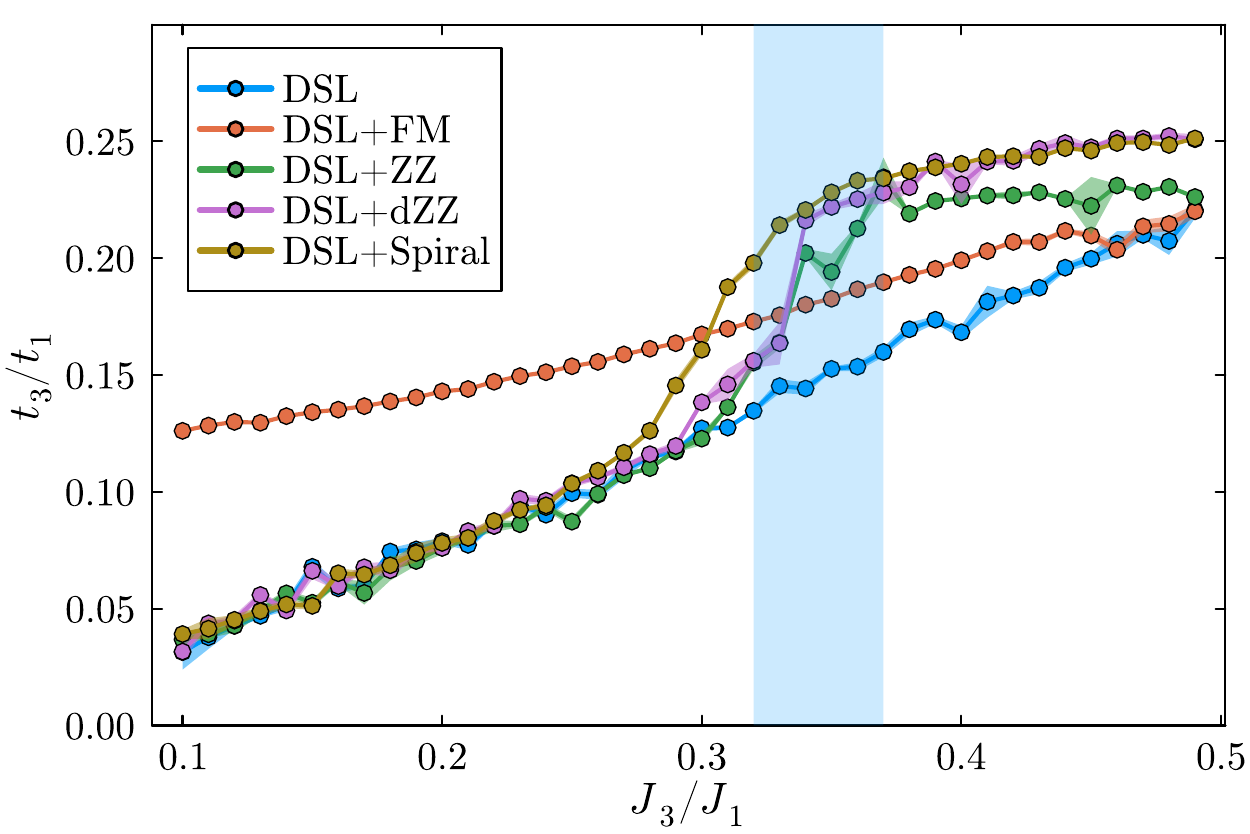}
 \caption{The ratio of third-neighbour hopping to first neighbour hopping \(t_3/t_1\) found in the optimum energy state in VMC. The optimal variational parameters were found through Stochastic reconfiguration gradient descent method over 250 steps.}
\label{fig:VMC t3}
\end{figure}

\begin{figure}[!ht]
 \centering
 \includegraphics[width=0.45\textwidth]{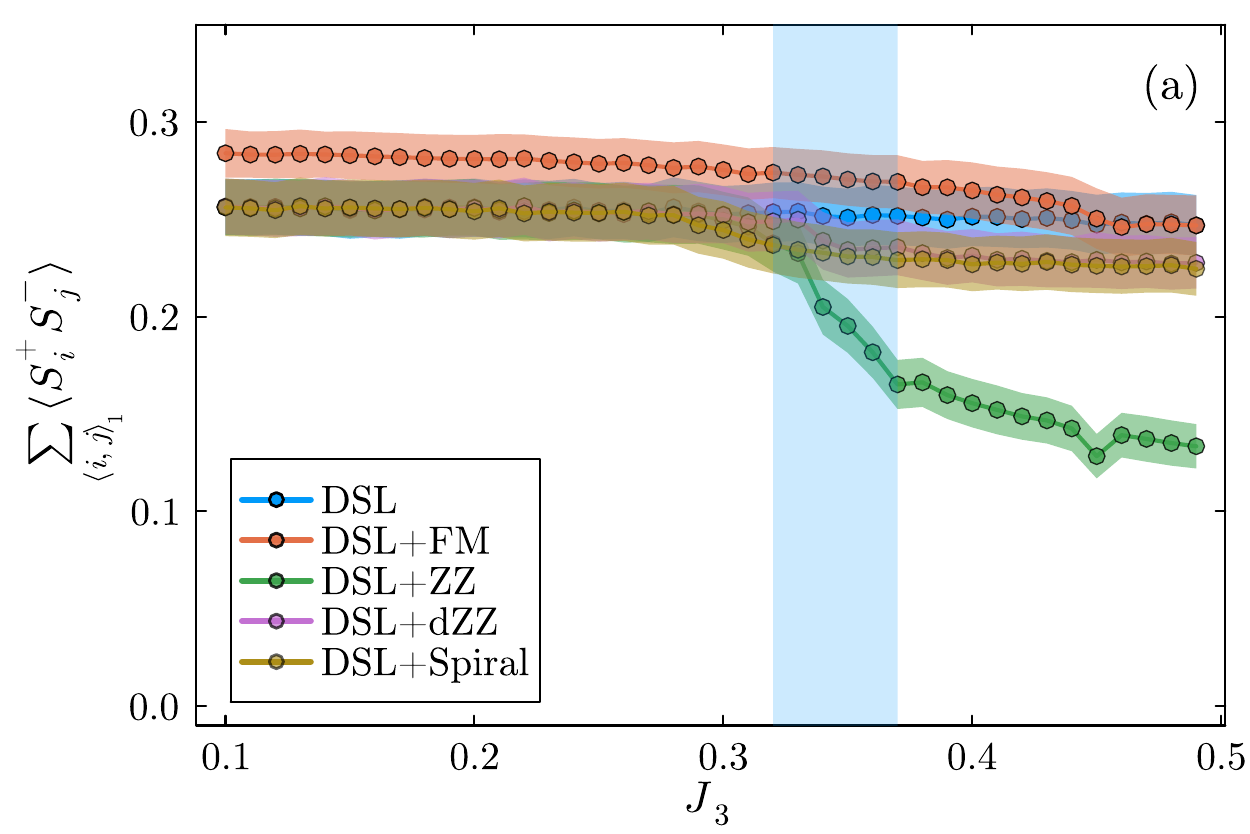}
 \includegraphics[width=0.45\textwidth]{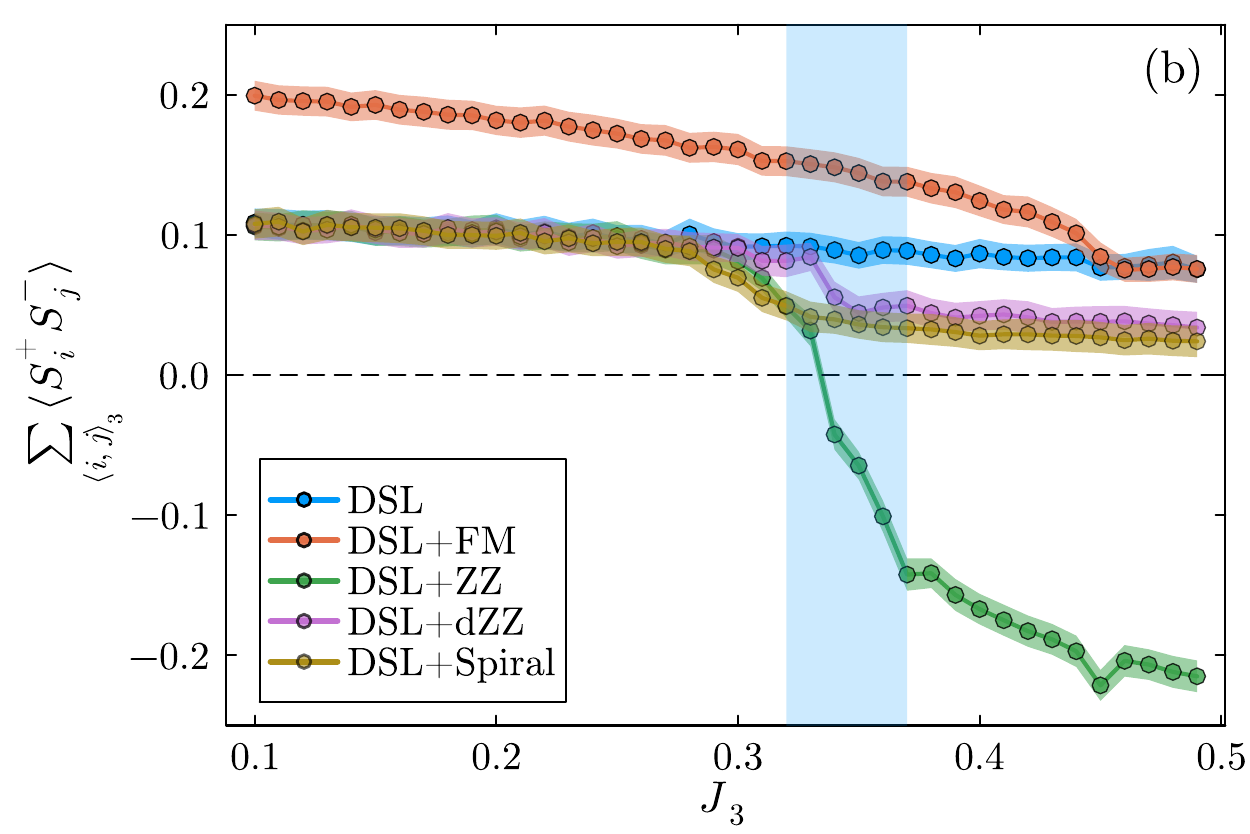}
 \caption{(a) nearest neighboring in-plane spin correlation, and (b) third-nearest neighbour in-plane spin correlation, measured in the optimum energy state found in VMC. All measurements were done over 10,000 measurement sweeps after 10,000 \emph{thermalization} sweeps.}
\label{fig:VMC corr}
\end{figure}

The mean-field ground state \(\ket{\psi_{\mathrm{MFT}}}\) can then be found by diagonalizing this quadratic Hamiltonian and filling states upto half-filling. This mean-field state variational state is then Gutzwiller projected to get the full many-body spin wavefunction \(\ket{\psi} = \ket{\psi(\{t_3, h_i, g_{ij}\})} = \exp\left(-\sum_{i, j} g_{ij} \hat{S}_i^z \hat{S}_j^z\right)\hat{P}\ket{\psi_{\mathrm{MFT}}(t_3, h_i)}\), where \(g_{ij}\) are Jastrow factors to further enhance spin-correlations. We then measure the expectation value of the full spin Hamiltonian in Eq.\eqref{easy plane Hamiltonian full} as \(E(\{t_3, h_i, g_{ij}\}) = \expval{\psi | \mh^{\ms} |\psi}/\expval{\psi|\psi}\) by sampling the many-body wavefunction through 10,000 (thermalization) + 10,000 (measurement) Monte Carlo sweeps. This energy is then optimized \textit{w.r.t} the variational parameters of the wavefunction using stochastic reconfiguration gradient descent \cite{becca_sorella_2017} for 250 steps. The optimal values of the third neighbour hopping is shown in Fig.\ref{fig:VMC t3}, which we find is roughly \(t_3/t_1\approx 0.1\)-\(0.2\) in the intermediate regime of interest. Furthermore, in-plane spin-spin correlation on first and third neighbours are reported in Fig.\ref{fig:VMC corr}.

\section{Mean-field DSL properties}\label{appendix:MFT}
\begin{figure}[!ht]
 \centering
 \includegraphics[width=0.48\textwidth]{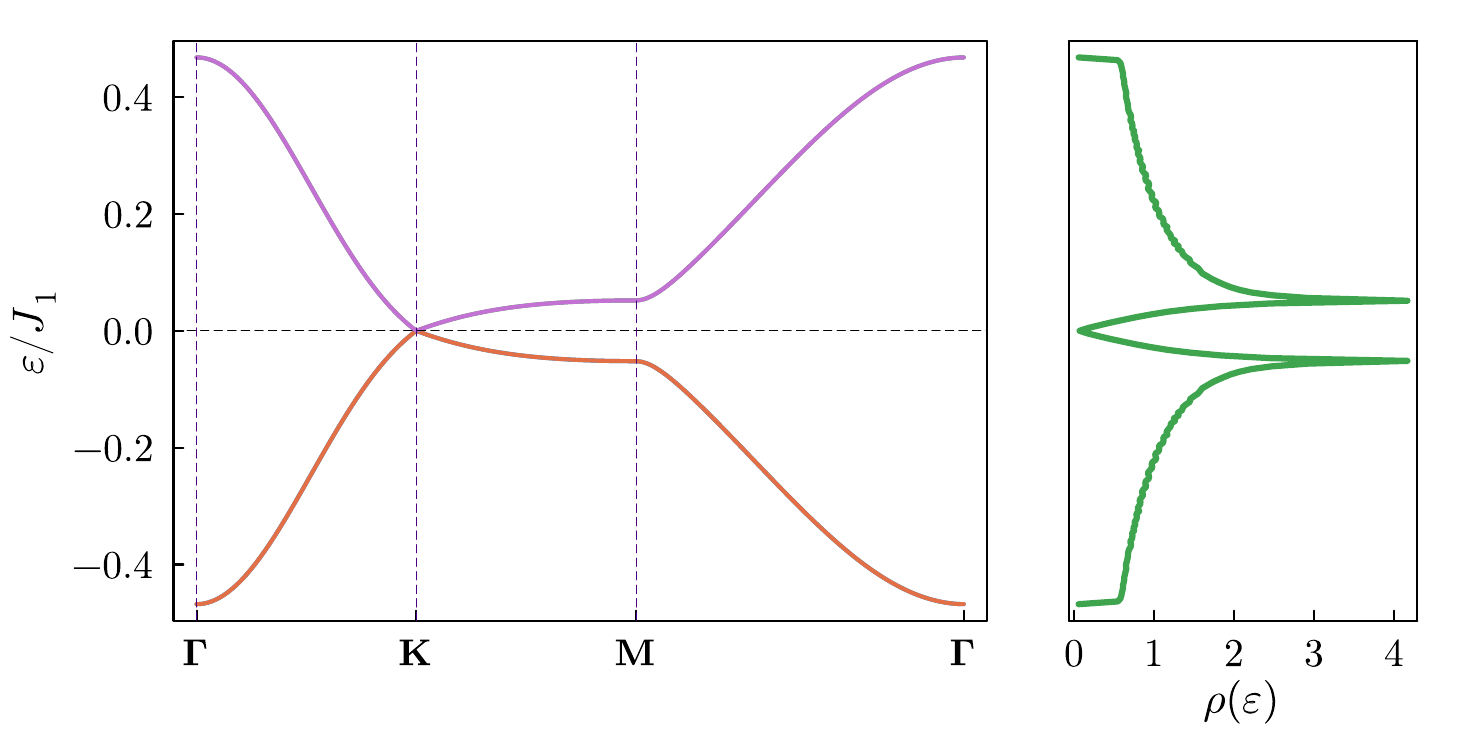}
  \includegraphics[width=0.48\textwidth]{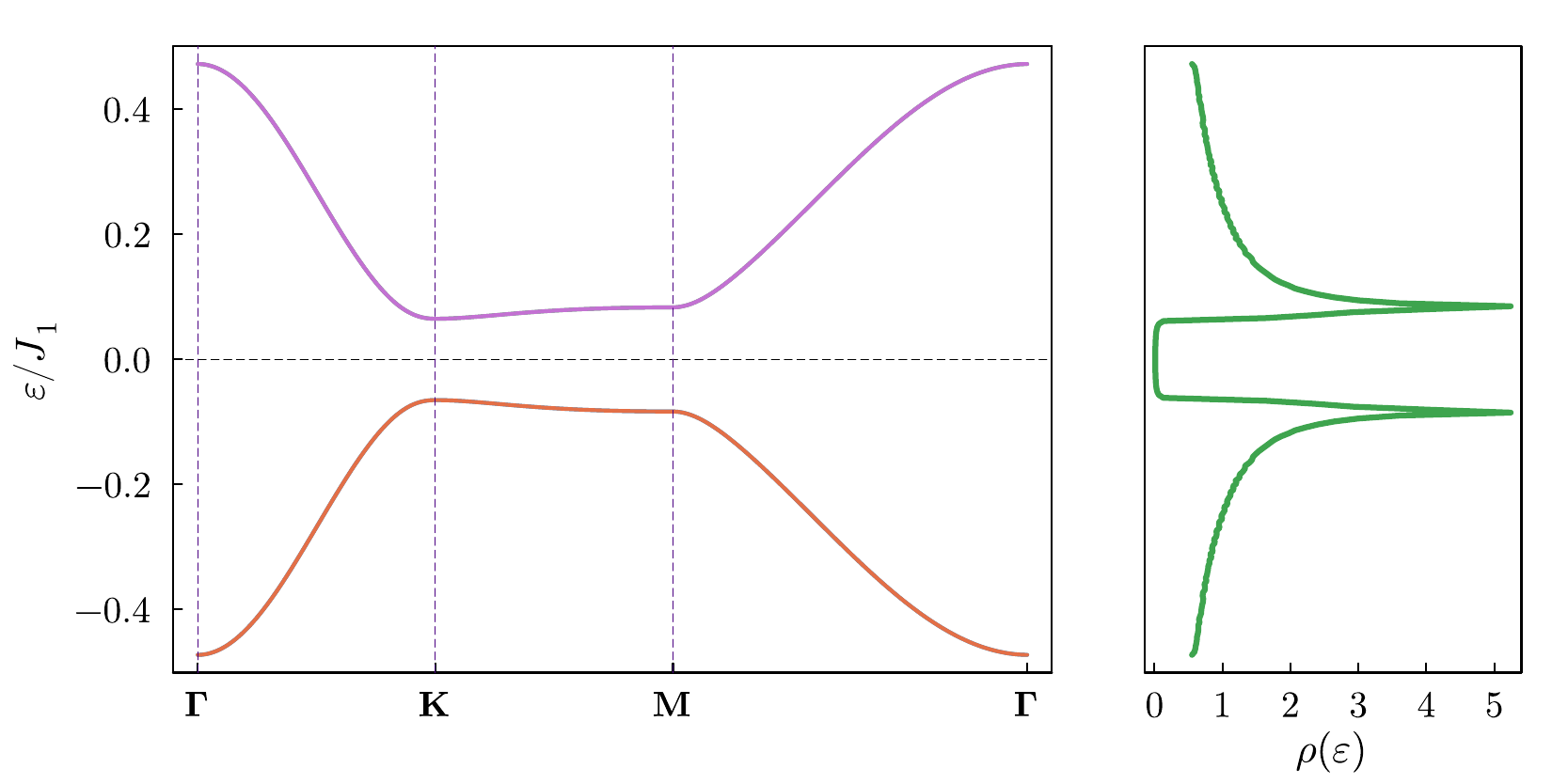}
 \caption{(a) Mean-field band structure of the DSL state with the Dirac nodes at the \(\bK\) points in the Brillouin zone. The density of states of the DSL state (right) which shows sharp Van-Hove peaks around \(\e(\bk)\approx \pm 0.1 J_1\) coming from the nearly flat dispersion along the boundary of the Brillouin zone. This leads to enhanced susceptibilities at \(\W\approx 0.1J_1\) seen in the THz response. (b) in the presence of an applied in-plane Zeeman field which gaps out the Dirac nodes with a gap \(\D\sim B_x\), but leaves the Van-Hove peaks in the density of states intact.}
\label{fig:DSL bands}
\end{figure}
\begin{figure}[!ht]
 \centering
  \includegraphics[width=0.4\textwidth]{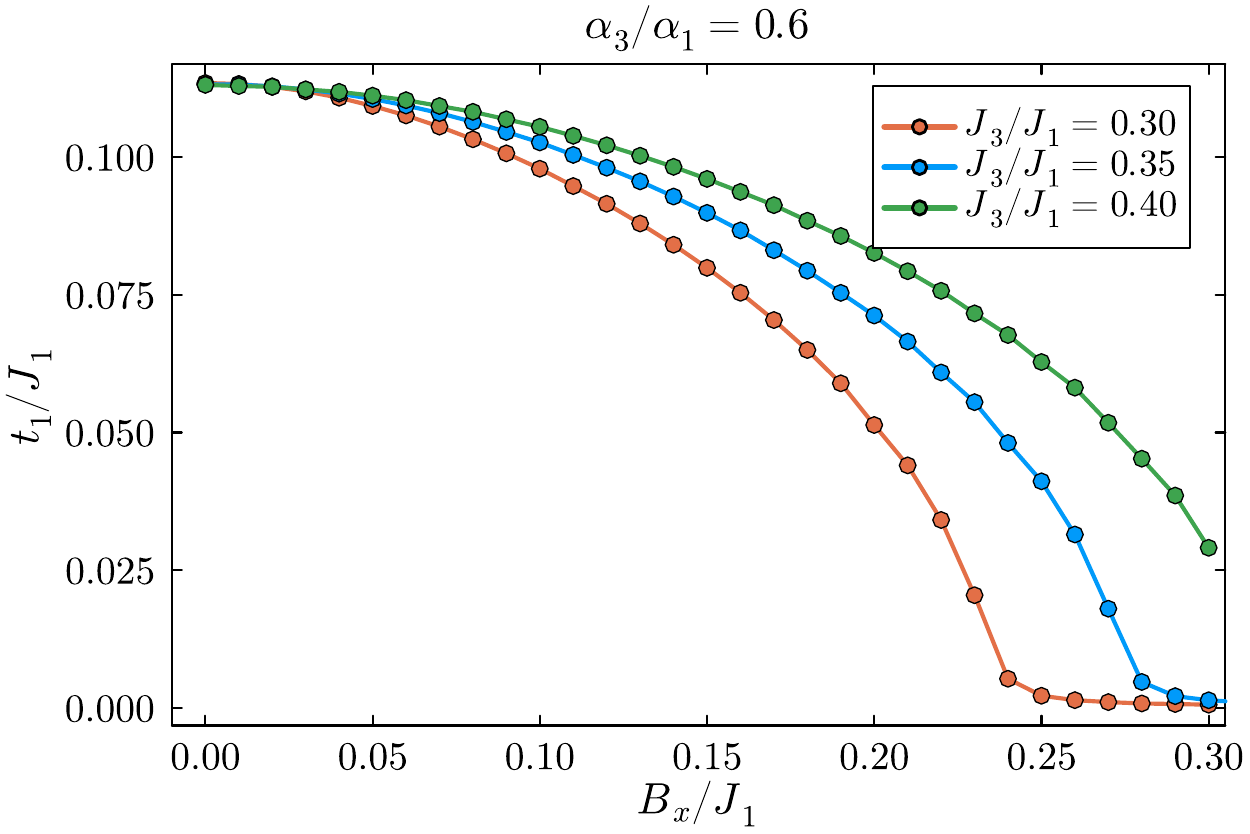}
  \includegraphics[width=0.4\textwidth]{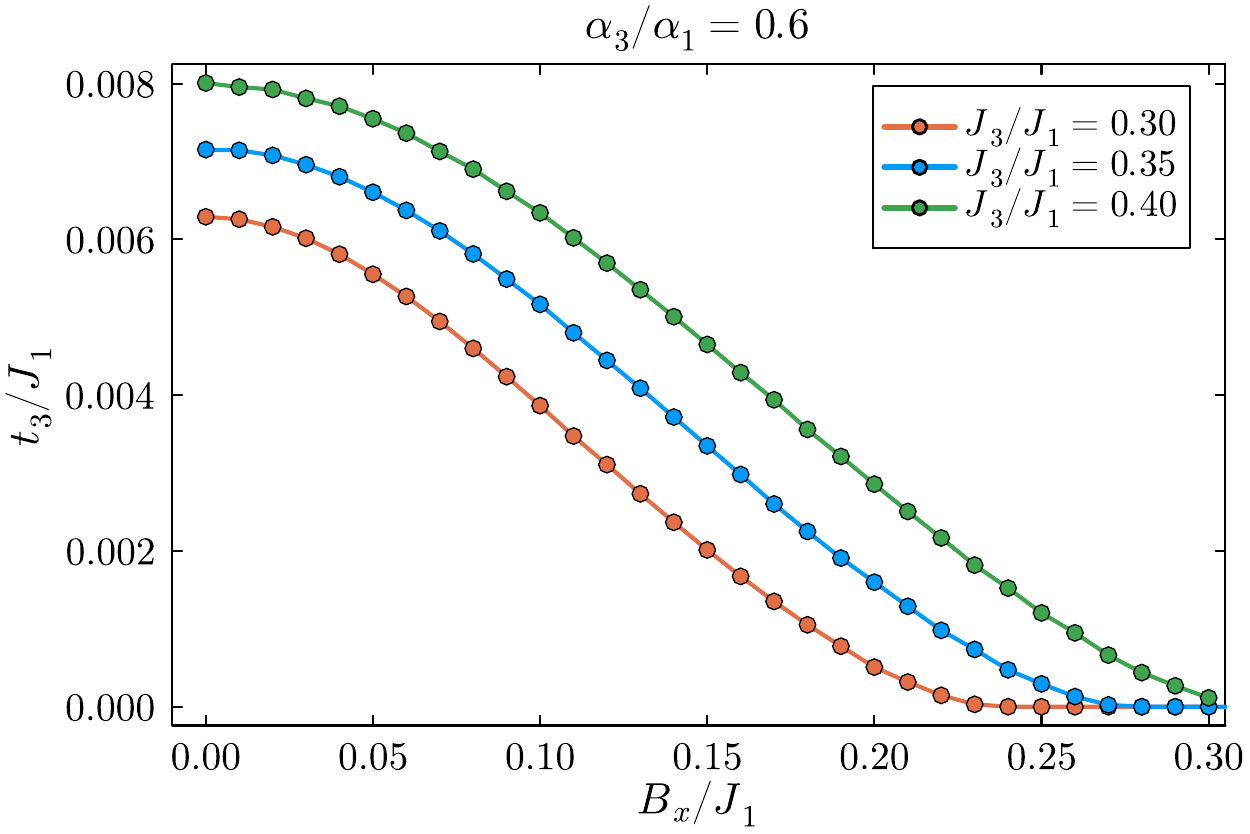}
 \caption{Mean-field hoppings as a function of applied Zeeman field in the intermediate regime near the spiral instability where \(\a_1\approx 0.7\).}
\label{fig:MFT with B field}
\end{figure}
\begin{figure}[!ht]
 \centering
  \includegraphics[width=0.4\textwidth]{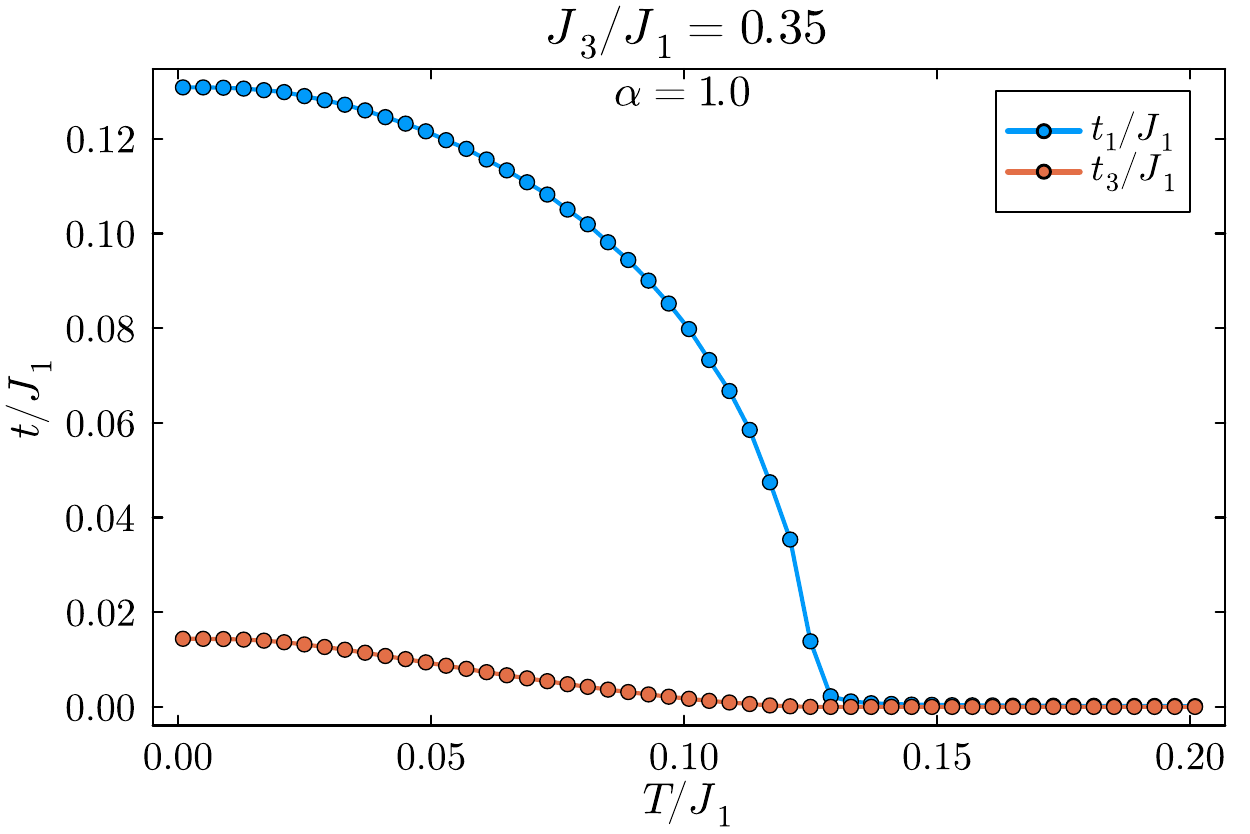}
 \caption{Mean-field hoppings \(t_1\) and \(t_3\) as a function of temperature in the intermediate regime.}
\label{fig:MFT with T}
\end{figure}

The mean-field Dirac state in the presence of an external in-plane Zeeman field \(B_x\) has the following Hamiltonian for the fermionic spinons
\begin{equation}
    \label{dirac Hamiltonian}
    \mh^{f} = -t_1\sum_{\expval{i, j}_1}\left(f^{\dagger}_{i, \a}\s^z_{\a\b}f_{j, \b}+\mathrm{h.c.}\right)-t_3\sum_{\expval{i, j}_3}\left(f^{\dagger}_{i, \a}\s^z_{\a\b}f_{j, \b}+\mathrm{h.c.}\right)-\frac{B_x}{2}\sum_{i} f^{\dagger}_{i, \a}\s^x_{\a\b}f_{i, \b}\,,
\end{equation}
where \(t_1\) and \(t_3\) can be found self-consistently in units of \(J_1\). For the pure quantum limit when \(\a_{ij} = 1\), we find that \(t_1\approx 0.13 J_1\), and \(t_3/t_1\approx 0.1\). The band structure and the density of states are shown in Fig.\ref{fig:DSL bands} both for \(B_x=0\), and \(B_x\neq 0\). We find Dirac nodes at the \(\bK\) and \(\bK'=-\bK\) points in the Brillouin zone, and interestingly, a very flat dispersion along the edge \(\bK\rightarrow\bM\). These low lying flat states lead to a Van-hove type singularity in the density of states at low energies, which persists even in the presence of an in-plane Zeeman field when the Dirac nodes get gapped out. 
The dependence of the hopping scales as a function of applied in-plane Zeeman field is also shown in Fig.\ref{fig:MFT with B field}. Lastly, as a function of temperature, we find the hoppings die off around \(T/J_1\approx 0.13\) as shown in Fig.\ref{fig:MFT with T}. We smoothen this decay and use that when calculating the NMR time \(1/T_1\) as a function of temperature in Fig.\ref{fig:NMR}.

\end{document}